\documentclass[preprint,showpacs,amsmath,amssymb,pre,aps,nofootinbibtwocolumn]{revtex4-1}
\usepackage{amsmath}
\usepackage{amssymb}
\usepackage{graphicx}
\usepackage{dcolumn}
\usepackage{bm}
\usepackage{xcolor}

\def\mean#1{\langle#1\rangle}
\newcommand{\av}[1]{\left\langle #1 \right\rangle}
\newcommand{\bracket}[1]{\left(#1\right)}

\begin{document}
\title{Analysis of aligning active local searchers orbiting around their common home position}

\author{J. Noetel$^1$ and L. Schimansky-Geier$^{1,2}$}
\affiliation{$^1$Institute of Physics, Humboldt University at Berlin,
  Newtonstr. 15, D-12489 Berlin, Germany\\
  $^2$Berlin Bernstein Center for Computational Neuroscience, Humboldt
  University at Berlin, Unter den Linden 6, D-10099 Berlin, Germany }

\begin{center}
\end{center}

\begin{abstract}
We discuss effects of pairwise aligning interactions in an ensemble of central place foragers or of searchers that are connected to a common home. In a wider sense, we also consider self moving entities that are attracted to a central place such as, for instance, the zooplankton Daphnia being attracted to a beam of light. Single foragers move with constant speed 
due to some propulsive mechanism. They explore at random loops the space around and return rhytmically to their home. In the ensemble, the direction of the velocity of a searcher is aligned to the motion of its neighbors. At first, we perform simulations 
of this ensemble and find a cooperative behavior of the entities. Above an over-critical interaction strength the trajectories of the searcher qualitatively changes and searchers start to move along circles around the home position. Thereby, all searchers rotate either clockwise or anticlockwise around the central home position as it was reported for the zooplankton Daphnia. At second, the computational findings are analytically explained by the formulation of transport
equations outgoing from the nonlinear mean field Fokker-Planck
equation of the considered situation. In the asymptotic stationary
limit, we find expressions for the critical interaction strength, the mean radial and orbital velocities of the searchers and their velocity variances. We also obtain the marginal spatial
and angular densities in the under-critical regime where the foragers  behave 
like individuals as well as in the over-critical regime where they rotate collectively around the considered home. We additionally elaborate the overdamped Smoluchowski-limit for the ensemble.
\end{abstract}
\maketitle

\section{Introduction}
\label{sec:extension_coll}
Some living entities with a home, a den or a nest perform local search around this localization playing the role of a hub in their 
life \cite{Klages}. We will call the central place "home" and the local search "homing" in agreement with \cite{Mittelstaedt}. In a broader 
concept, we also understand other central places as food sources, patches or hiding places, 
also attractive light sources, etc. as a home. Such searchers are often refereed to as central place foragers.

Animals leave often these points in order to inspect the environment but return rhythmically to this location. Such behavior is called local 
search in contrast to extended global search \cite{Bennichou2011}. It is found for foragers which are able to keep track of distances 
and orientations as they move and use these information to calculate their current position in relation to their home. If in a 2-dimensional 
landscape the distance and angle towards the home are known, the method is called path integration
\cite{Klages,Mittelstaedt,Cheng,Wang}. The specific exploration behavior might be based on some internal storage mechanism \cite{Seelig,Green}
or external cues \cite{Zeil} as for example special points of interest or pheromonic traces etc.

Research on this topic is often aligned on investigations concerned with ants, bees and flies  
\cite{Wehner_1981,Ronnacher2008,Collett2013,ElJundi_2017,Kim_Dickinson_2017}. For these animals and possibly 
for many others too, homing behavior is found during foraging. Especially, the various structured spatial patterns at which 
the entities move during their search and return have received a lot of interest and have been 
computed\cite{Wehner_et_al_1996,Vickerstaff_2005,Vickerstaff_2012,Waldner2018}. 
Typically such trajectories begin 
at the home. The motion is then comprised of loops that start and end at the target.

One might add, that a deeper insight of homing or local search gains an increasing technical significance. New ways of spatial exploration \cite{chien_2017} are developed in which robots are supposed to reach places which human beings have never approached \cite{Hook_2013,Girdhar_2011,Leonard_et_al_2007,Dubowsky_2005}. Homing 
behavior and local search are of high relevance to autonomous systems of surveillance, data collection, exploration, monitoring, etc. 
\cite{Leonard_et_al_2007,Duarte_et_al_2016} using internal path integration \cite{Nirmal,Moeller}. 
Especially, the development of devices at lower technical level might benefit from the knowledge how simpler organism operate.  

Recently, we introduced a stochastic minimal model for local search \cite{Noetel_2018,Noetel_2018c} for single searchers with constant
speed. During local search a forager, like a fly\cite{Kim_Dickinson_2017, ElJundi_2017}, or an ant \cite{Wehner_1981} explores 
the neighborhood of the given home. The position of the home defines the central location and the symmetry of the problem. 
Throughout the paper it will be  located at the origin of the coordinate system, for simplicity. In a wider sense such a home 
could also be the nest of the objects, it might be a target, or another food source or interesting 
place as, for example, a localized light source. Already
Okubo \cite{Okubo1986} in the $80$ies investigated swarms of midges. 
There the single animal of a group of males moves likewise being attracted by the common center of the swarm 
and escaping permanently after stochastic epochs. Further work on midges compared those swarms \cite{Gorbonos2016} with self-gravity 
and a possible velocity dependence of the central force was found \cite{Reynolds2017,Reynolds2018}, see also \cite{Schweitzer1994,Chavanis2014}.

In this work, we will be interested in general theoretical mechanisms of a cooperative behavior of the active searchers. We will 
investigate the dynamics of an ensemble of $N$ interacting searchers. The searchers will share their common home, or as explained 
above, a common target, points of common interest or other objects mimicking a home position for the moving units. 

Such swarming with a central place and moving aligning searchers around has many possible applications. 
We mentioned already the midgets studied by Okubo \cite{Okubo1986}. A broad variety of animal motions with this symmetry is 
listed in \cite{Delcourt}, for example. We add here the experiments on Daphnia Magna, a zooplankton about $3-5mm$, moving 
with speed $v_0=4-16mm/s$ and being attracted by a light shaft oriented vertically through the water wherein the Daphnia moves 
\cite{Ordemann2002,Ordemann,Ordemann2003Nova,ErdmannDaphnia}, see also \cite{Garcia_daphnia,Dees2008}. These 
water fleas create a massive unidirectionally  rotating swarm around the light shaft, as a result of the attraction, 
the motion and the interaction. The direction of the rotation depends on initial conditions and might change through external or 
internal perturbations.

This situation was multiple times simulated with moving and interacting particles 
\cite{Ordemann,Erdmann2003,ErdmannDaphnia,Vollmer,Mach_Schweitzer2007}. The simulations prove the selection of one rotation direction 
demonstrating the survival of a single peak in the distribution of angular momentum of the animals. 
Here, we discuss for a generic model the transition for single particle motion to collective motion through explicitly considering
interactions between searchers. Through the interactions individual trajectories qualitatively change. 

Different kind of interactions have been applied for this situation: short range aligning correlations \cite{Ordemann}, 
Morse like potentials \cite{ErdmannDaphnia} (see also\cite{Levine2001,Strefler2008,Thouma}), 
hydrodynamics Oseen interactions \cite{Erdmann2003,ErdmannDaphnia}, a Lennard-Jones potential \cite{Vollmer}, 
and avoidance potentials \cite{Mach_Schweitzer2007}. 

Here we introduce a local alignment often used in swarming models. In fact, many of the just mentioned different interactions have such 
aligning effect. However, as reported explicitly in the experimental studies on the Daphnia Magna \cite{Ordemann,ErdmannDaphnia}, 
the common motion of the animals induces a drag force of the surrounding fluid acting on the neighbors. For sufficiently high 
concentration of Daphnia, the water moves in the same direction as the animals. This experimental fact justifies the introduction of 
an mere alignment.

In detail, we assume the existence of a sensing radius (green area in
Fig. \ref{fig:schematic}). Inside this region  the heading directions of searchers shall synchronize. Such interaction was proposed 
in the seminal book by Okubo and Levin as an "arrayal force" \cite{Okubo2002} being one of the fundamental forces in the dynamics 
of animal groups. Such an alignment is frequently used in models of active Brownian particles to model swarming behavior 
\cite{Vicsek1995,Chate2008a,Chate2008b,Peruani2008,Romanczuk,Vicsek_2012,vortex} and 
in models of synchronization \cite{Kuramoto_1984,Pikovsky_et_al_2001}. It appears also for self-propelled particles at 
the micro-scale via hydrodynamic interactions \cite{Lauga2009,Stark,Mazza2018,Mazza2019}.  

We will investigate the thermodynamic limit, meaning large particle numbers and vanishing sensing radius. 
The limit indicates a break of ergodicity and asymptotic states will depend on initial configurations.
A few remarks concerning finite size effects are given in Appendix \ref{sec:limits}. The coupling strength should 
be moderate to avoid density instabilities that might break the circular symmetry. We study this system through simulations 
and through an analytical approach based on the kinetics of the probability density function (pdf) and its moments \citep{Romanczuk}. 
We will be mainly interested in the long time asymptotic state. We find the stationary marginal spatial particle density for the 
distance of the searchers from their home and the density of their orbital direction. From transport equations we obtain the averaged 
radial velocity and the orbital velocity as well as their variances.  

We give analytical expressions for the steady state of those quantities and find a phase
transition of the second kind for the orbital velocity, depending on the coupling strength of the alignment.  For small alignment the 
searchers will practically follow trajectories of the single particle model, while for large enough alignment the searchers will start 
rotating around the home. We compare the simulation results with the analytical expressions and find good agreement between the 
numerical and analytical results.

\section{Escape and return dynamics of an isolated searcher}
\label{sec:Model} 
\subsection{Path integration}
\begin{figure}
  \includegraphics[width=0.5\linewidth]{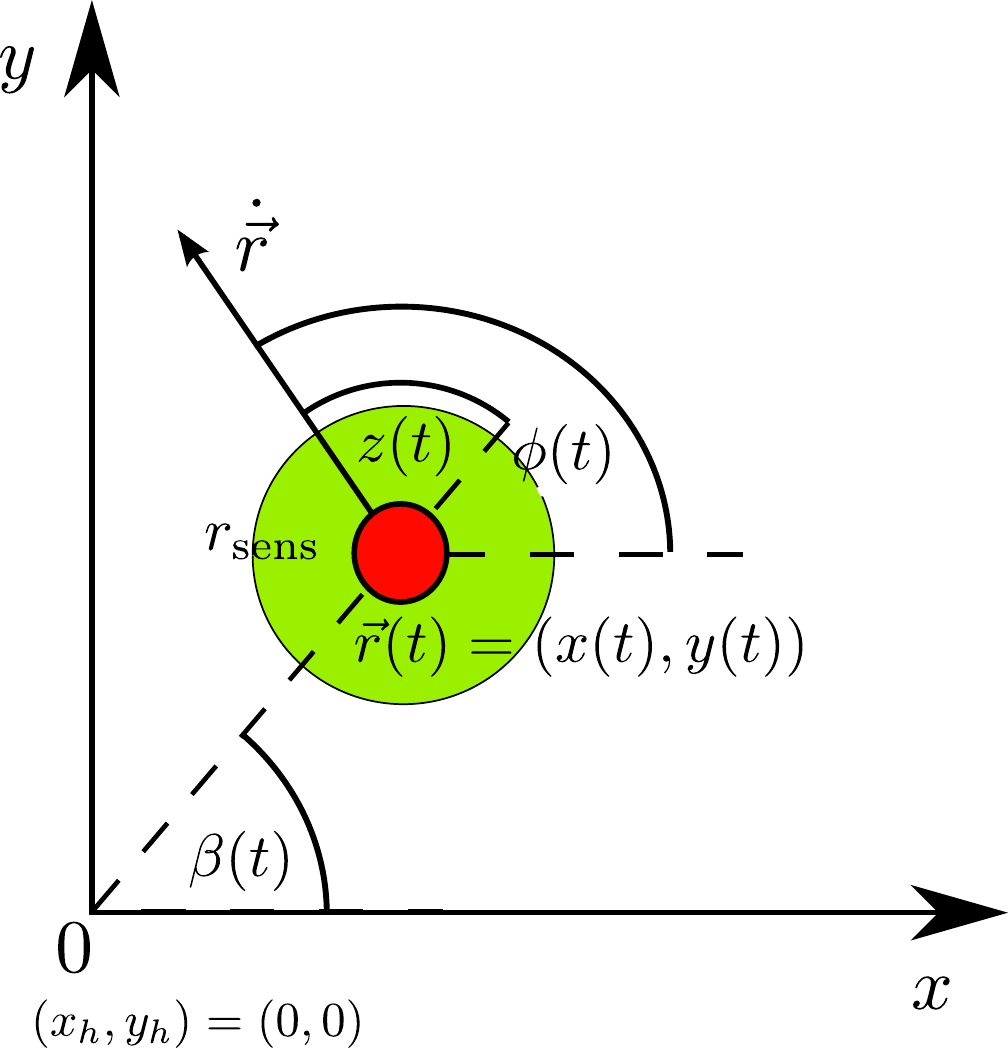}
    \caption{Schematic representation of coordinates and angles for
      the active searcher (red dot) in two dimensions. The position
      vector is $\vec{r}(t)$ with orientation $\beta(t)$, the velocity
      is $\vec{v}(t)=\dot{\vec{r}}(t)$. Its direction is the heading
      $\phi(t)$ and the angle $z(t)=\phi-\beta$ indicates the
      difference between the velocity and position vectors. The home is located at the origin and the green
      area stands for the interaction zone of the considered particle
      with surrounding particles.}
    \label{fig:schematic}
\end{figure}

The  model of the single searcher considers an active particle with position vector $\vec{r}(t)$ and moving with constant speed $v_0$ 
in two dimensions. Thus by definition it holds
\begin{eqnarray} 
\dot{\vec{r}}=\vec{v}(t)=v_0
\begin{pmatrix}
         \cos\phi(t) \\ \sin\phi(t)
\end{pmatrix}\,,
\label{eq:r_dot}
\end{eqnarray}
and $\phi(t)$ is the heading angle of the particle pointing
along the velocity vector as is depicted in Fig. \ref{fig:schematic}. As consequence of the constant speed the kinetic energy is 
conserved as it was proposed also in other models \cite{Mittelstaedt,Wehner_1981,Gallistel1990,Vickerstaff_2005}. The proposed
deterministic dynamics is similar to the Mittelstaedt bicomponent model \cite{Mittelstaedt_bico,Vickerstaff_2005}.

The main ingredient of the kinematics of the searcher is the dynamics
of the heading direction standing for the decision making step of the
searcher to choose a new direction.  In
\cite{Noetel_2018,Noetel_2018c}, we assumed
\begin{equation}
\dot{\phi} =  \kappa \sin(\phi-\beta) + \frac{\sigma}{v_0} \xi(t).
\label{eq:dottheta0}
\end{equation}
Therein, $\beta(t) \in[0,2\pi)$ is the direction of the position
vector in polar coordinates $\vec{r}(t)=r(t) \{\cos(\beta(t)),\sin(\beta(t))\}$ 
with $r(t)$ being the distance to the home. With respect to the Cartesian coordinates 
the angle is linked as
\begin{equation}
\beta(t) = \arctan{\frac{y(t)}{x(t)}}\,. 
\label{eq:beta}
\end{equation}
The first item of the r.h.s. of the heading dynamics \eqref{eq:dottheta0}
defines a search and return mechanism. If the searcher moves
outwardly, the directions of the position and the velocity vectors
differ less than $|\phi-\beta|\leq \pi/2$. The heading dynamics
predicts a repulsion of the heading direction from the position vector. 
It describes the search phase of the particle and searchers look around. Alternatively, 
returning to the home the two
angles behave as $|\phi-\beta|\geq \pi/2$. The heading dynamics
\eqref{eq:dottheta0} leads to an anti-alignment of the heading vector with the position vector. 
As a result, the searcher will return to its home.

In the last term of the heading dynamics, $\xi(t)$ stands for a source
of white noise with strength $\sigma$.  It grants the prediction of the new direction with an uncertainty at time $t$ 
and might be originated by a limited knowledge of the searcher concerning the precise relations between the two 
angles $\beta$ and $\phi$. In the previous work we took $\alpha$-stable Lev{\'y} noise modeling large rapid changes in the 
orientation of the velocity vector as reported for the fruit fly \cite{Kim_Dickinson_2017}. In the current study we fix $\alpha=2$ 
which makes $\xi(t)$ to Gaussian white noise changing the heading more smoothly in a diffusive manner. 

Our proposed model is strictly phenomenological. It means that we obtain the trajectories without having a neurophysiological 
manifestation and relations to external or internal cues. Hence, our model does not answer the questions regarding the reasons 
for the decisions made during the search. This decision finding \eqref{eq:dottheta0} and the path integration \eqref{eq:r_dot} is 
formulated as a stochastic dynamics which is a kind of instantaneous dead reckoning or vector handling based on an interaction between 
two directions. Knowing the current orientations of the position $\beta(t)$ and heading $\phi(t)$ vectors, the future motion is defined 
by calculating the new heading  $\phi(t+\delta t)$. How the two directions have been recovered at the given time $t$ remains open 
in the model. Whether this information is achieved through a earlier training, orientation flights or by internal counting processes 
is not answered. More detailed discussions on possible storage of spatial information and consequences for the kinematics of 
the searcher, the role of external cues, rewarding and different food sources is part of a larger 
literature \cite{Wehner_1981,Gallistel1990,Fourcassie1994,Freska1999,Vickerstaff2010,Collett2013}.

Otherwise, small variations of the basic dynamics \eqref{eq:r_dot} and \eqref{eq:dottheta0} might model more complex 
behavior such as is documented in the large biological literature. The enlargement of the excursions in case of an unsuccessful 
search \cite{Wehner_1981,Hoffmann1983} might be modeled by weakly lowering the coupling to the home, by decreasing slowly $\kappa$ 
as time elapses. In contrast, 
a straight trajectory directly to the nest \cite{Wehner2002Cal,Collett2010} could be caused by a sudden switch 
to high values of $\kappa$. Similar modifications would allow to describe the change from an Archimedian search to the random loop. 
Narrow hairpin-like shaped excursions as observed for foraging search of honeybees and bumblebees \cite{Capaldi2000,Osborne2013} 
could be reflected by considering $\kappa$ dependent on the distance and an increase of the velocity. More difficult is the modeling 
of shifts of position of the searcher creating multiple fictive homes \cite{Wehner2002Cal,Reynolds2007b} as well as a scale free 
flights of bees \cite{Reynolds2007}\cite{Reynolds2007b,Klages2013}. Such detailed biological analysis of our search and return model 
might be the contents of a future work in a biological journal.

\subsection{Resume of dynamical properties of the model}
The analytical tractability of the model becomes visible if transforming to a polar presentation.
Let $r(t)=\sqrt{x^2(t)+y^2(t)}$ be the distance from the origin being the localization of the home. Further on, we 
introduce the difference between the two directions of the position and heading vector as $z(t)=\phi(t)-\beta(t)$. The third variable 
is the direction $\beta(t)$ as in Eq. \eqref{eq:beta}. For these new variables the equations of motion read
\begin{eqnarray}
&&\dot{r}\,=\,v_0 \cos(z)\,,\nonumber\\
&&\dot{z}\,=\, \bracket{\kappa - \frac{v_0}{r}} \sin(z)\,+\, \frac{\sigma}{v_0} \,\xi(t)\,,\nonumber\\
&&\dot{\beta}\,=\,\frac{v_0}{r}\sin(z)\,.
\label{eq:polar_single}
\end{eqnarray}
It is immediately seen that the $\beta(t)$ variable separates from the two others. 
The last equation in \eqref{eq:polar_single} can be integrated after having defined the couple $r(t)$ and $z(t)$. 

\begin{figure}
  \includegraphics[width=0.49\linewidth]{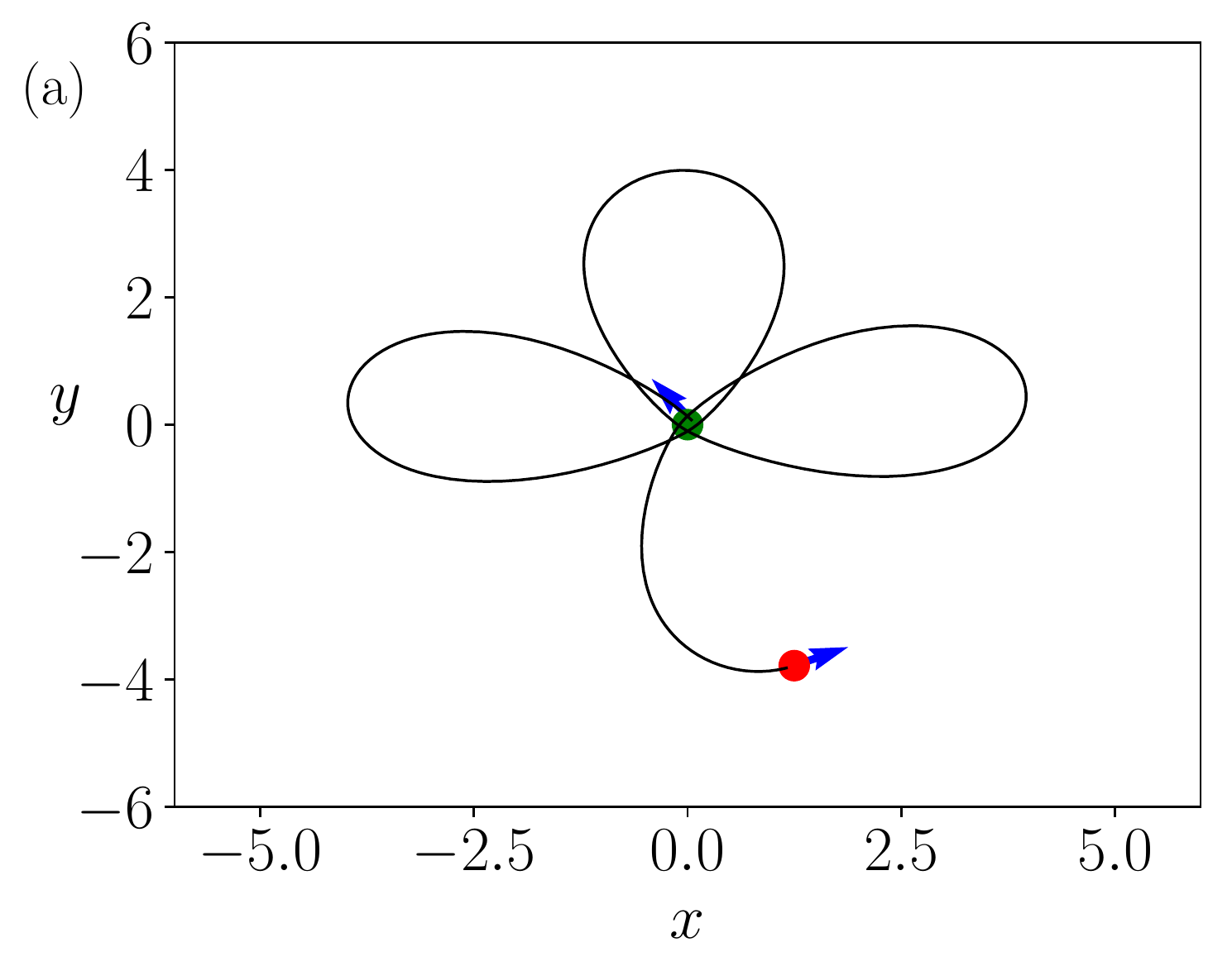}~~~~\includegraphics[width=0.49\linewidth]{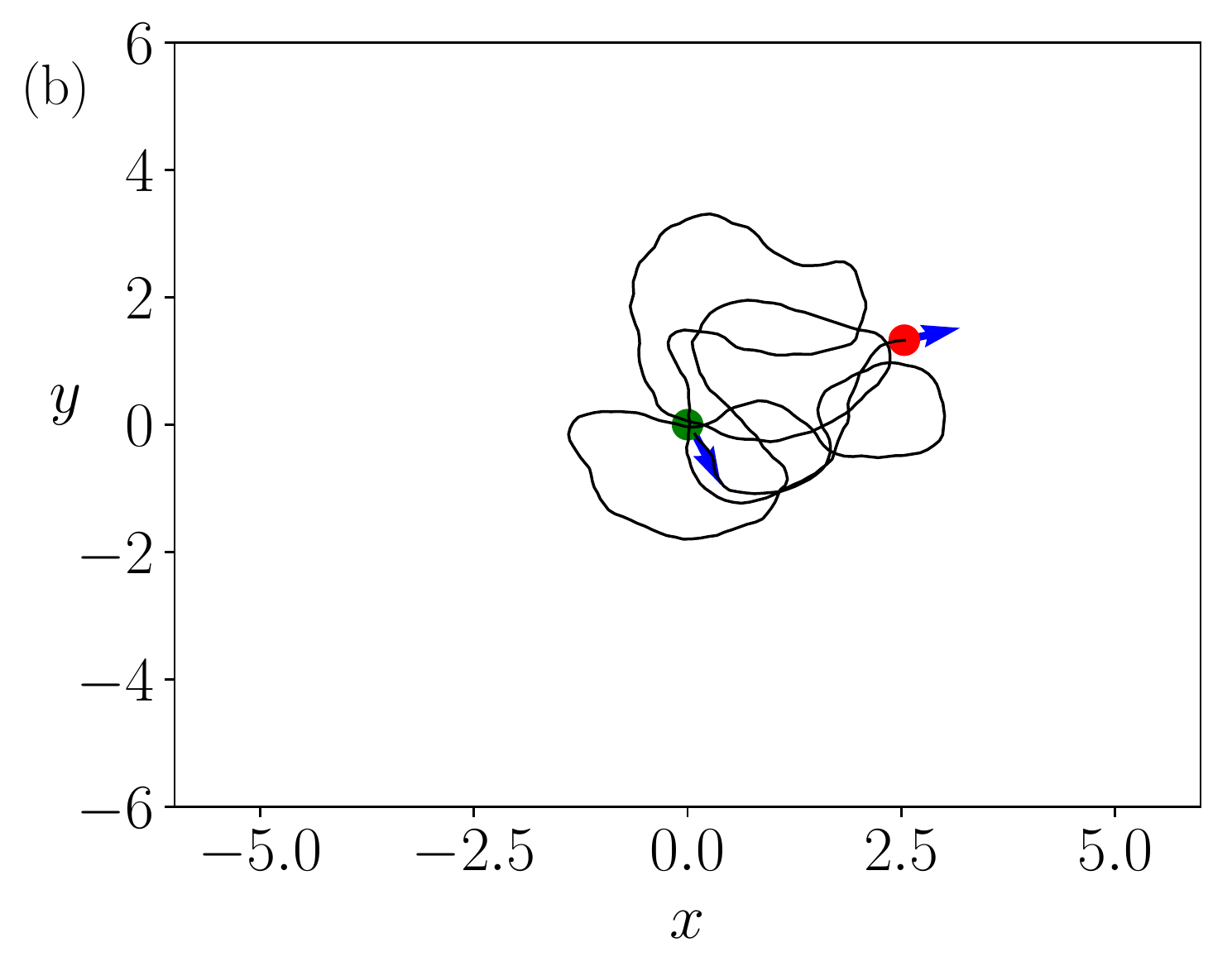}
  \caption{Sample paths of active searchers originated by Eqs.\eqref{eq:r_dot} and \eqref{eq:dottheta0}. 
  (a): The deterministic case without noise $\sigma^2=0$. 
  The dynamics exhibits trajectories which periodically return to the home. 
  The shape is similar to a rosette. 
  (b): The stochastic dynamics with Gaussian noise and $\sigma^2=0.1$ creates random loops around the home. 
  Sample paths also return after finite times to the home. Parameters: $v_0=1$, $\kappa=1$.  }
  \label{fig:sample_pathes}
\end{figure}

Remarkably, the two equations for $r,z$ define without noise $\sigma=0$ a deterministic oscillatory motion in the $(r,z)$ space. 
Stationary fixed points of the dynamics are saddles or centers as it was reported in \cite{Noetel_2018}. Trajectories 
perform an oscillatory loops with alternating search and return epochs. In Fig. \ref{fig:sample_pathes} we present 
deterministic and stochastic trajectories as computed from Eqs.\eqref{eq:r_dot} and \eqref{eq:dottheta0}. 
The deterministic dynamics in the $(x,y)$ space  creates trajectories similar to a rosette with leaves, 
called also random loops. The searchers 
return periodically to the nest. The leaves are located around the center with a fixed precession of the perihelion.
Such trajectories reflect the stereotypical behavior of central place foragers, which start at the home and then 
move in loops from and back to the home.
In \cite{Noetel_2018} we have calculated the period as well as the precession. We mention, that a similar model was introduced 
by Waldner and Merkle\cite{Waldner2018} who computed similar trajectories.  
  
With noise the leaves become random loops. The stochastic sample path also returns permanently to the vicinity of the home 
in finite times. Inspection of the autocorrelation function of the distance exhibits a behavior of a damped oscillation with characteristic 
time $\tau_{\phi}$ from \eqref{eq:time_relax}. Such random loops have been reported  for the fruit fly \cite{Kim_Dickinson_2017} and 
the dessert ants \cite{Wehner_1981} and isopods \cite{Hoffmann1983}. Also it was reported, that foraging honeybees 
\cite{Capaldi2000,Collett2000} and bumblebees \cite{Osborne2013,Makinson2018} move along random loops.

Without noise the dynamics is conservative, meaning there exist an integral of motion $X$ which ins constant along the deterministic trajectories $r(t),z(t)$ with initial states $r_0$ and $z_0$.  It reads
\begin{equation}
X(r,z)\,=\,v_0 \sin(z) \exp \bracket{-\frac{\kappa}{v_0}r}\,=\,X(r_0,z_0)\,.
\label{eq:integral}
\end{equation}
Since the sign of the integral $X$ is also conserved, there is no trajectory which might cross the straight lines $z=0$ or $z=\pm\pi$ in the $r,z$-space. This means that the deterministic angular momentum
\begin{equation}
L(r,z)\,=\,r^2\,\dot{\beta}=v_0\,r\,\sin(z)\,=\,v_0\,\exp\left(\frac{\kappa}{v_0} r\right) X(r,z),
\label{eq:L}
\end{equation}
along a trajectory never changes its sign. In the deterministic model a single particle either moves in a clockwise ($z>0$) or a 
counterclockwise $(z<0)$ fashion around the home in the Cartesian coordinate system. 
This behavior of single searchers is in agreement with the observations  for the Daphnia Magna 
\cite{Ordemann,ErdmannDaphnia,Mach_Schweitzer2007}. Attracted by the light the singular Daphnia moves persistently a certain number 
of orbits clockwise or anti-clockwise before random influences induce a switch of the rotation sense.

\begin{figure}[h]
    \includegraphics[width=0.4\linewidth]{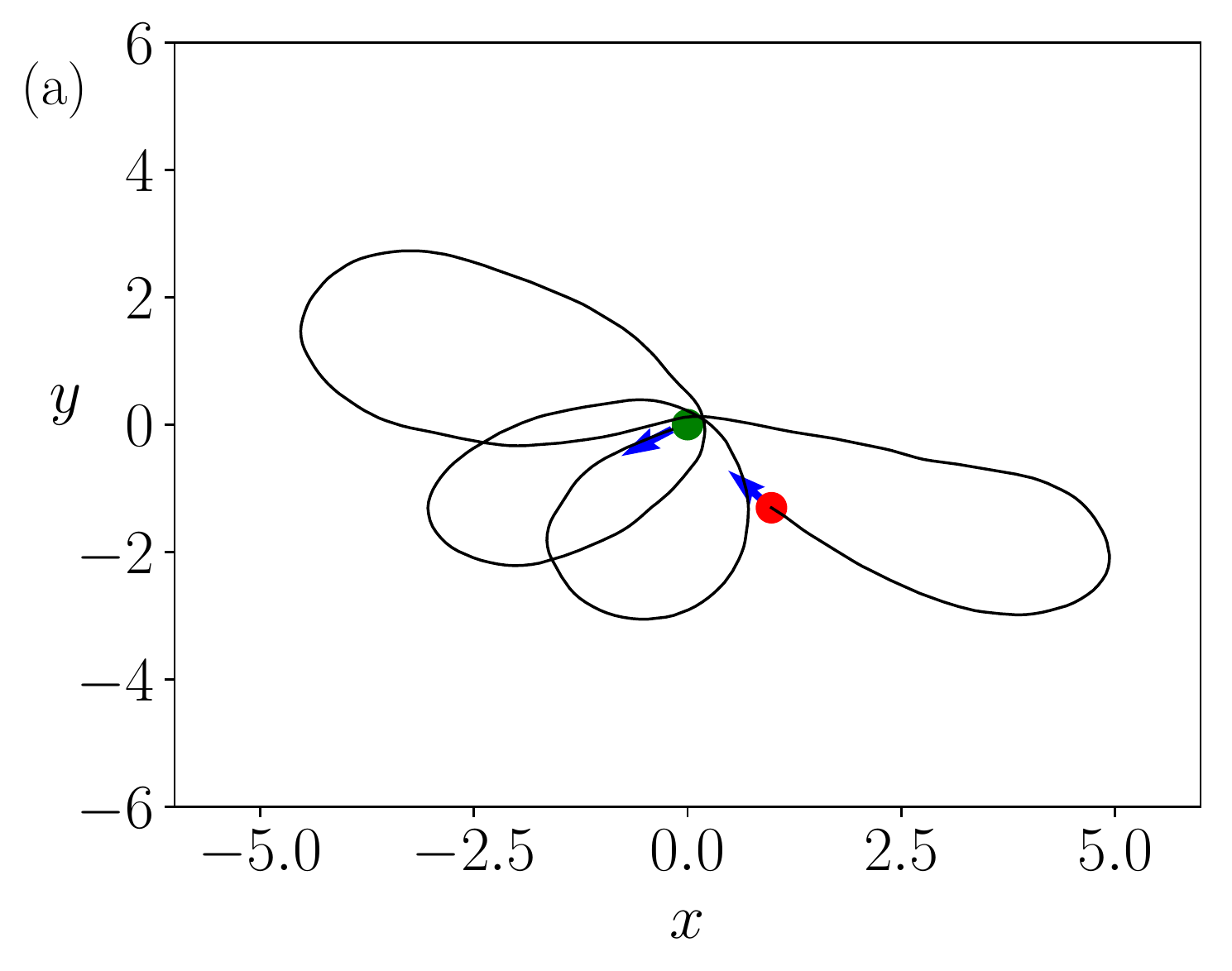}
    \includegraphics[width=0.4\linewidth]{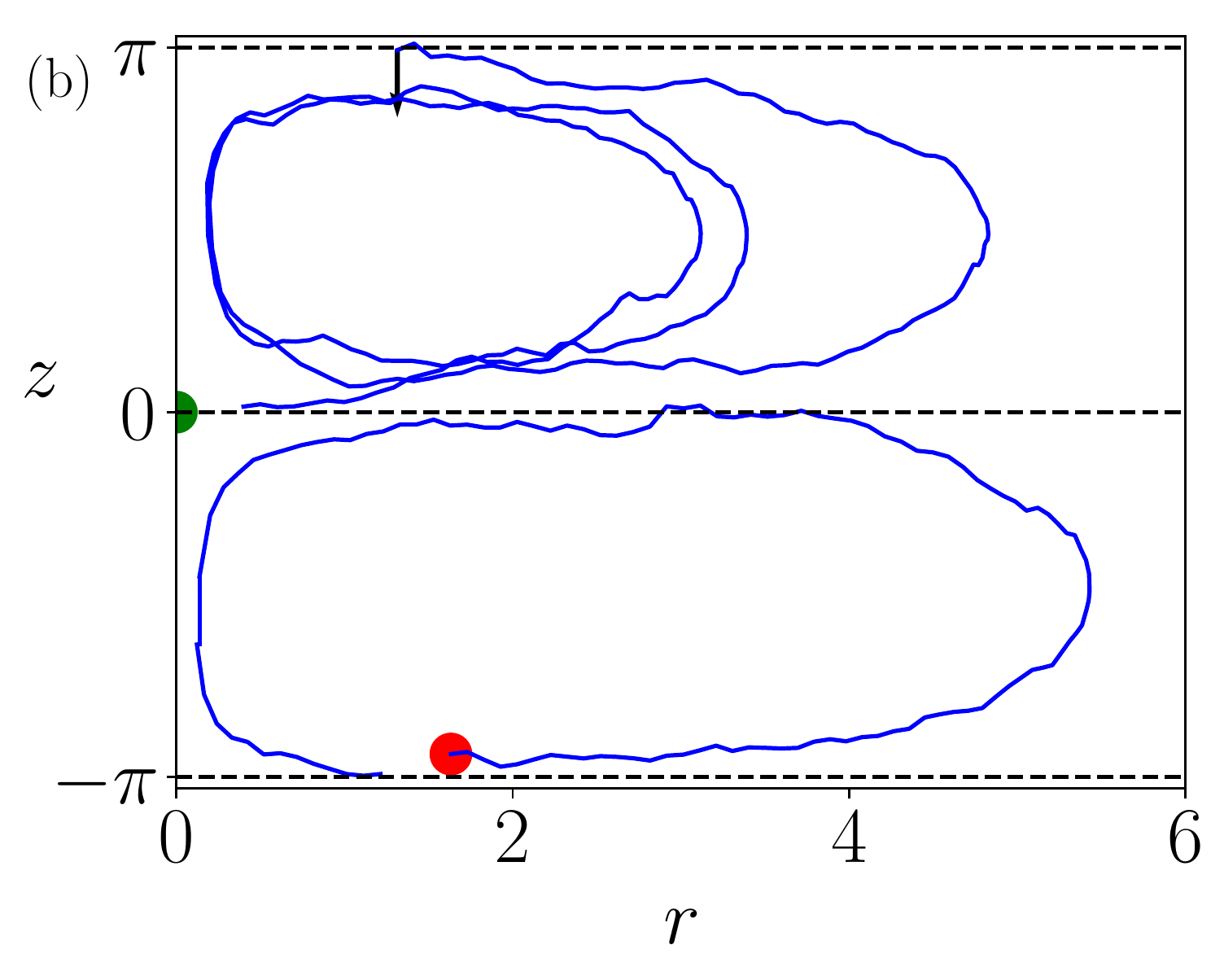}
    \caption{Sample trajectories with small noise $\sigma^2=0.01$.
      (a): Trajectory in the $(x,y)$ plane. Green dot marks the home.
      The particle starts close to the origin (green
      dot). (b): The corresponding motion in the $(r,z)$ plane. The
      crossing of the $z= \pm \pi$ line (indicated by
      the arrow) changes the sign of the angular momentum. It is
      caused by the action of the noise in the $z$ dynamics.  Other
      Parameters: $v_0=1$, $\kappa=1$.}
    \label{fig:plane_lownoise}
\end{figure}

The existence of the noise in the model makes the full dynamics \eqref{eq:r_dot} and \eqref{eq:dottheta0} irreversible. 
There exist a relaxation time
\begin{equation}
\tau_\phi\,=\,\bracket{\frac{v_0}{\sigma}}^2\,.
\label{eq:time_relax}
\end{equation}
After $\tau_\phi$ all angular dynamics becomes uniform and
the integral of motion is damped out. At this time scale the
noise also changes the directions of rotation. This behavior is
illustrated in Fig. \ref{fig:plane_lownoise} wherein we show a
sample trajectory in the $(x,y)$ plane as well as in the $r,z$
one. The particle starts at the home, marked as green dot at
$(x,y)=(0,0)$. A time frame of $\Delta t=30$ is shown. The trajectory
is under the influence of rather continuous random disturbances
which is the Gaussian case. We see that the noise creates transitions
between the two half-planes in the $(r,z)$ dynamics. At values $z=0$ and
$z=\pm \pi$ the deterministic part in the $z$-dynamics \eqref{eq:polar_single}
vanishes. However, the noise can cause transitions to
$z$ values with different sign corresponding to a change of the
rotation mode.

As follows from Eq. \eqref{eq:L}, the angular momentum is linked
directly to the integral of motion $X$. With noise this integral
becomes stochastic, i.e. $X_\xi(t)$ and the angular momentum
$L_\xi(t)$, as well. In \cite{Noetel_2018} we showed that in case with
noise the mean value of the integral $\av{X_\xi(t)}_{X_0}$ conditioned
to a certain initial condition $X_0$ decays after the time
$\tau_\phi$. Consequently, we might conclude that the average angular
momentum will vanish as well which reflects the permanent switches of
the sign of $L_{\xi}(t)$.

With noise in the angular dynamics the spatial dynamics of a single searcher
  becomes stationary. Initial conditions $x_0,y_0,\phi_0$ are lost and
  the corresponding pdf $P(x,y,\phi| x_0,y_0,\phi_0,t)$ approaches for
  $t\to \infty$ a stationary shape $P(x,y,\phi)$ being the unique
  asymptotic attractor of the stochastic dynamics. Since after
  $\tau_\phi$ the angles $\phi$ and $\beta$ are equi-distributed in
  $[0,2\pi]$, the further evolution of the system takes place in
  space. This overdamped evolution can be characterized by the time
at which the particle spreads by diffusion over the characteristic
length $r_c=v_0/\kappa$ of this problem, i.e.
\begin{equation}
\tau_r\,=\,\frac{\sigma^2}{v_0^2\kappa^2}\,.
\label{eq:time_diff}
\end{equation}
Therein we have used as diffusion coefficient $D_{\text{eff}}=
v_0^2\tau_\phi/2$ known for an active micro-swimmer with constant speed
\citep{mikhailov,Noetel:2017}. In case of coupling, this diffusion coefficient changes as can seen below (see \ref{sec:mathe4}).

Both time scales \eqref{eq:time_relax} and \eqref{eq:time_diff} are determined by the value of the noise 
intensity but scale differently with $\sigma^2$. This different scaling causes an optimal noise for the mean time 
which is needed to find new food source in the neighborhood of the home as was reported earlier \cite{Noetel_2018}. 

In \cite{Noetel_2018} we also derived the Smoluchowski-equation for the marginal spatial density $\rho(x,y,t)$. 
Its stationary solution was found to be a exponential function
\begin{equation}
\rho(x,y)=\left(\frac{\kappa}{v_0}\right)^2 \exp\bracket{-\frac{\kappa}{v_0}\,\sqrt{x^2+y^2}}
\label{eq:marg_density}
\end{equation} 
This marginal density agrees pretty well with experimental data from the fruit fly \cite{Kim_Dickinson_2017}. 
Surprisingly, the marginal spatial density function to find a particle at a specific point $(x,y)$ is independent of 
the noise strength $\sigma$.

\section{Ensembles of coupled active searchers}
\subsection{The model of $N$ active searchers with alignment}
The interacting $N$ searchers will move in two dimensions with
constant speed $v_0$. Their common home is situated at
$(x_h,y_h)=(0,0)$. The position vectors are given by
$\vec{r}_i(t)=(x_i(t),y_i(t)),\, i=1,2,\ldots,N$. The direction of
motion, the heading direction, is given by the angle $\phi_i(t)$. A
schematic representation of the coordinates of a single particle is
shown in Fig.\ref{fig:schematic}.

The equations of motion for the position are:
\begin{eqnarray}
  \dot{x}_i &&= v_0 \cos(\phi_i(t))\,\nonumber\\
  \dot{y}_i &&= v_0 \sin(\phi_i(t))\,.
  \label{eq:xy_dot_coll}
\end{eqnarray}
The local search and the alignment dynamics happen in the time evolution
of the heading direction:
\begin{equation}
  \dot{\phi}_i = -\kappa \sin(\beta_{i} - \phi_i)+\frac{\mu}{N_i}\sum_{j \in \Omega_i}\sin(\phi_j-\phi_i)
  + \frac{\sigma}{v_0} \xi_i(t)\,.
  \label{eq:phi_dot_coll}
\end{equation}
The first term on the right hand side causes local search
\cite{Noetel_2018,Noetel_2018c} as explained above.  There $\kappa$ is
the common coupling strength towards the home and the
$\beta_i(t)=\arctan(y_i/x_i)$ stands again for the position angle.  

The second term on the right hand side is the new alignment interaction
between neighboring searchers. For simplicity, we assume a
Kuramoto-like or Vicsek-like interaction of the heading directions of
the searchers \cite{Kuramoto_1984,Vicsek1995,Romanczuk}. With the assumed constant speed, 
it coincides with the arrayal force defined in \cite{Okubo2002}. The sum is taken 
over $\Omega_i$ which lists the current numbers $j$ of particles inside the 
sensing radius of particle $i$, i.e. with distances $r_{ij}=\sqrt{(x_i-x_j)^2+(y_i-y_j)^2}$ 
that obey $r_{ij} \leqslant r_\text{sens}$. The $N_i$ is the overall number of particles 
inside the sensing radius, respectively, the number of items in the list $\Omega_i$.  

The coupling strength for the alignment is given
by $\mu>0$.  We divide the coupling strength $\mu$ by the number of
neighbors to restrict the influence of the alignment on the overall
motion.  The searcher balances now the wish of coupling towards the
home and alignment with its neighbor. The particles maintain in
average a non vanishing distance as noise is present in the model and tends to
uniform the particles, although we do not consider explicitly
repulsion between particles. As already noticed, the couplings strength $\mu$ should 
be moderate to avoid density instabilities that might break the circular symmetry. 
In particular, we consider always $\mu \leqslant \kappa$.

The third term on the right hand side
represents Gaussian white noise with $\langle \xi_i(t) \rangle=0$ and
$\langle \xi_i(t)\xi_j(t^\prime)
\rangle=2\delta_{ij}\delta(t-t^\prime)$. It describes an uncertainty
in the decision of the new direction. Noise sources of different
particles are considered to be uncorrelated. The parameter $\sigma$
denotes an unique noise strength for all particles.

\subsection{Numerical results of aligning active searchers}
\begin{figure}
  \includegraphics[width=0.49\linewidth]{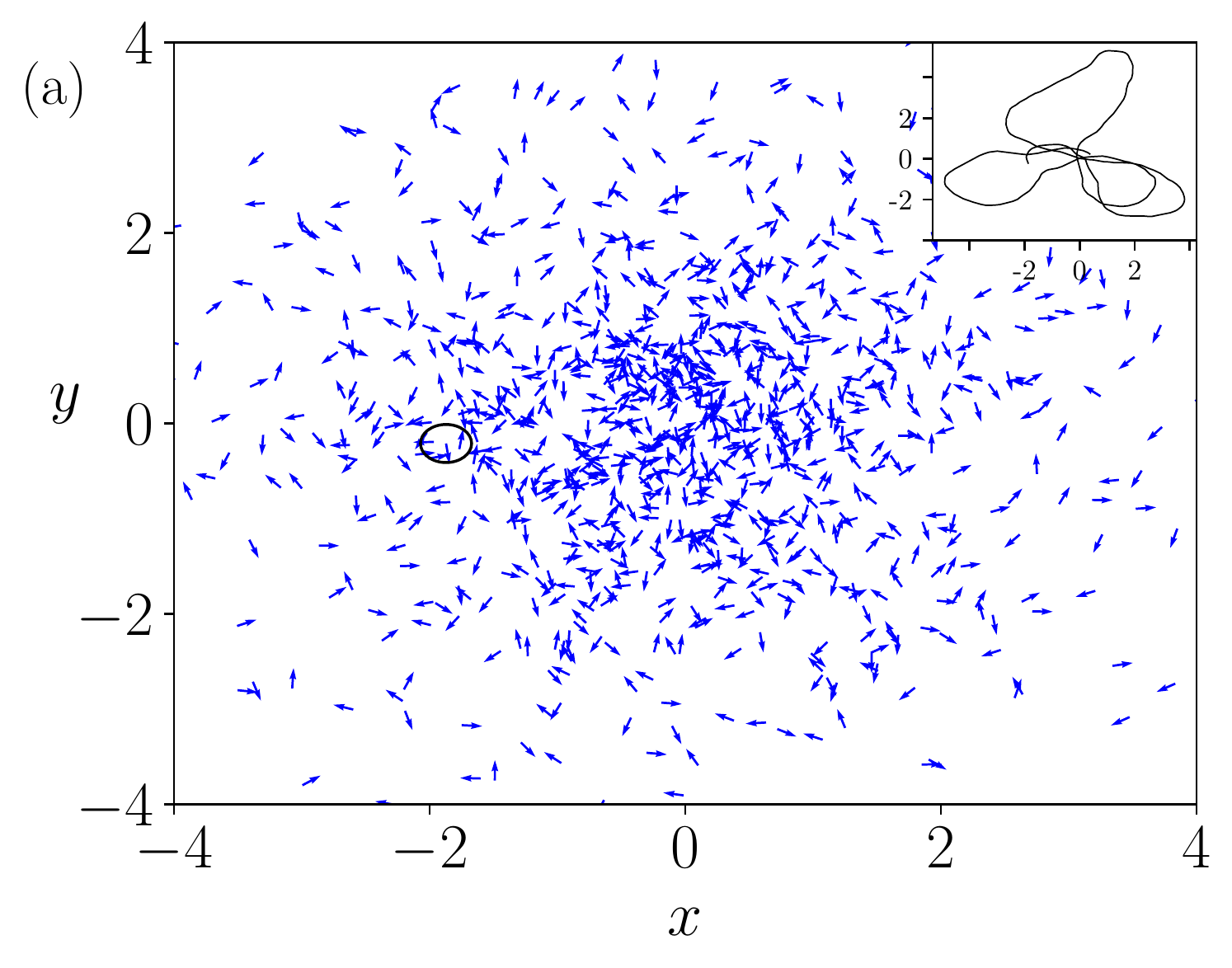}
  \includegraphics[width=0.49\linewidth]{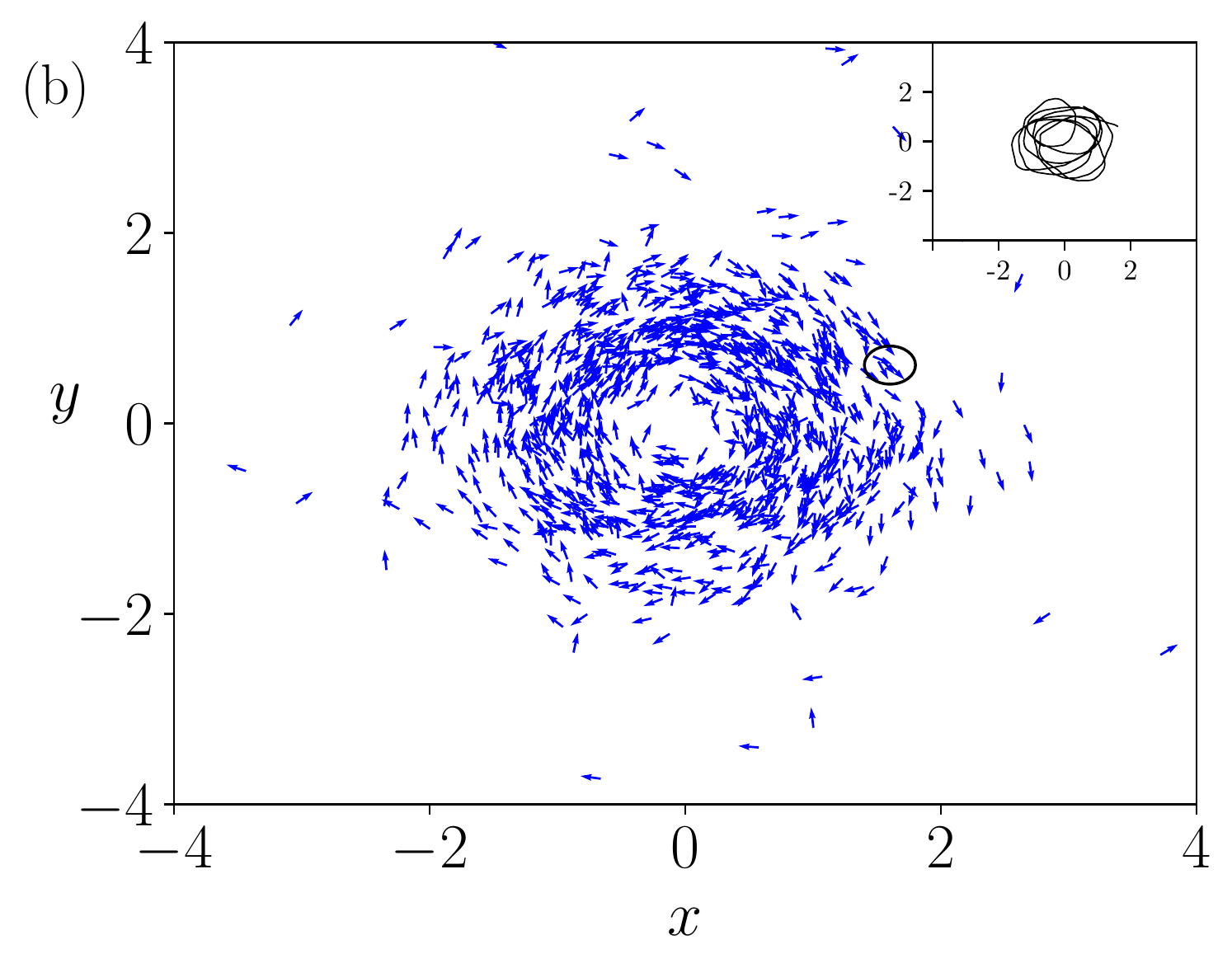}
   \caption{(a): Snapshot of searchers with small alignment $\mu=0.1$
    according to Eqs.
    \eqref{eq:xy_dot_coll} and \eqref{eq:phi_dot_coll}. Searchers will follow their individual strategies 
    and no cooperative behavior is observed. (b): Snapshot
    of searchers with alignment $\mu=0.7$.  For sufficient alignment
    strength the searchers form a rotating circle around the home. Rotations might be clockwise or 
    anticlockwise in dependence on initial values. Black lines in the insets indicate one individual trajectory over a 
    longer period of time. The random loops in the undercritical regime turn into circular motion for overcritical 
    coupling $\mu$. Circles indicate the size of the sensing radius.
    Parameters: $N=1000$, $v_0=1$, $\kappa=1$,
    $r_{\text{sens}}=0.2$, $\sigma^2=0.1$. }
    \label{fig:traj_align}
\end{figure}

Fig.\ref{fig:traj_align} shows a snapshot of the positions and headings of $N=1000$ searchers. 
The trajectory of an individual searcher is shown in the insets of the graph. For picture
(a) a coupling strength of $\mu=0.1$ and for (b) a coupling
of $\mu=0.7$ was chosen.
While the searchers in (a) perform individual search motion in case of weak $\mu$, 
they collectively rotate around their common home or target in (b). Therefore, we
expect to find a transition at a critical value of the coupling strength $\mu_{\text{crit}}$ between individual 
motion $\mu< \mu_{\text{crit}}$ and
collective motion $\mu>\mu_{\text{crit}}$. We also note that the foragers in (a) are more spread out over the space than
the foragers in (b). The insets show typical individual trajectories. 
The individual trajectories
change from central place foraging motion (a) to circular motion (b).

We will investigate how the 
asymptotic marginal densities of the distance from the home as well as the marginal density of the difference between the heading and the direction of the position vector change if varying the coupling strength $\mu$. 
    A critical behavior is predicted for
  the mean orbital velocity of the active local searchers in
  dependence on the relation between $\mu$ to the ratio between the
  value $v_0$ and the noise intensity $\sigma$. In consequence, also the variances of the radial 
  and orbital velocities will exhibit different behavior for small and large coupling
\begin{figure}
  \includegraphics[width=0.48\linewidth]{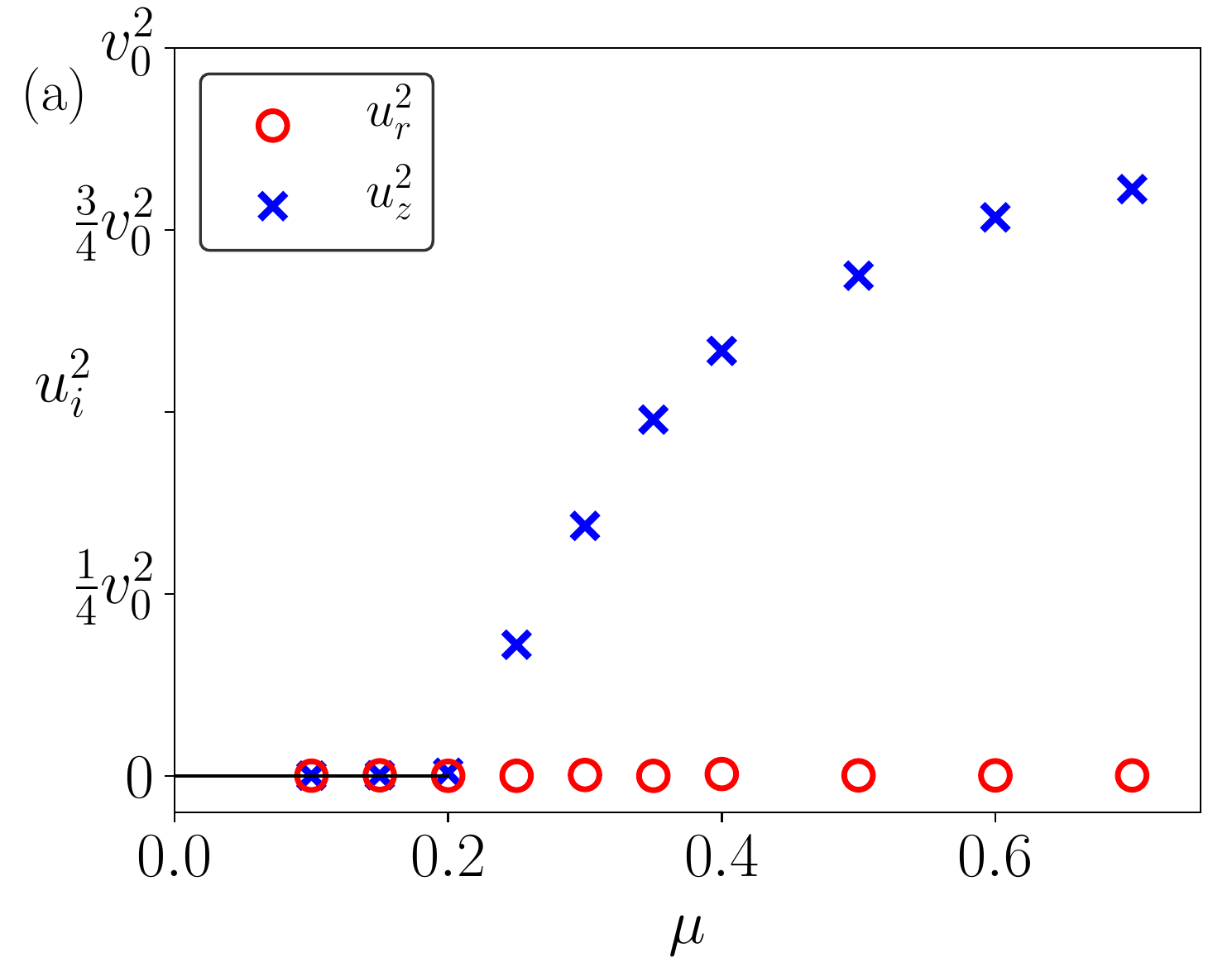}
  \includegraphics[width=0.49\linewidth]{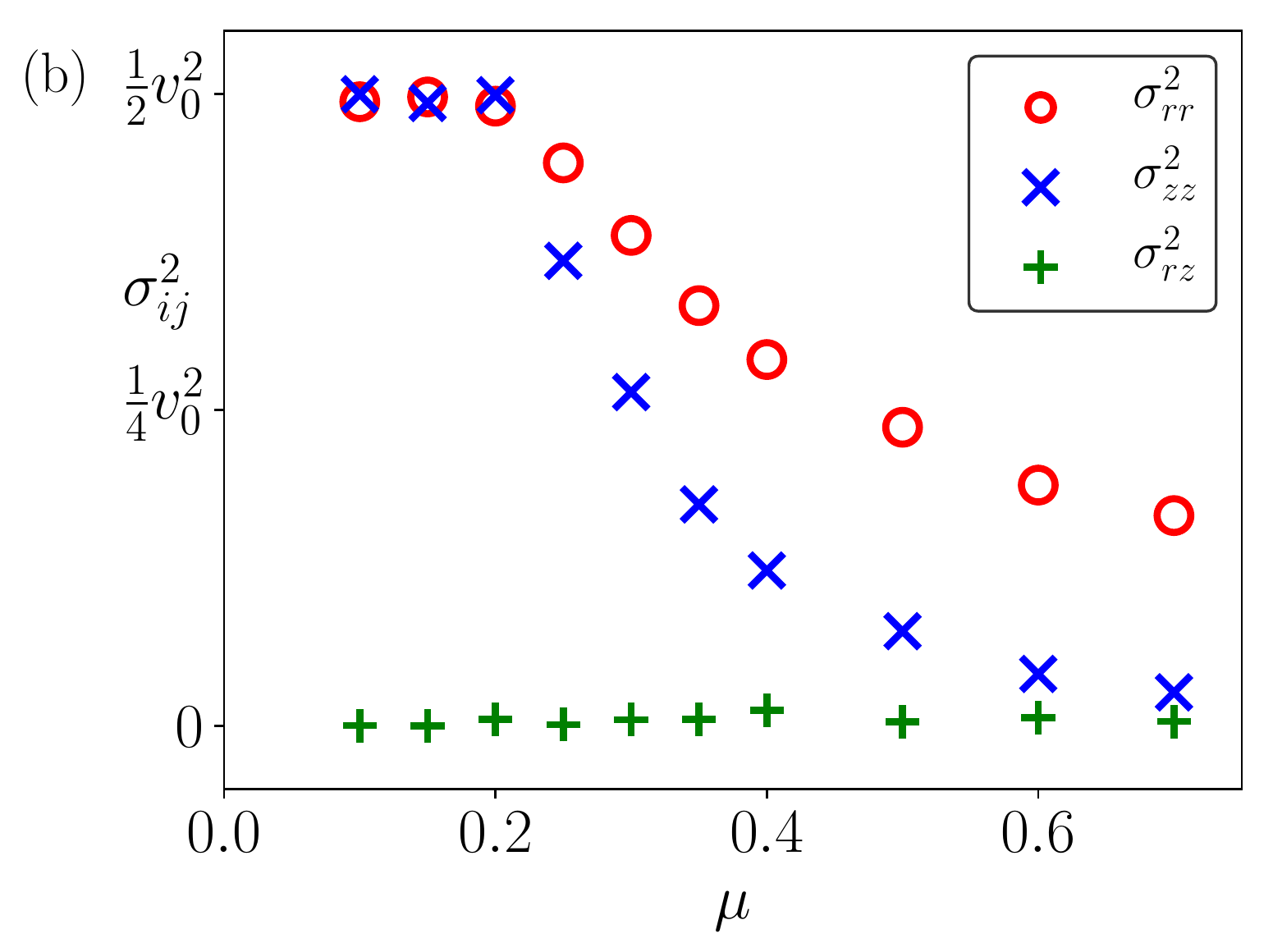}
  \caption{(a): Squared average radial and orbital velocities from
    simulations as symbols. Beyond the critical coupling strength
    $\mu_{\text{crit}} \approx 0.2$ a cooperative rotating of the
    searchers around the home is instigated which gains all energy for
    strong coupling. (b): The covariance and the variance of the
    orbital and radial velocity fluctuations as function of the
    coupling strength $\mu$. The covariance vanishes at arbitrary
    coupling. In the cooperative regime both variances decay with different strength and
    disappear for large coupling strength. Parameters: $N=1000$,
    $v_0=1$, $\kappa=1$, $\sigma^2=0.1$, $r_{\text{sens}}=0.2$.  }
  \label{fig:vel_var_sim}
\end{figure}

In Fig.\ref{fig:vel_var_sim} we present results in the asymptotic long
time state which quantify the two different regimes. We measured
during the simulations of $N=1000$ particles the mean radial $u_r$ and
orbital $u_z$ velocities and the variances and the covariance
$\sigma^2_{rr},\sigma^2_{zz}$, and $\sigma^2_{rz}$ of their fluctuations. We
point out that all presented values in pictures (a) and (b)
are independent on space. These velocity characteristics loose possible initial dependencies 
on coordinates in the asymptotic limit. They become homogeneous in space at least at the places 
where a sufficiently large number of particles is present during the measurements.
On  the left hand side the stationary mean velocities are presented as
functions of the coupling strength $\mu$. The asymptotic radial velocity vanishes always
independently from $\mu$. In difference, the squared angular velocity exhibits a soft 
transition at $\mu_{\text{crit}}\approx 0.2$. For lower coupling strength it vanishes as well, 
for $\mu$-values larger the critical one an orbital motion is created. The
orbital squared velocity grows with stronger coupling and saturates at
$v_0^2$ for infinitely large coupling where all energy is in the
circular drift. The direction (clockwise or anti-clockwise) of the
rotation around the home depends on initial values and with noise even
changes in the direction of the rotations are possible.

Fig. \ref{fig:vel_var_sim}(b) shows asymptotic of the second
central moments. One sees that the covariance $\sigma_{rz}^2$ of radial and 
orbital velocity fluctuations disappears, thoroughly. For under-critical coupling strength the variances
 $\sigma_{rr}^2$ and $\sigma_{zz}^2$ share the same value which is
half of the possible kinetic energy.  In case of an over-critical coupling
strength both variances shrink and vanish as $\mu \to \infty$. The
decay of the orbital velocity fluctuations is stronger than the radial fluctuations decrease with growing $\mu$.

\begin{figure}
  \includegraphics[width=0.49\linewidth]{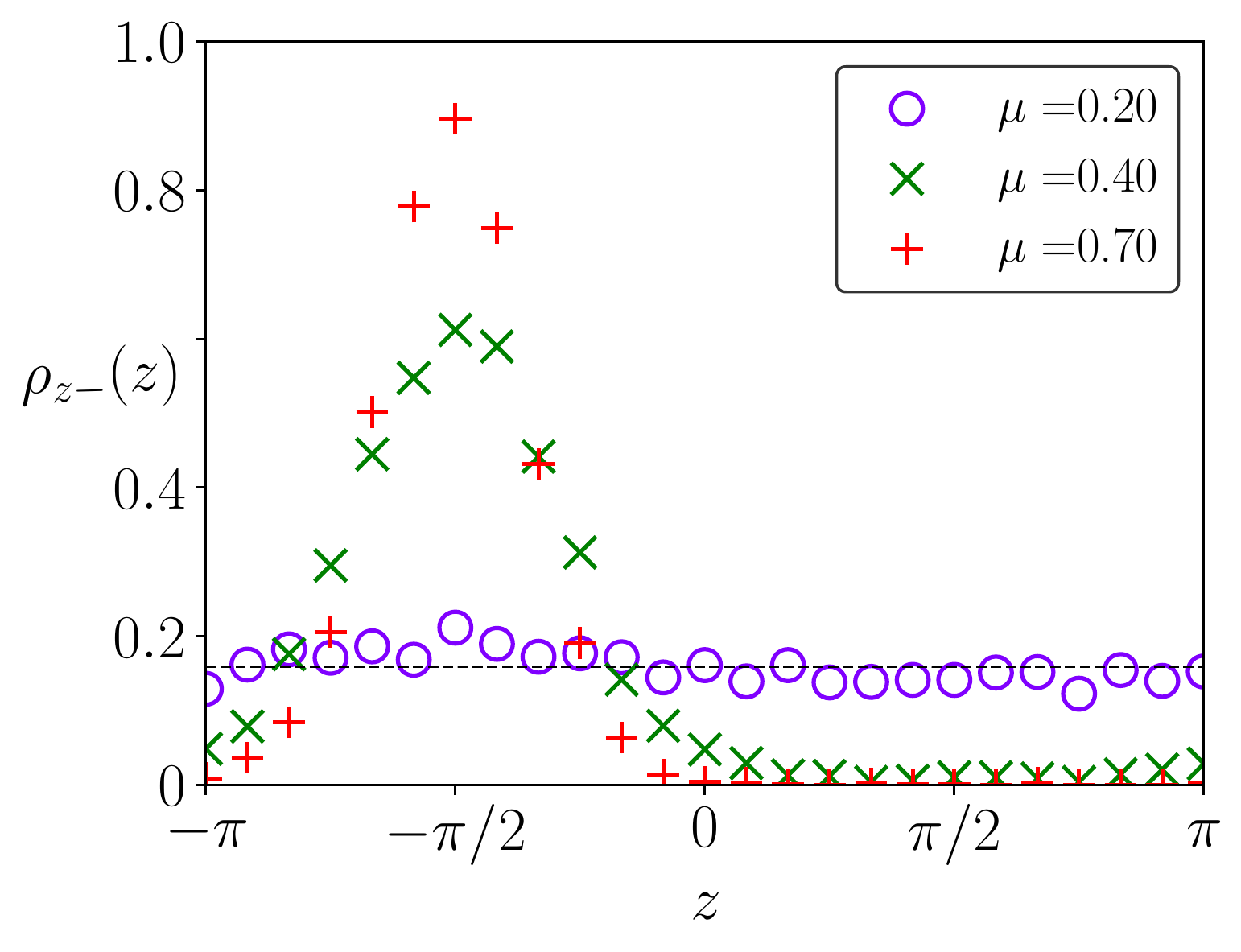}
  \caption{ Simulation results for the marginal density of the angle
    $z$ as symbols for three different values of the coupling
    $\mu$. The simulation results were chosen for $u_z<0$, meaning
    clockwise rotation in the $(x,y)$ plane. For counterclockwise
    rotation the densities are shifted by $\pi$.  All three values for
    the coupling are chosen $\mu>\mu_{\text{crit}}=0.2$. The dashed
    line corresponds to a uniform density. This density is approached
    for $\mu=\mu_{\text{crit}}$. Parameters: $N=1000$, $v_0=1$, $\kappa=1$, $\sigma^2=0.1$,
    $r_{\text{sens}}=0.2$.  }
  \label{fig:rhoz_exp}
\end{figure}

The occurrence of ordered circular motion is accompanied by a change in the distribution of the $z$ angle. 
The latter is uniform for an under-critical coupling strength. The distributions of the position angle $\beta$ and 
of the heading $\phi$ remain uniform also with over-critical couplings strength. The  marginal pdf $\rho_z(z)$ of 
their difference $z$ exhibits a maximum at a fixed position if the coupling becomes over-critical. It is 
presented in Fig.\ref{fig:rhoz_exp} where the marginal $z$-density as obtained from numeric simulations 
is shown for various values of $\mu$. The particular situation with a peak at $z=-\pi/2$ 
corresponds to the clockwise rotation of the aligning units around the home. 
Alternatively, the anticlockwise motion might collect maximal probability above $z=\pi/2$. Symmetry in the $2\pi$ 
periodic $z$ space is established with respect to $\pm \pi/2$, 
meaning $\rho_z(\pm \pi/2 +z')=\rho_z(\pm \pi/2 -z')$.

For a system with a finite particle number $N$ as defined by the set of Langevin equations in our model, 
the asymptotic marginal $z$ pdf is always the sum of two pdf's, one with a clockwise and the other with a anti clockwise rotations. 
Due to the symmetry of this bistable situation connected with possible transitions between the two directions of rotation, 
both pdf's contribute equally to the steady pdf.
Alternatively, in the thermodynamic limit as considered in the following analytic approach, ergodicity is broken and as final 
state we get one of the two possible pdf's in dependence on the initial state. Later on in the discussion of the 
analytical results, we restrict the consideration to the particular type of clock wise rotations.

\subsection{Polar presentation of the particle dynamics}
\label{sec:mathe}
We change to polar coordinates,  i.e., to the distance
  $r_i(t)=\sqrt{x_i^2(t)+y_i^2(t)}$ and the direction
  $\beta_i(t)=\arctan(y_i(t)/x_i(t))$ of the position vector
  $\vec{r}_i(t)$ (see Fig \ref{fig:schematic}).  Furthermore, we
introduce the angle $z_i(t)=\phi_i(t)-\beta_i(t)$, being the
  difference between the heading and the position. A value of $z=
  \pi/2$ indicates counterclockwise rotation, a value of $z=-\pi/2$
  clockwise rotation. Alternatively, values around $0$ and $\pi$ stand
  for motion pointing outwards, respectively, towards the home. Simulations in 
  Fig. \ref{fig:traj_align}(b) show a clockwise rotation 
of the aligned collective motion, hence maximal probability in the $z$-distribution 
is expected for values around $z\approx - \pi/2$.

We derive for the radial velocity:
\begin{equation}
\dot{r}_i = v_0 \cos(z_i)\,,
\label{eq:r_dot_align}
\end{equation}
and for the angular velocity:
\begin{equation}
\dot{\beta}_i = \frac{v_0}{r_i} \sin(z_i)\,.
\label{eq:beta_dot_align}
\end{equation}
The dynamics for the angle $z$ becomes:
\begin{equation}
  \dot{z}_i =  -\left(\frac{1}{r}-\frac{1}{r_c}\right) v_0 \sin(z_i)+\frac{\mu}{N_i}\sum_{j\in\Omega_i}\sin(z_j-z_i+\beta_j-\beta_i) + \frac{\sigma}{v_0} \xi_i(t)\,,
  \label{eq:z_dot_align}
\end{equation}
where again $r_c=v_0/\kappa$ and $\Omega_i$ contains the
indices of the $N_i$ particles within the sensing radius of the particle
$i$.

For the third equation \eqref{eq:z_dot_align}, we assume that the
position angles $\beta_i\approx\beta_j$ cancel each other. The $z$
dynamics becomes:
\begin{equation}
  \dot{z}_i =  -\left(\frac{1}{r}-\frac{1}{r_c}\right) v_0 \sin(z_i)+\frac{\mu}{N_i}\sum_{j\in\Omega_i}\sin(z_j-z_i) + \frac{\sigma}{v_0} \xi_i(t)\,.
\label{eq:z_dot_align_app}
\end{equation}
This approximation is valid if having a sufficient small sensing
radius $r_{\text{sens}}\ll r_c=v_0/\kappa$. Therein $r_c$
  appears as the distance where searchers reside with maximal
  probability which will be shown, later on. The approximation fails
  obviously for small distances from the home. Hence, the typical length
  scales of \eqref{eq:z_dot_align_app} should be much larger than the interaction radius,
  i.e. $r \gg r_{\text{sens}}$.

And thus, the searchers are described by the actual values of the set
$r_i(t), \beta_i(t)$ and $z_i(t)$. For a single independent searcher
$(\mu=0)$ the $(r(t),z(t))$ dynamics separates from the $\beta(t)$
dynamics \cite{Noetel_2018}.  With alignment present this is not the
case as the position angle enters the $z_i$ dynamics through the
neighborhood $\Omega_i$. Nevertheless, as will be discussed now, at larger time scales 
$t\gg \tau_\phi$ the $\beta_i$ become uniform distributed as well as the heading 
directions $\phi_i$ do. Only their binary relation expressed by $z_i(t)$ will matter for the behavior.

\section{Probability density function and transport equations}
\label{sec:mathe2}
\subsection{The nonlinear Fokker-Planck equation with effective aligning force}
We consider the many particle pdf $P_N(r_1,z_1,\beta_1, \ldots ,r_N,
z_N, \beta_N,t)$ in polar presentation. We consider the pdf to be well approximated by the product of the one
particle PDF $P(r,z,\beta,t)$ having in mind the limit of infinite
particle numbers and assuming the validity of a mean field
approximation. In consequence, we formulate the alignment
force $f_{al}(r,\beta,z,t)$ which a particle experiences at position $r,\beta$ from a second 
particle with coordinates $r',\beta'$ inside the sensing radius of the first one $\Omega(r,\beta)$ as:
\begin{eqnarray}
  f_{al}(r,z,\beta,t)\, = \,\frac{\mu}{N_{\text{sens}}(r,\beta,t)} \,N \,\int_{(r',\beta')\in \Omega(r,\beta)}{\rm{d}} r'{\rm{d}}\beta' \int_{-\pi}^{\pi}{\rm{d}}z'\,\sin(z'-z)\,P(r',\beta',z',t)\,.\nonumber\\
  \label{eq:interact_1}
\end{eqnarray}
The particle density in the sensing region reads accordingly
\begin{eqnarray}
  N_{\text{sens}}(r,\beta,t)\,=\,N\,\int_{(r',\beta') \in \Omega(r,\beta)}{\rm{d}} r' {\rm{d}}\beta' \int_{-\pi}^{\pi}{\rm{d}}z'\,P(r',\beta',z',t)\,.
 \label{eq:interact_11} 
\end{eqnarray}

For the further evaluation of \eqref{eq:interact_1}, we introduce the marginal spatial pdf
$\rho(r,\beta,t)$ and the mean velocities $u_r(r,\beta,t)$,
$u_z(r,\beta,t)$ in dependence of the polar coordinates $r,\beta$ as
\begin{eqnarray}
  \rho(r,\beta,t)&&=\int_{-\pi}^{\pi} {\rm{d}}z
  P(r,\beta,z,t)\,, \label{eq:def_rho}\\ u_r(r,\beta,t)\rho(r,\beta,t)&&=v_0\int_{-\pi}^{\pi}
  {\rm{d}}z
  \cos(z)P(r,\beta,z,t)\,, \label{eq:def_ur}\\ u_z(r,\beta,t)\rho(r,\beta,t)&&=v_0\int_{-\pi}^{\pi}
      {\rm{d}}z \sin(z)P(r,\beta,z,t)\,.
  \label{eq:def_uz}
 \end{eqnarray}
 Introducing \eqref{eq:def_rho}-\eqref{eq:def_uz} into the interaction
 term \eqref{eq:interact_1} and performing the $z'$ integration, leads
 to:
\begin{eqnarray}
  &&f_{al}(r,\beta,t)
  \,=\,\\ &&\frac{\mu}{N_{\text{sens}}(r,\beta,t)} N \int_{(r'\beta')\in \Omega(r,\beta)} {\rm{d}}r'{\rm{d}}\beta'\,\bracket{u_z(r',\beta',t)\cos(z)-u_r(r',\beta',t)\sin(z)}\,\rho(r',\beta',t)\,. \nonumber
\label{eq:interact_2}
\end{eqnarray}
With the above assumptions about a small sensing radius
$r_{\text{sens}}$ and that distances are with high probability around
$r \approx r_c$ one can simplify the expressions \eqref{eq:interact_1}
and \eqref{eq:interact_11}. We equalize positions and
directions $r \approx r'$ and $\beta \approx \beta'$ inside the
integral. It means that for sufficient small sensing radius the
average velocities as well as the marginal density are constant inside
this radius. It results in $u_r(r',\beta',t) \approx u_r(r,\beta,t)$
and $u_z(r',\beta',t) \approx u_z(r,\beta,t)$. The same approximation
is done in the expression for the $N_{\text{sens}}(r,\beta,t)$ in
Eq. \eqref{eq:interact_11}. The latter drops together with the density
$N\,\rho(r',\beta',t)\, \pi r_{\text{sens}}^2 \approx N_{\text{sens}}(r,\beta,t)$ in
Eq.\eqref{eq:interact_1}. In consequence, it follows for the effective
aligning force \eqref{eq:interact_1}:
\begin{eqnarray}
  f_{al}\,\approx\,\mu
  \left(u_z(r,\beta,t)\cos(z)-u_r(r,\beta,t) \sin(z) \right)\,.
  \label{eq:interact_3}
\end{eqnarray}

Note, that $u_z$ is defined as an orbital velocity being orthogonal to
$u_r$.  It is neither an average angular velocity for the position
angle $\beta$ nor is it an average angular velocity derived from the
$z(t)$ dynamics given by Eq. \eqref{eq:z_dot_align}.

Fig. \ref{fig:traj_align}(b) shows the particles rotating homogeneously
around the home. Also the disordered motion on the l.h.s. possesses
rotational symmetry. Both snapshots have equilibrated around the
center which is true as $t\geqslant \tau_\phi$. The noise has created this
rotationally symmetric shape of the heading directions $\phi$ as well
as the directions of the position vector $\beta$. Hence, in the
following we will assume rotational symmetry for the $\beta$ and
$\phi$ dynamics. We will not üpay attention to the relaxation of
angular inhomogeneities of these two variables being more interested
in the properties of the asymptotic state as functions of $r$.
  
In consequence, the marginal density \eqref{eq:rho} becomes
$\rho(r,\beta,t) \approx \rho(r,t)$, and the mean velocities
\eqref{eq:def_ur} and \eqref{eq:def_uz} are approximated as
$u_r(r,\beta,t) \approx u_r(r,t)$ and $u_z(r,\beta,t) \approx
u_z(r,t)$.

As we approximated the alignment term \eqref{eq:interact_3}
independent of the position angle the $(r,z)$ dynamics separates from
the angular $\beta$ dynamics. Also the one particle pdf looses the
$\beta$ dependence. Hence the Fokker-Planck equation (FPE) is given by:
\begin{eqnarray}
  \label{eq:fpe_align_2}
  \frac{\partial}{\partial t}P&&=-v_0\frac{\partial}{\partial
    r}\cos(z)P+\\ &&+v_0\frac{\partial}{\partial
    z}\left(\left(\frac{1}{r}-\frac{1}{r_c}+\frac{\mu}{v_0^2} u_r
  \right)\sin(z) -\frac{\mu}{v_0^2} u_z\cos(z)\right)P
  +\left(\frac{\sigma}{v_0}\right)^2 \frac{\partial^2}{\partial
    z^2}P\,,\nonumber
\end{eqnarray}
and $P$ is the transition pdf
$P=P(r,z,t|r_0,z_0,t_0)$ describing the reduced $(r,z)$ dynamics.  This
FPE is valid for large particle numbers $N \to
\infty$, vanishing sensing radius $r_{\text{sens}} \ll r_c$ and at
time scales $t\geqslant \tau_\phi$. The equation is nonlinear in $P$ since the $u_r$ and $u_z$ are functions of the pdf.

\subsection{Transport equations}
In order to find approximate solutions, we derive transport equations
for the first three moments \eqref{eq:def_rho}, \eqref{eq:def_ur} and
\eqref{eq:def_uz} being now independent of $\beta$. By respective
standard multiplication of the FPE and integration over $z$ we obtain
relations between the reduced moments of trigonometric functions
depending on space and time. The marginal density of the distance
obeys the continuity equation
\begin{eqnarray}
  &&\frac{\partial}{\partial t}\rho=-\frac{\partial}{\partial r}\rho u_r\,.
  \label{eq:rho}
\end{eqnarray}

The equations of the first moments read for the mean radial velocity
\begin{eqnarray}  
\label{eq:trans_r}
  &&\frac{\partial}{\partial
    t}u_r\rho=\\ &&-\frac{\partial}{\partial
    r}\rho\bracket{u_r^2+\sigma_{rr}^2}+\left(\frac{1}{r}-\frac{1}{r_c}
  \right)\rho \bracket{u_z^2+\sigma_{zz}^2}-\frac{\mu}{v_0^2}\rho u_z \sigma_{rz}+\bracket{\frac{\mu}{v_0^2}\sigma_{zz}^2-\left(\frac{\sigma}{v_0}\right)^2}
  \rho u_r\,, \nonumber
\end{eqnarray}
and for the mean orbital velocity
\begin{eqnarray}   
  &&\frac{\partial}{\partial
    t}u_z\rho=\\ &&-\frac{\partial}{\partial
    r}  \rho \bracket{u_r u_z+\sigma_{rz}^2}-\left(\frac{1}{r}
  -\frac{1}{r_c}
  \right)\rho \bracket{u_r u_z+\sigma_{rz}^2}-\frac{\mu}{v_0^2}\rho u_r\sigma_{rz}^2+\bracket{\frac{\mu}{v_0^2} \sigma_{rr}^2 -\left(\frac{\sigma}{v_0}\right)^2 }\rho u_z \nonumber\,.
  \label{eq:trans_u}
\end{eqnarray}
Therein,  we have introduced the variances as 
\begin{eqnarray}
  \sigma^2_{rr}(r,t)\rho(r,t)&&=\int_{-\pi}^{\pi} {\rm{d}}z\,
  v_0\cos^2(z)\,P(r,z,t)\,-\,u_r^2\,\rho,
  \nonumber\\ \sigma^2_{zz}(r,t)\rho(r,t)&&=\int_{-\pi}^{\pi}
              {\rm{d}}z \,v_0\sin^2(z)\,P(r,z,t)\,-\, u_z^2\,\rho\,.
  \label{eq:def_varz}
\end{eqnarray}
Note that due to the constant speed of the particles the variances and
mean velocities are not independent
\begin{eqnarray}
\sigma^2_{rr}+u^2_r+\sigma^2_{zz}+u^2_z=v_0^2\,. 
\label{eq:speed}
\end{eqnarray}
One of these values might be expressed by the three other moments. 

We have also introduced the covariance as
\begin{eqnarray}
  \sigma^2_{rz}(r,t)\rho(r,t)&&=\int_{-\pi}^{\pi} {\rm{d}}z \,v_0^2\cos(z)\,\sin(z)\,P(r,z,t)\,-\,u_ru_z\rho\,. 
  \label{eq:def_cov}
\end{eqnarray}
In our theoretical consideration we will assume that the two velocity deviations 
from their mean $\delta u_r=v_0\cos(z)-u_r$ and $\delta u_z=v_0\sin(z)-u_z$ are not correlated. 
It was shown numerically for the asymptotic stationary limit as presented above (see Fig.\ref{fig:vel_var_sim}). 
We simplify our model here and will neglect the correlation by putting $\sigma_{rz}(r,t) = 0$, further on.

We derive for the second moments (see Eqs.\eqref{eq:def_varz}):
\begin{eqnarray}
\frac{\partial\rho\bracket{u_r^2 +\sigma_{rr}^2}}{\partial t}&&\,=\,
-\frac{\partial}{\partial r} \mean{v_0^3\cos^3(z)}\,+\,2\left(
\frac{1}{r}-\frac{1}{r_c}\right)\mean{v_0^3\cos(z)\sin^2(z)}
\\ &&+\frac{2\mu}{v_0^2}\bracket{u_r\mean{v_0^3\cos(z)\sin^2(z)}-u_z
  \mean{v_0^3 \cos^2(z)\sin(z)}}+\frac{4\sigma^2}{v_0^2}\rho
\bracket{\frac{v_0^2}{2}-(u_r^2+\sigma_{rr}^2)},\nonumber
\label{eq:cov_dyn1}
\end{eqnarray}
and 
\begin{eqnarray}
\frac{\partial \rho\bracket{u_z^2+\sigma_{zz}^2}}{\partial t}&&\,=\,
-\frac{\partial}{\partial r} \mean{v_0^3\cos(z)\sin^2(z)}\,-\,2\left(
\frac{1}{r}-\frac{1}{r_c}\right)\mean{v_0^3\cos(z)\sin^2(z)}
\\ &&-\frac{2\mu}{v_0^2}\bracket{u_r\mean{v_0^3\cos(z)\sin^2(z)}-u_z
  \mean{v_0^3
    \cos^2(z)\sin(z)}}+\frac{4\sigma^2}{v_0^2}\rho\bracket{\frac{v_0^2}{2}-\bracket{u_z^2+\sigma_{zz}^2}}\,,\nonumber
\label{eq:cov_dyn2}
\end{eqnarray}
where we made use already of the assumed simplification.

In subsection \ref{ana_stat_asym} we will decouple higher moments in equations for
second velocity moments. One usual way would be a Gaussian approximation of this higher moments. 
This case and its results are presented in the Appendix \ref{sec:Gauss}. 
Here in the next subsection \ref{sec:Mises} we will derive an expression for the marginal 
density $\rho_z(z)$ of the angular variable $z$ outgoing from the nonlinear FPE \eqref{eq:fpe_align_2}. 
It will be a von Mises distribution. The resulting decoupling scenario yields better agreement 
between the transport theory and the numeric findings.   

\subsection{The marginal $z$-density as von Mises distribution}
\label{sec:Mises}
The marginal density $\rho_z(z.t)$ of the angular variable $z$ is defined by the integral
\begin{eqnarray}
  &&\rho_z(z,t)=\int_0^\infty {\rm{d}}r P(r,z,t)\,.
  \label{eq:def_rho_z}
\end{eqnarray}
Likewise for the spatial marginal density, we derive here a dynamics for $\rho(z,t)$ starting from the FPE \eqref{eq:fpe_align_2} and integrating away this time the distance $r$. The corresponding FPE reads:
\begin{eqnarray}
  \label{eq:fpe_align_rhorhoz3}
  \frac{\partial}{\partial t} \rho_z= v_0 \frac{\partial}{\partial z}
 \bracket{ \left\langle \frac{1}{r}-\frac{1}{r_c}+\frac{\mu}{v_0^2} u_r\right\rangle_r \sin(z)
  -\frac{\mu}{v^2_0} \langle u_z \rangle_r
  \cos(z)}\rho_z+\left(\frac{\sigma}{v_0}\right)^2 \frac{\partial^2}{\partial
    z^2}\rho_z\,.
\end{eqnarray}
In this equation we formally abbreviated 
$\left\langle\,.\,\right\rangle_r=\int{\rm d}r\, .\, P(r,t|z)$ and $P(r,t|z)$ is the conditional pdf of a distance $r$ for a fixed $z$. 
In simulations this dependence on $z$ appeared to be weak. In order to obey the numerically found asymptotic $z$-symmetry, 
the bracket in front of the $\sin(z)$ function in \eqref{eq:fpe_align_rhorhoz3} should disappear for arbitrary $z$-values. 
Extrema of the density $\rho_z$ have been found at angles where the $\cos(z)$-function vanishes. 
Therefore, it holds  $\mean{1/r-1/r_c+\mu u_r/v_0^2}_r=0$.  
The velocity $u_z$ becomes asymptotically constant and homogeneous in space as well. The bracket can be dropped and 
we write $\mean{u_z}_r=u_z$. With these settings we can  integrate the stationary FPE. The marginal angular pdf  \eqref{eq:def_rho_z}
yields a von Mises distribution \cite{stratonovich,ihle} which reads
\begin{eqnarray}
  \rho_z(z) \,=\, \frac{1}{2\pi I_0\bracket{\frac{\mu u_z v_0}{\sigma^2}}}\exp\left(\frac{\mu u_z}{\sigma^2}v_0\sin(z)\right)\,.
  \label{eq:ansatz_rhoz2}
\end{eqnarray}
Therein the $I_n$ stands for the modified Bessel function of the first kind. 

In dependence on the sign of $u_z$ the von Mises distribution peaks
above one of the values $z= \pm \pi/2$ and has the minimum at $z=\mp
\pi/2$. Since we deal with a nonlinear FPE, ergodicity is broken in the
over-critical solution and the asymptotic solution depends on initial
states. With under-critical coupling strength no mean orbital drift
exists $u_z=0$ and the von Mises distribution collapses into the
uniform distribution.

The von Mises distribution allows a self-consistent definition of
the stationary mean orbital velocity $u_z$. The latter is given as the
solution of the expression:
\begin{eqnarray}
  \label{eq:consist}
  u_z =\frac{1}{2\pi I_0\left(\frac{\mu u_z v_0}{\sigma^2}
    \right)}\int_{-\pi}^\pi {\rm{d}}z\,\,
  v_0\sin(z)\,\exp\left(\frac{\mu u_z
  }{\sigma^2}v_0\sin(z)\right)\,=\,v_0\frac{I_1\left(\frac{\mu u_z
      v_0}{\sigma^2} \right)}{I_0\left(\frac{\mu u_z v_0}{\sigma^2}
    \right)}\,.
\end{eqnarray}
Such self-consistent relations are well known from many disciplines in physics. 
Exemplarily we remind on atom physics \citep{Slater1959}, on plasma physics \citep{Balescu1960}, 
chemical physics \citep{Mukamel1978}, the theory of equilibrium\citep{Stanley1971,Kadanoff2009} and 
nonequilibrium  \citep{Vandenbroeck1997,Sagues2007}  phase transitions, and on synchronization 
phenomena of phase oscillators\citep{Kuramoto_1984,Strogatz_2000}. 
The usual way of finding the solution is the geometric construction of the l.h.s and the r.h.s. 
Their intersection(s) yield the solution of \eqref{eq:consist}. It is known that this solution $u_z$
undergoes a pitchfork bifurcation at
\begin{eqnarray}
  \mu_{\text{crit}}=2\sigma^2/v_0^2\,. 
  \label{eq:mucrit}
\end{eqnarray} 
It is just the value of the coupling strength at which the slope of
the r.h.s. of \eqref{eq:consist} taken as function of the velocity $u_z$
coincides with the one from the l.h.s. in the limit of vanishing
velocity. For smaller coupling a single solution $u_z=0$ exists,
only. For over-critical coupling $\mu > \mu_{\text{crit}}$ there
are three intersections in \eqref{eq:consist}. One of them again
vanishes and appears to be unstable with respect to small
perturbations. The other two are stable solutions with different sign
and identical non-vanishing absolute value.

Later on, in the discussion part we present also the results obtained by means 
of the von Mises distribution. In the next subsection, the latter is used for decoupling the higher 
moments in the transport equations.


\subsection{Analysis of the stationary asymptotic state}
\label{ana_stat_asym}
In this section we study the stationary states as they follow from the
asymptotic limit of the transport equations and use the results of
the last chapter.

\subsubsection{Under-critical coupling strength}
First, we take a look upon the stationary limit for the case $\mu
\leqslant \mu_{\text{crit}}$. There no cooperation in the motion exists and the 
orbital velocity disappears $u_z(r)=0$. The density $\rho(r,t) \to \rho(r)$
becomes stationary as well.  The radial flux $u_r$ disappears 
which is a consequence of  the continuity equation
\eqref{eq:rho}. This disappearance can be confirmed by calculating 
the stationary mean flux using the von Mises distribution.

At second, with $u_z=0$ the orbital angle distribution $\rho_z(z)$
squeezes to the periodic equidistribution on $z\in[-\pi,\pi)$. 
It causes the disappearance of all third order moments in the balance equations. 
Therefore, in agreement with Eqs. \eqref{eq:cov_dyn1} and \eqref{eq:cov_dyn2} the  variances 
share the same stationary energies with
\begin{eqnarray}
\sigma^2_{rr}\,=\,\sigma^2_{zz}\,=\,\frac{v_0^2}{2}\,.
\label{eq:var_under_ss}
\end{eqnarray}

The equation for the orbital flux reads
\begin{equation}
0\,=\, \frac{\partial}{\partial t}u_z\,=\, \bracket{\frac{\mu}{2}-\bracket{\frac{\sigma}{v_0}}^2}\,u_z
\label{eq:orbital_under}
\end{equation}
where we have dropped the stationary density and inserted the radial variance 
from \eqref{eq:var_under_ss}. Obviously, it possesses the solution $u_z=0$   which is homogeneous 
in space as also the variances in \eqref{eq:var_under_ss} are.

Eq. \eqref{eq:orbital_under} also defines the border of stability of the present solution. 
The bracket at the r.h.s. vanishes at the critical coupling strength $\mu_{\text{crit}}$ 
from \eqref{eq:mucrit}. Stability of the non-cooperative behavior is given only for under-critical values $\mu$. 

The single value which depends on coordinates is the marginal density
$\rho(r)$. The equation which determines its functional dependence is
the stationary equation for the mean radial flux $\rho u_r$. It
becomes with the findings from above
\begin{eqnarray}
0=\frac{\partial}{\partial t} \rho
u_r\,=\,-\,\frac{v_0^2}{2}\bracket{\frac{\partial}{\partial r}\rho
  +\bracket{\frac{1}{r}-\frac{1}{r_c}}\rho(r)} \,.
\label{eq:rho_under}
\end{eqnarray} 
As solution for the weakly coupled ensemble, we obtain the same stationary
marginal density as it was derived for a single independent particle \cite{Noetel_2018, Noetel_2018c}. There we have found
\begin{eqnarray}
 \rho(r)\,=\,\frac{r}{r^2_c} \exp\bracket{-\frac{r}{r_c}}\,.
\label{eq:rho_under_ss}
\end{eqnarray}
which we obtain here for the marginal density of weakly coupled searchers.
We note that all moments with three trigonometric functions
disappear. Hence, with the given variances \eqref{eq:var_under_ss}
and vanishing fluxes all stationary transport equations are exactly
fulfilled. Also the average $\mean{{1}/{r}-{1}/{r_c}+\mu u_r/v_0^2}_r$ vanishes
with \eqref{eq:rho_under_ss} which we used to obtain the
von Mises distribution as in \eqref{eq:ansatz_rhoz2}.

\subsubsection{Over-critical coupling strength}
Secondly, we formulate decoupling relations for the moments including
three trigonometric functions if $\mu \geqslant \mu_{\text{crit}}$
with $u_z\ne 0$ in agreement with the simulations and the von Mises 
distribution. Here, we follow a partial integration scheme of the stationary von Mises distribution. 
Doing so and requiring $u_z\ne 0$ we obtain the identity 
\begin{equation}
u_z \int_0^{2\pi} {\rm d}z\, f(z) v_0 \cos(z) \rho_z(z) \,=\,-\, \frac{\sigma^2}{\mu} \,\int_0^{2\pi}{\rm d}z \frac{{\rm d}f(z)}{{\rm d}z} \rho_z(z) 
\label{eq:part_int}
\end{equation}
with $\rho_z(z)$ from \eqref{eq:ansatz_rhoz2} and $f(z)$ as polynomial of $\cos(z)$ and $\sin(z)$ 
functions. Note, that at the r.h.s. the single $\cos(z)$ function has disappeared and  the order of the moment is reduced. 

Using \eqref{eq:part_int} and properties of the trigonometric functions one easily verifies that 
the expressions vanish for $f(z)=\cos^2(z)$ and $ \sin^2(z)$. It also vanishes for $f(z)=v_0$ meaning again 
that  the stationary radial flux disappears. i.e. $u_r(r)=0$, and the density becomes 
stationary $\rho(r,t) \to \rho(r)$.

And thus, the two remaining third order moments in our theory are linked to the second order moments. It holds
\begin{eqnarray}
&&u_z \mean{v_0^3 \sin(z)\cos^2(z)}= \rho
  \frac{\sigma^2}{\mu}\bracket{u_z^2+\sigma_{zz}^2-u_r^2-\sigma_{rr}^2}\,,\\ &&u_z\mean{v_0^3
    \sin^3(z)}=\rho v_0^2 u_z^2 - \rho
  \frac{\sigma^2}{\mu}\bracket{u_z^2+\sigma_{zz}^2-u_r^2-\sigma_{rr}^2}\,.
\label{eq:thirdmoments}
\end{eqnarray} 

Afterwards, insertion of those expressions into the r.h.s. of the balance
equations for the mean radial and orbital energies \eqref{eq:cov_dyn1}
and \eqref{eq:cov_dyn2}, we derive exactly
\begin{eqnarray}
\rho \frac{\partial}{\partial t} \sigma_{rr}^2\,=\, 0\,,~~~~~~~~~\rho \frac{\partial}{\partial t} \bracket{u_z^2+\sigma_{zz}^2}\,=\,0\,.
\label{eq:station_var}
\end{eqnarray}
The transport equations confirm the decoupling by means of the von Mises distribution and 
both radial and orbital energies become stationary.

The stationary equation for the mean orbital velocity yields the same condition as in the under-critical case 
\begin{eqnarray}
0=\frac{\partial}{\partial t}u_z\,=\,\bracket{\frac{\mu}{v_0^2}
  \sigma^2_{rr}-\left(\frac{\sigma}{v_0}\right)^2} \,u_z\,.
\label{eq:mean_orb}
\end{eqnarray}    
Again the density drops out in the stationary state and this equation possesses spatially homogeneous steady states. 
One solution is $u_z=0$ which appears to be stable for under-critical coupling $\mu
\leqslant \mu_{\text{crit}}$ as discussed above. Since by assumption $u_z\neq 0$ we select as solution for the cooperative regime
\begin{equation}
\sigma^2_{rr}\,=\,\frac{\sigma^2}{\mu}\,.
\label{eq:rr_over}
\end{equation}
One might add here that the same expression could be obtained by determining the stationary mean radial energy via 
averaging with the von Mises distribution and setting $u_r=0$.

The equation for mean radial velocity $u_r$ becomes the equation for determining the marginal density $\rho(r)$. 
Insertion of \eqref{eq:rr_over} into \eqref{eq:trans_r} gives
\begin{eqnarray}  
  &&0=\rho \frac{\partial}{\partial t}u_r\,=\, -\frac{\partial}{\partial
    r}\frac{\sigma^2}{\mu}\rho+\left(\frac{1}{r}-\frac{1}{r_c}\right)\bracket{v_0^2-\frac{\sigma^2}{\mu}}\rho\,.
 \label{eq:rho_over}   
\end{eqnarray}
This equation can be integrated and we get the steady state of the spatial marginal density in case of $\mu\geqslant \mu_c$:
\begin{eqnarray}
  &&\rho(r)=C\left[r\exp\left(-\frac{\kappa}{v_0}r\right) \right]^{\frac{v_0^2 \mu}{\sigma^2}-1}\,,
  \label{eq:rho_f}
\end{eqnarray}
with $C$ being the normalization constant. We mention that in case of
the critical coupling strength, i.e. if $\mu=\mu_c$ the exponent in
this distribution becomes unity. Hence, the marginal density
\eqref{eq:rho_f} at critical coupling strength coincides with the
marginal density in the under-critical situation \eqref{eq:rho_under_ss}
reminding the soft character of the pitchfork bifurcation. 

In the cooperative regime the marginal density also depends on 
the coupling strength and on the noise intensity. In 
comparison to \eqref{eq:rho_under_ss} the density in \eqref{eq:rho_f} appears more narrow above its maximal state $r_{c}$ as can be inspected in Fig.\ref{fig:traj_align}. 
That is why, the exponent is larger than unity in the over-critical situation. 
We also remark, that again the spatial average of $\mean{1/r-1/r_c+\mu u_r/v_0^2}_r$ vanishes.

To close the analysis, we have to determine the mean orbital velocity and its variance. 
The latter can be obtained using the conservation of speed \eqref{eq:speed} as  
variance
\begin{eqnarray}
\sigma^2_{zz}\,=\, v_0^2-\sigma^2_{rr}-u^2_z
\label{eq:orbvariance}
\end{eqnarray}
and it remains to determine the stationary $u_z$. 

We could rely here to the result of the self-consistent solution \eqref{eq:consist} 
using the von Mises distribution. But we will follow again the transport equations to obtain an analytic expression.
For this purpose we inspect the balance for the second mixed moment $\rho u_r u_z$. In
case of under-critical coupling $u_r=0$ and $u_z=0$ and both sides of
the balance equation vanish. In case of an over-critical coupling the
equation reads
\begin{eqnarray}
\frac{\partial \rho u_r u_z}{\partial
  t}&&\,=\,-\frac{\partial}{\partial r}
\mean{v_0^3\cos^2(z)\sin(z)}\,+\,2\left(
\frac{1}{r}-\frac{1}{r_c}\right)\bracket{\mean{v_0^3\sin^3(z)}-\mean{v_0^3\cos^2(z)\sin(z)}}
\nonumber
\\ &&+\frac{2\mu}{v_0^2}u_r\bracket{\mean{v_0^3\sin^3(z)}-\mean{v_0^3\cos^2(z)\sin(z)}}-\frac{4\sigma^2}{v_0^2} \rho u_r u_z.
\label{eq:cov_dyn3}
\end{eqnarray}
wherein we start to decouple as we did before. We obtain ($u_z\neq 0$)
\begin{eqnarray}
0= u_z \frac{\partial }{\partial
  t} \rho u_r u_z\,=\,-\bracket{v_0^2-\frac{2\sigma^2}{\mu}}\frac{\partial}{\partial r}\frac{\sigma^2}{\mu} \rho-\bracket{\frac{1}{r}+\frac{1}{r_c}}\rho\bracket{v_0^2 u_z^2-\frac{2\sigma^2}{\mu}\bracket{v_0^2-\frac{2\sigma^2}{\mu}}}\nonumber \\
\label{eq:cov_ss}
\end{eqnarray}
To be in agreement with the equation for the stationary marginal density \eqref{eq:rho_over} we have to require that the mean orbital velocity is homogeneous in space and becomes
\begin{eqnarray}
 u_z^2\,=\, v_0^2-\frac{\sigma^2}{\mu}\bracket{1+\frac{2\sigma^2}{\mu v_0^2}}\,.
\label{eq:mean_orb_ss}
\end{eqnarray}
Afterwards, we can complete the set of stationary characteristics. We find for the orbital variance 
\begin{eqnarray}
\sigma^2_{zz}\,=\, 2\frac{\sigma^4}{\mu^2 v_0^2}
\label{eq:orb_var_ss}
\end{eqnarray}
which is smaller than the radial variance in the cooperative regime. Again we underline that all asymptotic 
characteristics of the velocity are homogeneous in space. Spatial dependence is determined over the dependence of 
the marginal density $\rho(r)$, only.

The orbital velocity exactly vanishes at the critical value of the
coupling strength. For infinitely large coupling all energy is
concentrated in the orbital flow and the variances vanish.  In
comparison with the Gaussian decoupling it approximates the integral
expression \eqref{eq:consist} and fits better the data from the
simulations. Surprisingly, this velocity as well as the variances do
not depend on the interaction strength $\kappa$ with the home.

\subsection{Marginal spatial densities: Smoluchowski equations}
\label{sec:mathe4}
In order to find a kinetic equation for the radial distribution
$\rho(r,t)$,  we take a look upon the dynamics of the mean radial velocity $\rho u_r$ before it vanishes. It reads
\begin{eqnarray}  
 \frac{\partial}{\partial t} u_r\rho = \left(-\frac{\partial}{\partial
   r}\left(\sigma^2_{rr}+u_r^2\right)+\left(\frac{1}{r}-\frac{1}{r_c}\right)\left(\sigma^2_{zz}+u_z^2\right)+
 \left( \frac{\mu}{v_0^2}
 \sigma^2_{zz}-\left(\frac{\sigma}{v_0}\right)^2\right) u_r
 \right)\rho\,.
 \label{eq:radial1}
\end{eqnarray}
Therein we again have neglected  the covariance $\sigma_{rz}$. The radial flux $\rho u_r$ can be eliminated by 
taking the derivative with respect to $r$ on both sides and using the continuity equation \eqref{eq:def_rho}. 
As result the radial flux disappears and we are left with the telegraph equation
\begin{eqnarray}  
  \frac{\partial^2}{\partial t^2} \rho + \frac{1}{\tau_\phi}\left(1-\frac{\mu}{\sigma^2} \sigma_{zz}^2 \right) \, \frac{\partial}{\partial t}
 \rho\,= \,\frac{\partial^2}{\partial r^2}\bracket{u_{rr}^2+\sigma_{rr}^2}\rho-\frac{\partial}{\partial r}
 \left(\frac{1}{r}-\frac{1}{r_c}\right)\bracket{u_z^2+\sigma_{zz}^2}\rho \,.
 \label{eq:radial3}
\end{eqnarray}
and $\tau_\phi$ is from \eqref{eq:time_relax}. This equation reflects still inertia and is of hyperbolic type. The reduction 
to the overdamped Smoluchowski equation as presented below consists in a transition to a parabolic diffusive behavior at 
larger time scales. A profound discussion of similar transitions was recently considered in \cite{Bonilla2018}.

Further on, we distinguish again between under- and over-critical coupling strengths. 
For the expressions of the mean radial and orbital energies in \eqref{eq:radial3} 
we will use the steady state homogeneous expressions as obtained for the two cases in the last section. 
The latter are valid in the asymptotic stationary limit, for which we eventually  find the stationary densities of the two regimes. 

First, if $\mu < \mu_c$ the second moments of the radial and orbital
energy share the same kinetic energy $v_0^2/2$. We get instead
of \eqref{eq:radial3}:
\begin{eqnarray}  
 \tau_\phi  \frac{\partial^2}{\partial t^2} \rho + \left(1-\frac{\mu}{\mu_{\text{crit}}} \right) \, \frac{\partial}{\partial t}
 \rho\,= \frac{v^2_0 \tau_\phi}{2}\bracket{\frac{\partial^2}{\partial r^2} \rho- \frac{\partial}{\partial r}\left(\frac{1}{r}-\frac{1}{r_c}\right)\rho}\,.
 \label{eq:radial2}
\end{eqnarray}

The prefactor of the first temporal derivative at the l.h.s. is always positive for
under-critical coupling. With assumed small the relaxation time $\tau_\phi$ at time scales $t \geqslant \tau_\phi$ 
we can eliminate the inertia in the problem which is presented by
the second temporal derivative. We obtain
the Smoluchowski equation 
\begin{eqnarray}  
 \, \frac{\partial}{\partial t} \rho\,= \,D_{\text{eff}}\,\left( \frac{\partial^2}{\partial r^2}
 \rho-\frac{\partial}{\partial r}
 \left(\frac{1}{r}-\frac{1}{r_c}\right)\rho \right )\,,
 \label{eq:radial4}
\end{eqnarray}
for the radial evolution at time scales $t \geqslant \tau_\phi$. Therein is 
\begin{eqnarray}
D_{\text{eff}}=\frac{v_0^2}{\mu_{\text{crit}}-\mu}
\label{eq:diff_under}
\end{eqnarray}
the effective diffusion coefficient 
of the marginal density of particles coupled with under-critical strength. 
We note that $D_{\text{eff}}$ is always positive for under-critical coupling strength 
but increases with the coupling strength. The stationary solution $\rho(r)$ of Eq. \eqref{eq:radial4} 
is the marginal density as presented in \eqref{eq:rho_under_ss}. Remarkably, it is faster approached with 
increased coupling but exhibits no dependence on $\mu$ and $\sigma^2$.

Let us now concern with the spatial density for over critical-coupling
strength $\mu \geq \mu_c$. We proceed in a similar way but the
occurrence of the mean orbital velocity $u_z$ as derived in
\eqref{eq:mean_orb_ss} changes the derivation significantly. After taking the derivative \eqref{eq:radial1} 
with respect to $r$ we obtain another telegraph equation which looks this time as:
\begin{eqnarray}  
 \tau_\phi\frac{\partial^2}{\partial t^2} \rho + \bracket{1-\frac{\mu_{\text{crit}}}{\mu}} \, \frac{\partial}{\partial t} \rho\,=
 \,\frac{v_0^2}{\mu}\,\left(\frac{\partial^2}{\partial r^2}
 \rho-\frac{\partial}{\partial r} \rho \bracket{\frac{1}{r}-\frac{1}{r_c}} \bracket{\frac{\mu v_0^2}{\sigma^2}-1} \right) \,.
 \label{eq:radial66}
\end{eqnarray}
We notice that the expression in front of the first derivative is always positive
in the over-critical situation. This gives us again the possibility to eliminate the 
inertia in the problem at time scales $t \geqslant \tau_\phi$. We obtain the Smoluchowski equation for 
the over-critical regime which reads this time 
\begin{eqnarray}  
 \frac{\partial}{\partial t} \rho\,= \,D_{\text{eff}}\,\left(\frac{\partial^2}{\partial r^2}
 \rho-\frac{\partial}{\partial r} \left(\frac{1}{r}-\frac{1}{r_c}
\right)\bracket{\frac{v_0^2 \mu}{\sigma^2}-1}\,\rho \right) \,.
 \label{eq:radial6}
\end{eqnarray}
The effective diffusion coefficient reads
\begin{eqnarray}
D_{\text{eff}}=\frac{v_0^2}{\mu-\mu_{\text{crit}}}
\label{eq:diff_over}
\end{eqnarray}
and decays starting from the value of the critical coupling strength. 
The stationary solution of the Smoluchowski equation \eqref{eq:radial6} was presented in the last section as Eq. \eqref{eq:rho_f}.

\section{Discussion}
\label{sec:disc}

\begin{figure}
  \includegraphics[width=0.48\linewidth]{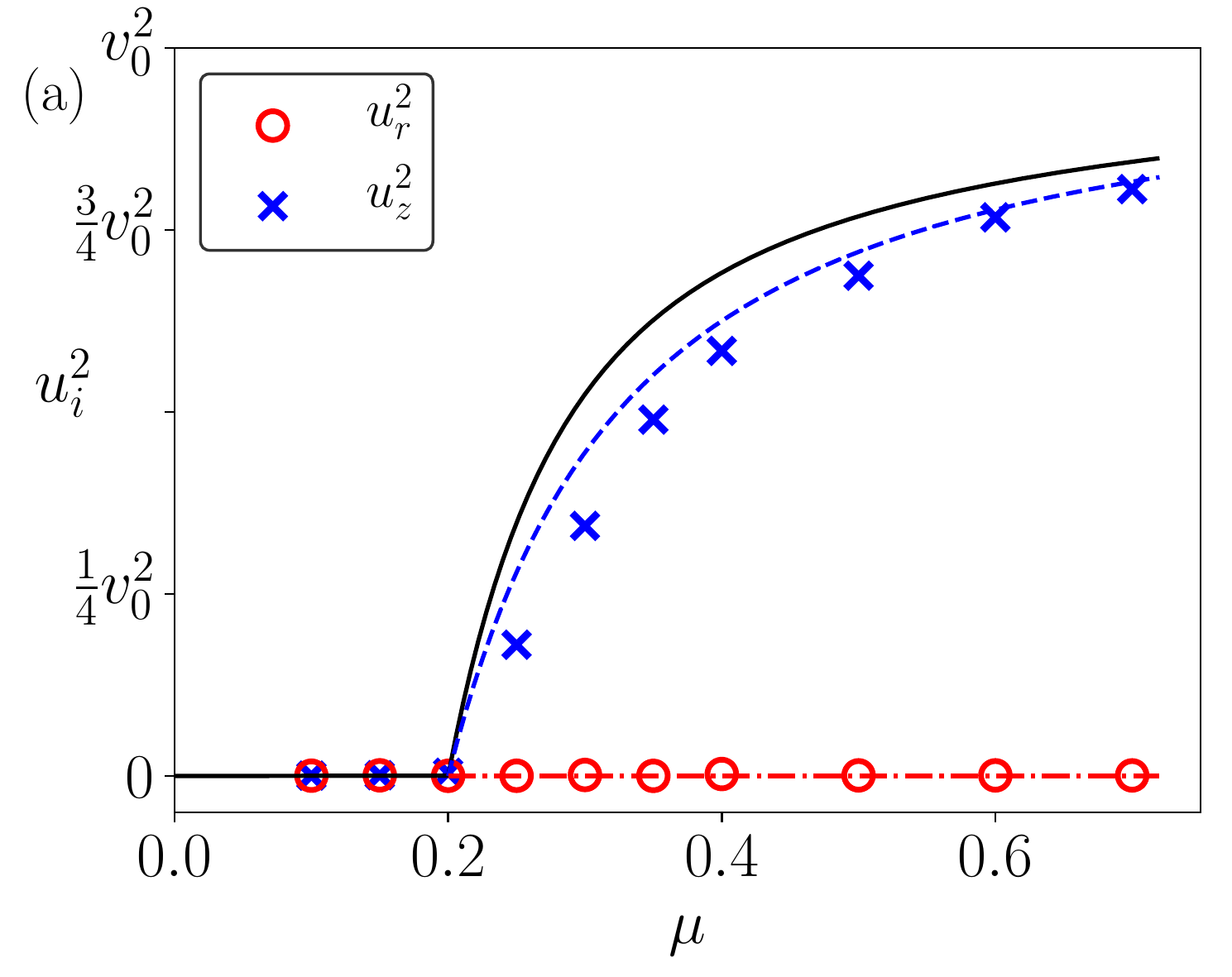}
  \includegraphics[width=0.49\linewidth]{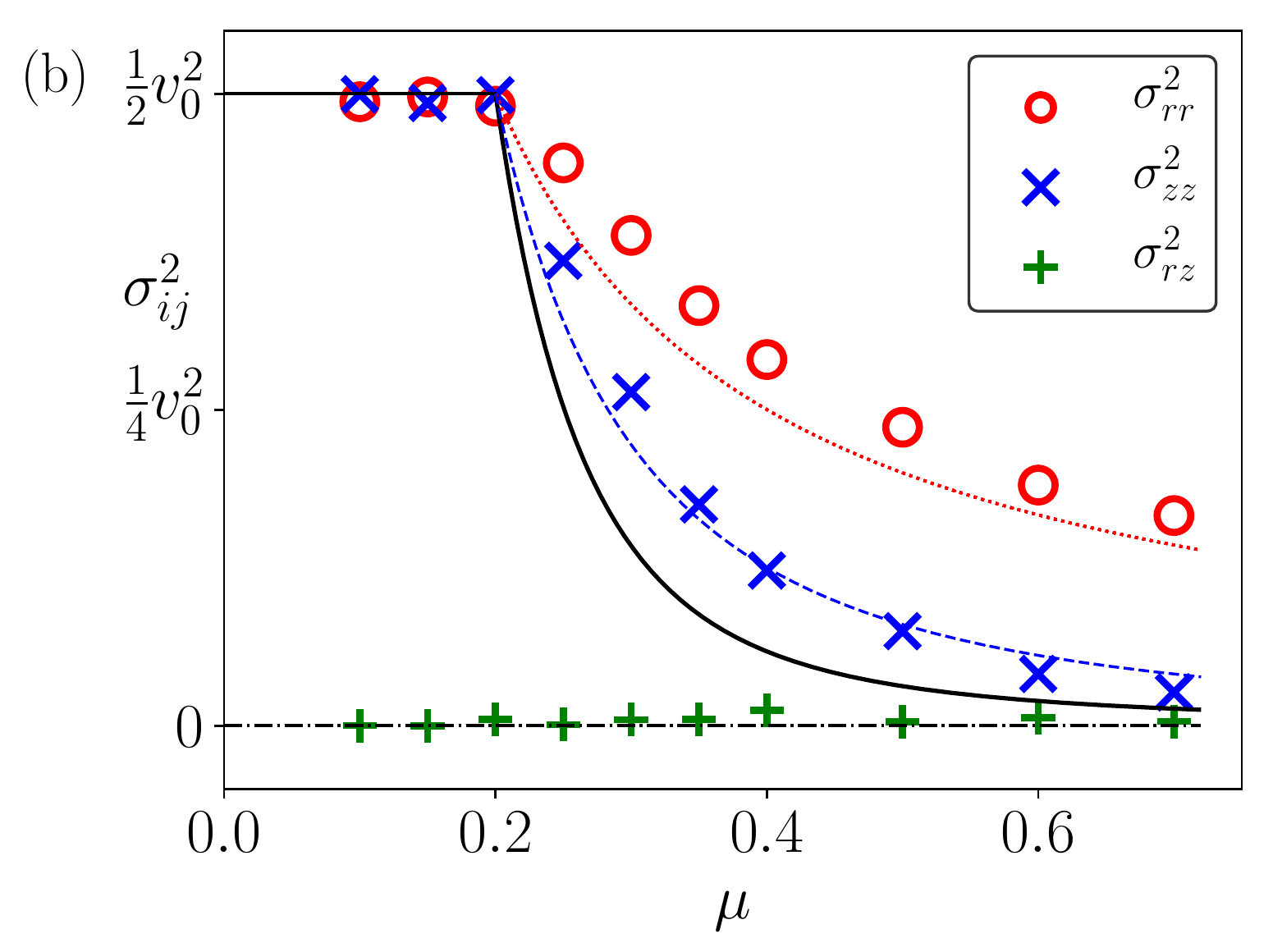}
  \caption{(a): Squared average velocities from simulations as
    symbols and according to Eq.\eqref{eq:consist} as black line and according 
    to Eq.\eqref{eq:mean_orb_ss} as blue dashed line. The red dashed dotted line 
    corresponds to $u_r=0$. The
    critical value for the bifurcation is $\mu_{\text{crit}}=0.2$.
    (b): Variances and covariance from simulations (symbols). 
    Black line is the expectation value of the orbital variance obtained from the von Mises density and according to 
    Eq.\eqref{eq:orb_var_ss} as blue dashed line, according to
    Eq. \eqref{eq:rr_over} as red dotted line as dashed line and according to
    $\sigma^2_{rz}=0$ as dashed dotted line.  Parameters:
    $N=1000$, $v_0=1$, $\kappa=1$, $\sigma^2=0.1$,
    $r_{\text{sens}}=0.2$.  }
  \label{fig:vel_var}
\end{figure}

In this section, we compare our analytical results with simulations and
discuss our findings.  We start by discussing the average velocities.
In Fig. \ref{fig:vel_var}(a), we compare simulation results
for the average velocities $u_z$ and $u_r$ as symbols with the
analytical results from Eq.\eqref{eq:mean_orb_ss} as blue dashed line, the von Mises distribution 
\eqref{eq:consist} as black solid line and the assumption $u_r=0$ as red dashed dotted line.

For $\mu>\mu_{\text{crit}}$, here $\mu_{\text{crit}}=0.2$, the
solution $u_z=0$ becomes unstable and two symmetric solutions
$u_z\neq0$ are stable. The individual foraging motion turns to cooperative circular motion. In the two dimensional $(x,y)$ plane the
searchers start rotating around the home corresponding to the right of
Fig.\ref{fig:traj_align}. While the mean orbital velocity $u_z$ seems to be better
approximated by the dashed line than by the solid line derived from the von Mises density,
we point out, that this might be only the consequence from the simulation of a limited number of particles ($N=1000$).

In Fig. \ref{fig:vel_var}(b), we show
that our analytical results for the variances Eq.
\eqref{eq:rr_over} as red dotted line, Eq.\eqref{eq:orb_var_ss} as dashed line,
the result obtained from the von Mises density as black solid line 
and the covariance as dashed dotted line at zero
approximate the simulation results as symbols reasonably well. At the
critical value of the coupling for the alignment the variances start
decaying. Here, also the variance $\sigma^2_{zz}$ seems to be 
better approximated by the blue dashed dotted line than the black solid line, but this might be a result
of simulations with finite numbers of particles.

Due to the bifurcation there exist two densities $\rho_z$. One for
$u_z<0$ and the second one for the opposite sign of the velocity. We
write $\rho_{z-}$ for the density with the maximum at $z=-\pi/2$.

\begin{figure}
  \includegraphics[width=0.49\linewidth]{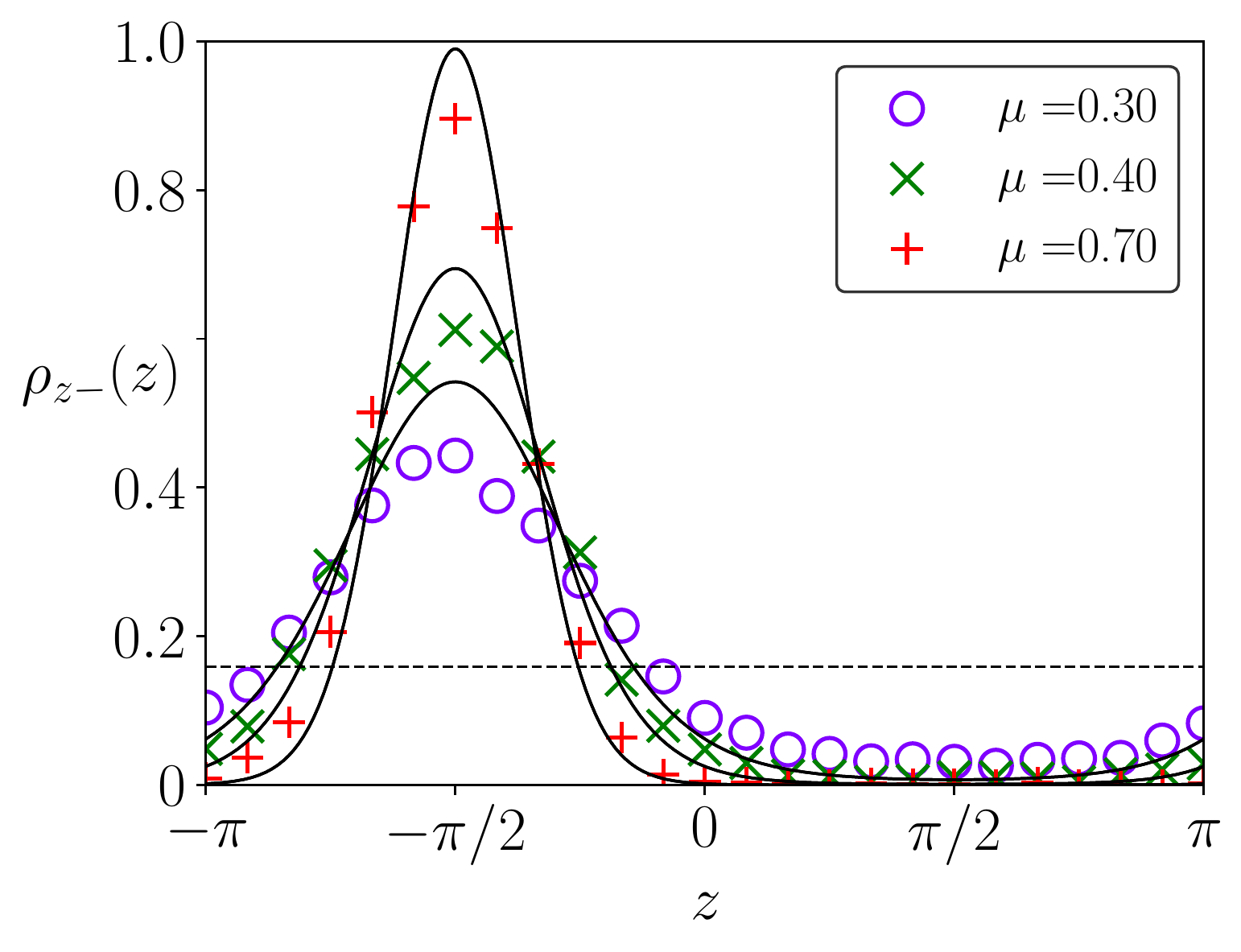}
  \caption{Simulation results for the marginal density of the angle
    $z$ as symbols and in comparison as lines the approximation
    \eqref{eq:ansatz_rhoz2} for three different values of the coupling
    $\mu$. The simulation results were chosen for $u_z<0$, meaning
    clockwise rotation in the $(x,y)$ plane. For counterclockwise
    rotation the densities are shifted by $\pi$.  All three values for
    the coupling are chosen $\mu>\mu_{\text{crit}}=0.2$. The dashed
    line corresponds to a uniform density. This density is approached
    for $\mu=\mu_{\text{crit}}$. Deviation from the theoretical value
    are largely to limited number $N$ of simulated searchers.
    Parameters: $N=1000$, $v_0=1$, $\kappa=1$, $\sigma^2=0.1$,
    $r_{\text{sens}}=0.2$.  }
  \label{fig:rhoz}
\end{figure}
We show in Fig.\ref{fig:rhoz} simulation results for the density
$\rho_{z-}$ as symbols for three different values of the coupling $\mu$.
For the simulations results the positions of the particles are ignored as our analytical also
is space independent.
As lines  we present the corresponding von Mises density \eqref{eq:ansatz_rhoz2}.
The dashed line corresponds to a uniform density $\rho_z=1/(2\pi)$.
This limit is achieved for $\mu=\mu_{\text{crit}}=0.2$ and can be seen in Fig.\ref{fig:rhoz_exp}.
The simulation results are chosen only from steady states where the
searchers rotate clockwise in the $(x,y)$ plane. The density
\eqref{eq:ansatz_rhoz2} was chosen correspondingly. For
counterclockwise rotation the densities are shifted by $\pi$.
The von Mises density approximates the overall space independent angular density from the simulation well.

\begin{figure}
  \includegraphics[width=0.47\linewidth]{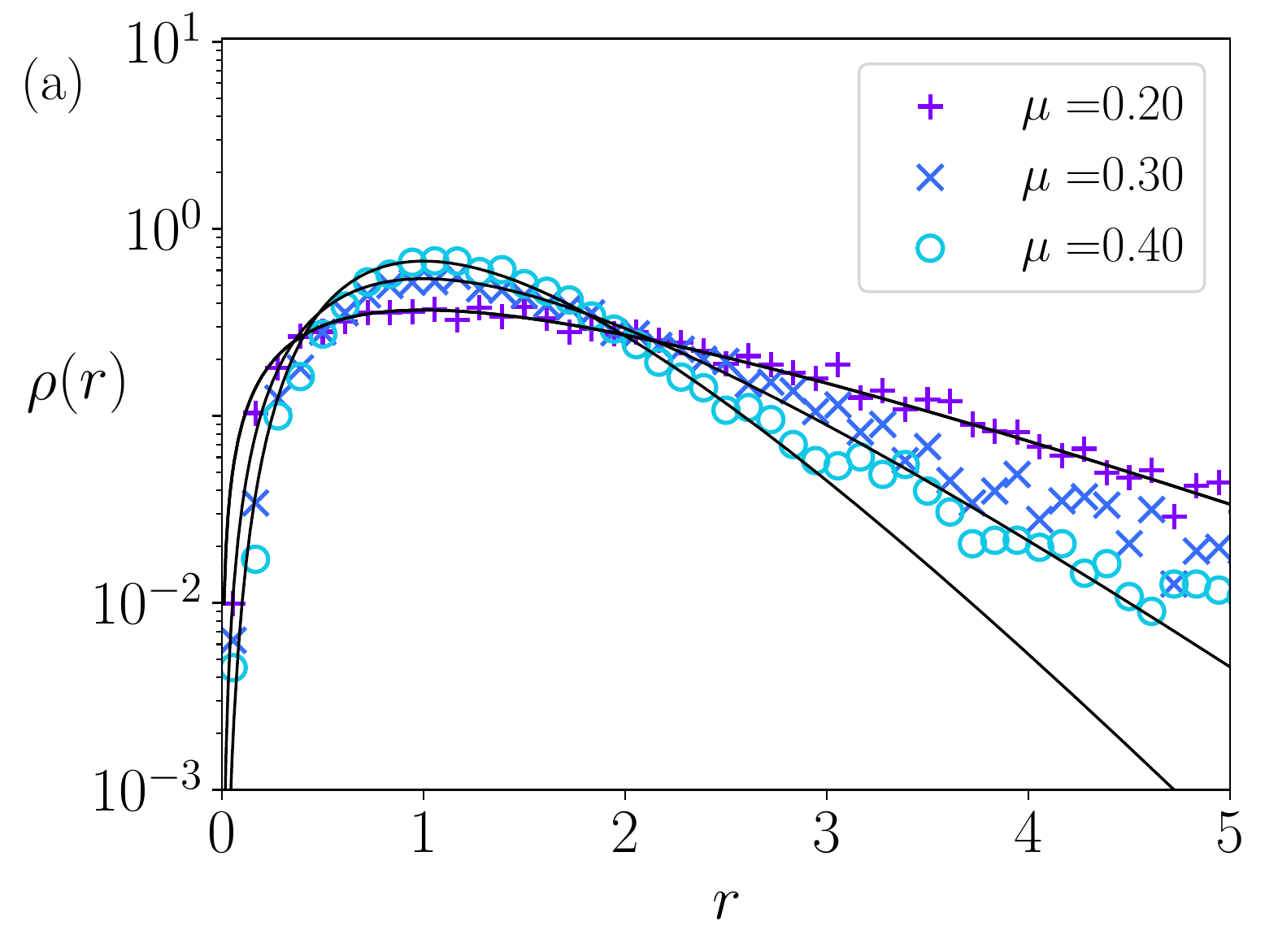}
  \includegraphics[width=0.52\linewidth]{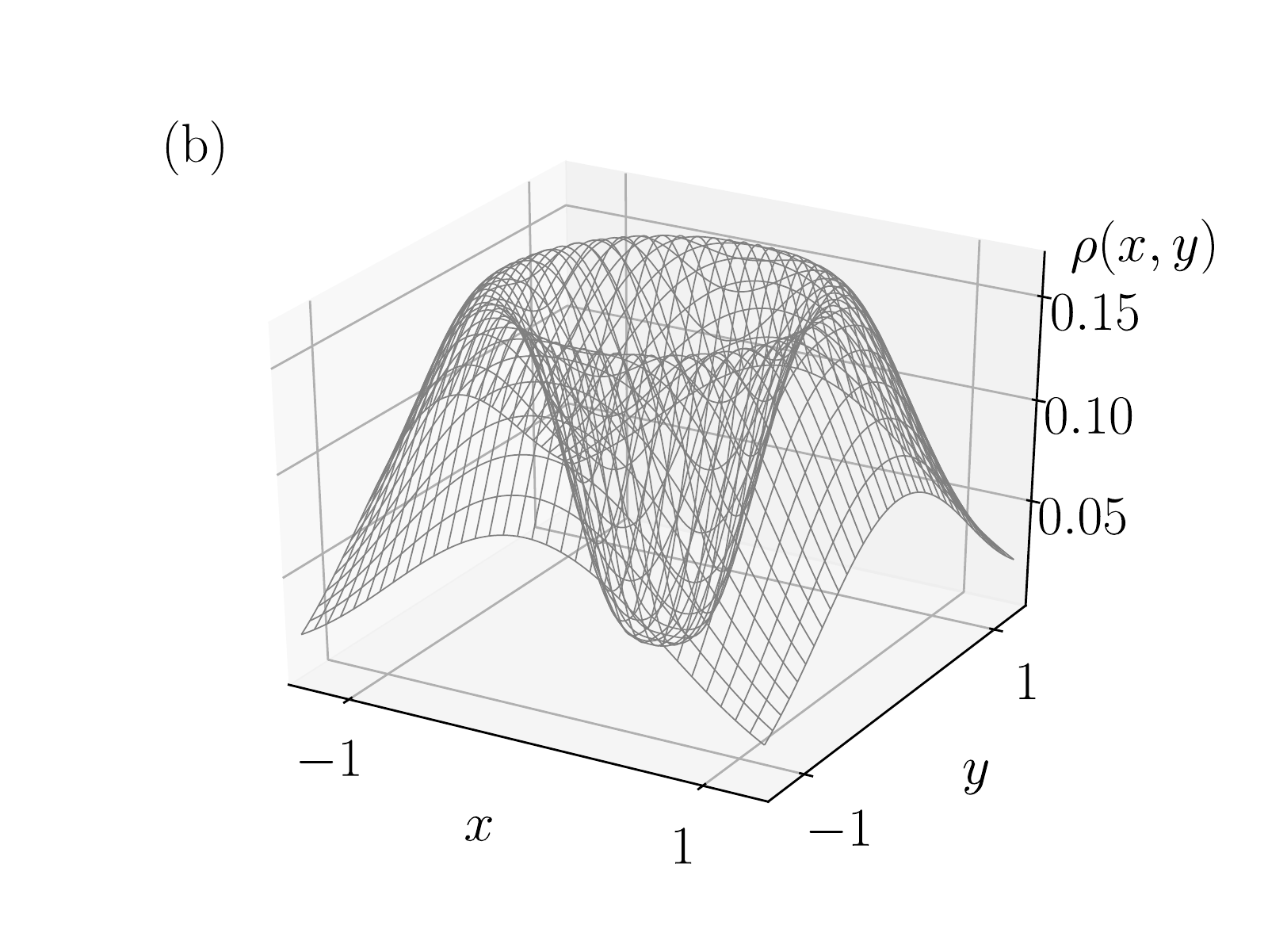}
  \caption{(a): Steady state spatial density for searchers with
    alignment for different different alignment strength $\mu$.  The
    critical value is $\mu_{\text{crit}}=0.2$.  Symbols according to
    simulations of $N=2000$ searchers. Lines according to Eq.
    \eqref{eq:rho_f}. (b): Spatial density for the $(x,y)$ plane.
    For the coupling a value of $\mu=0.7$ was chosen. While for
    vanishing alignment the density in the $(x,y)$ plane is maximal at
    the origin, with an alignment strength large than the critical
    value the density vanishes at the origin.  Parameters: $v_0=1$,
    $\kappa=1$, $\sigma^2=0.1$, $r_{\text{sens}}=0.2$.  }
  \label{fig:rv_uz}
\end{figure}

Finally, we show in Fig.\ref{fig:rv_uz} the spatial density derived from
simulations as symbols and the analytical result from \eqref{eq:rho_f} as line. 
The value
for the alignment strength of $\mu=0.2=\mu_{\text{crit}}$ corresponds
transition value.  Simulation and theory agree well for all displayed
values of alignment strength up to a distance of approximately
$r\approx3$. At this distance the simulation results deviate from the
theory. This is due to the finite number of simulated searchers, i.e.
$N=2000$. At distances $r>3$ the searchers rarely interact with each
other due to low density, while for the theory we assumed infinite
particle numbers.  Parameters are given in the capture of the figure.
In Fig. \ref{fig:rv_uz}(b), we show the result for the
marginal density of the position in the $(x,y)$ plane, pointing out
that for $\mu>\mu_{\text{crit}}$, here chosen
$\mu=0.7>\mu_{\text{crit}}=0.2$, the density at the center decays
with increasing $\mu$. The parameters for this figure are as given in
the caption.

\section{Conclusion}
\label{sec:conc}
We have studied the transition from individual motion to collective motion for an ensemble of coupled local searchers. Individually,
each forager follows a simple version of a path integrating scenario
given by a recently proposed search and return dynamics
\citep{Noetel_2018}. The searchers of the ensemble move with constant
speed and were coupled via an interaction which aligns their heading
directions. The latter models the possibility of the foragers to avoid
collisions and might for instance result from a drag force of the comoved fluid around
the searcher. As a experimental situation for which our model can apply we mention Daphnia moving around an 
attracting light shaft playing the role of a common home. Former studies on the transition from individual to cooperative
trajectories were restricted to 
computer simulations \cite{Erdmann2003,Vollmer,Mach_Schweitzer2007,Levine2001,Thouma}.
Studies describing midges swarms \cite{Gorbonos2016,Reynolds2017,Reynolds2018} do not consider transitions between individual
and collective behavior.

Here we give a profound analytical analysis based on a kinetic theory of an ensemble of searchers. 
Outgoing from a nonlinear FPE \eqref{eq:fpe_align_2} we derive nonlinear transport equations and make a detailed bifurcation 
analysis of the latter.

The situation with a central position also reminisces of models with
binary attractive forces between active particles
\cite{Erdmann2005,Strefler2008,Romanczuk} where the common center of
mass assumes the role of the central position. But with binary
interactions, this position is not fixed in space and might diffuse
due the existence of a Goldstone mode, whereas the home in the current
investigation is fixed.

The important different point in our investigation compared to the large
  number of previous analytical, generic studies on collective swarming behavior 
  of self moving units is the existence of a fixed central
  position, similar to a central field created by the particles but the central place does not shift. 
  The existence of such a central
  attractive location is quite similar to the existence of external fields  \cite{Enculescu, Gorbonos2016,Reynolds2017,Reynolds2018}. 
  It  prevents a propagation of a swarm along fixed directions as often found in swarming situations. 
  As consequence of the aligning interaction a rotation of the entities around their central position 
  is obtained as collective behavior.

In the paper we found a good agreement for the nonequilibrium phase
transition between the numerical simulations and the analytic
expressions. Qualitatively, the undercritical situation does
  not reflect the interactions between particles. Nevertheless it
  might be worth to investigate correlations between the searchers despite the
  absence of collective order as recently reported and discussed for swarms of
  midges \cite{Attanasi2014,Attanasi2014b,Reynolds2018}. In the undercritical situation qualitatively maintain trajectories
  of the single forager model, while for overcritical coupling trajectories change to circular motion.
  
  In our analysis the
marginal density of the distance $\rho(r)$ is the single value
which exhibits asymptotically a space dependence. The other
characteristics as the mean orbital and radial velocity and their
variances get homogeneous in space. The covariance between the two
vector components was neglected in agreement with the results of
simulations. Except the latter one and the mean radial velocity, all mentioned
  characteristics change qualitatively their behavior at the critical
  coupling strength.

We discussed also differences between a decoupling of higher order moments with a von Mises distribution 
and using a Gaussian decoupling. The first one exhibits a better agreement with the simulations since it was 
able to reflect the periodicity of the angular variable. Consistently, the two variances of the velocity components 
differ in their value in case of the von Mises decoupling. We also were able to derive this von Mises distribution for 
the marginal angular velocity outgoing from the mean field FPE. 

The transition from the periodic uniform distribution to a single
peaked at critical coupling strength had shown good agreement
between simulations and analytics. Similar good agreement was
reported for the marginal spatial density $\rho(r)$ of distances from the
home. As effect of the alignment this density becomes more narrow
around its most probable distance $r_{\text{max}}$. The latter is determined by
$r_{\text{max}}=r_c=v_0/\kappa$ in both under- and over-critical coupling
situations. The localization of this peak appears to be independent of
the strength $\mu$ between the entities. The contraction of the pdf around $r_{\text{max}}$
reflects the transition from random loops to circular motion of the single searcher.
 
We also discussed the derivation of a Smoluchowski-equation with
under- and the over-critical coupling strength's. Usually this notion
is applied for the kinetics of the pdf of spatial coordinates
describing the diffusion of Brownian particles \cite{Kramers} in
overdamped situations. Therein, the characteristic time scale is
larger than the corresponding brake time $\tau_\gamma$ of the Brownian
particle \cite{Hwalisz}. In our situation, the role of the brake time
is taken over by $\tau_\phi$. At time scales larger than the angular
relaxation time, the inertia of the entities can be neglected which
has resulted in the elimination of the angular dynamics. Along these
time scales the approximate description of the spatial pdf is given as
a diffusion in an external field which is the topic of a Smoluchowski
equation.

Our analysis was restricted to values $
\mu\leqslant\kappa$. Simulations suggest that for $ \mu\geqslant\kappa$ another
transition to more exotic flocking patterns appears, where the particles no longer move in circles 
around the home and the circular symmetry is broken \cite{Thouma}. Research in such direction also with more 
complex interaction rules is done in \cite{Bonilla2019}.

\section{Acknowledgement}
The authors thank DFG for funding the International research training
group 1740 ``Dynamical Phenomena in Complex Networks: Fundamentals and
Applications'' wherein part of the research has been performed. LSG
thanks W. Ebeling (Rostock), M. Mazza ( Loughborough) and B. Ronacher
(Berlin) for fruitful discussions.
  
\appendix
\section{Limits of the approximation}
\label{sec:limits}
While we focused our analysis on the continuum limit of infinite
particle numbers and vanishing sensing radius, we numerically
investigated the limits of our approximation for the stationary
density \eqref{eq:rho_f}.  For this purpose, we varied the particle
numbers as well as the sensing radius. An important limiting factor is
the number of interacting particles, that means the number of
particles within one sensing radius $r_{\text{sens}}$.  Even for a
rather small total number of simulated particles a transition to a
collective rotation around the home can be found.  The particles are
initially randomly distributed and form over time a persistent
rotating cluster.
\begin{figure}
  \includegraphics[width=0.47\linewidth]{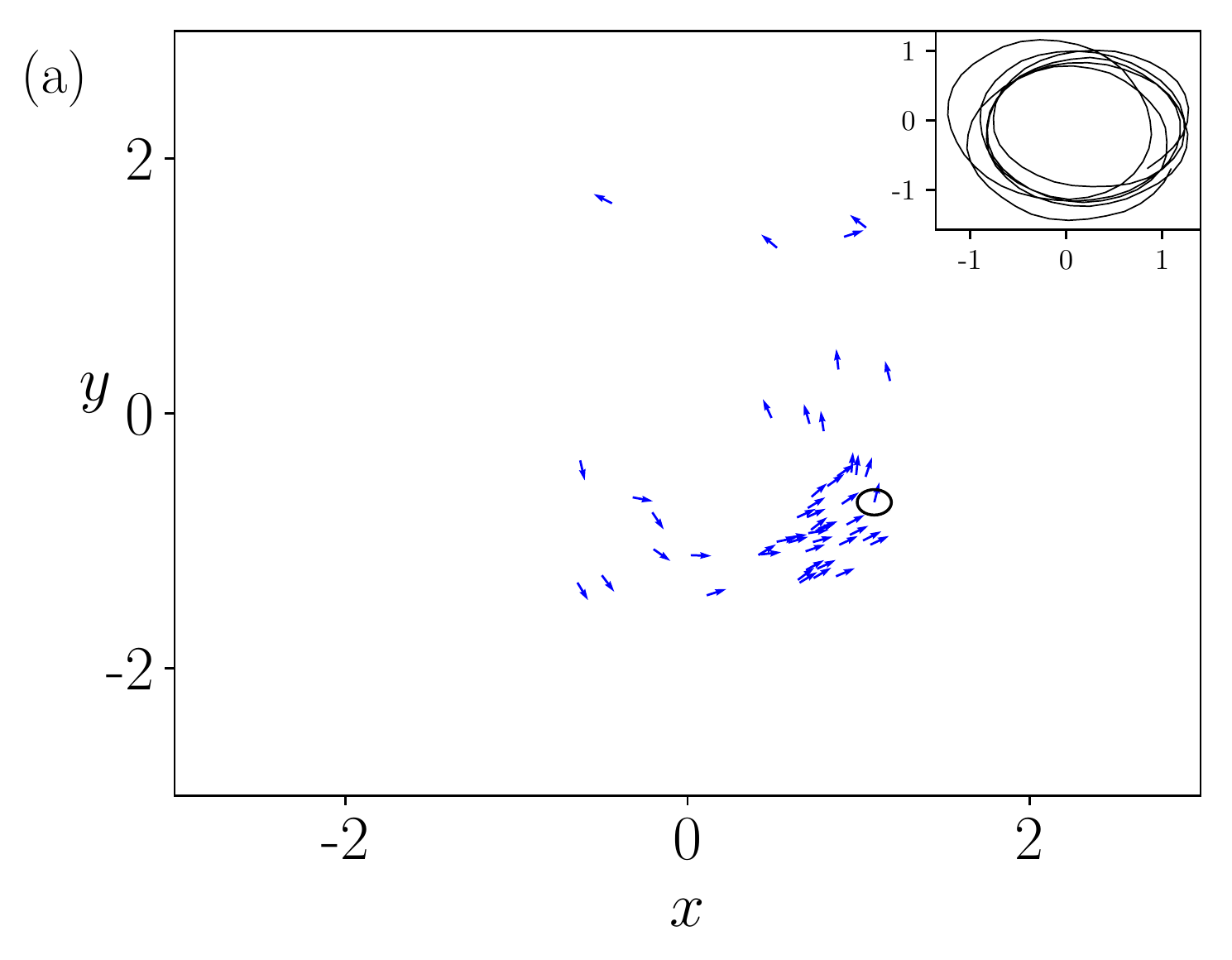}
  \includegraphics[width=0.50\linewidth]{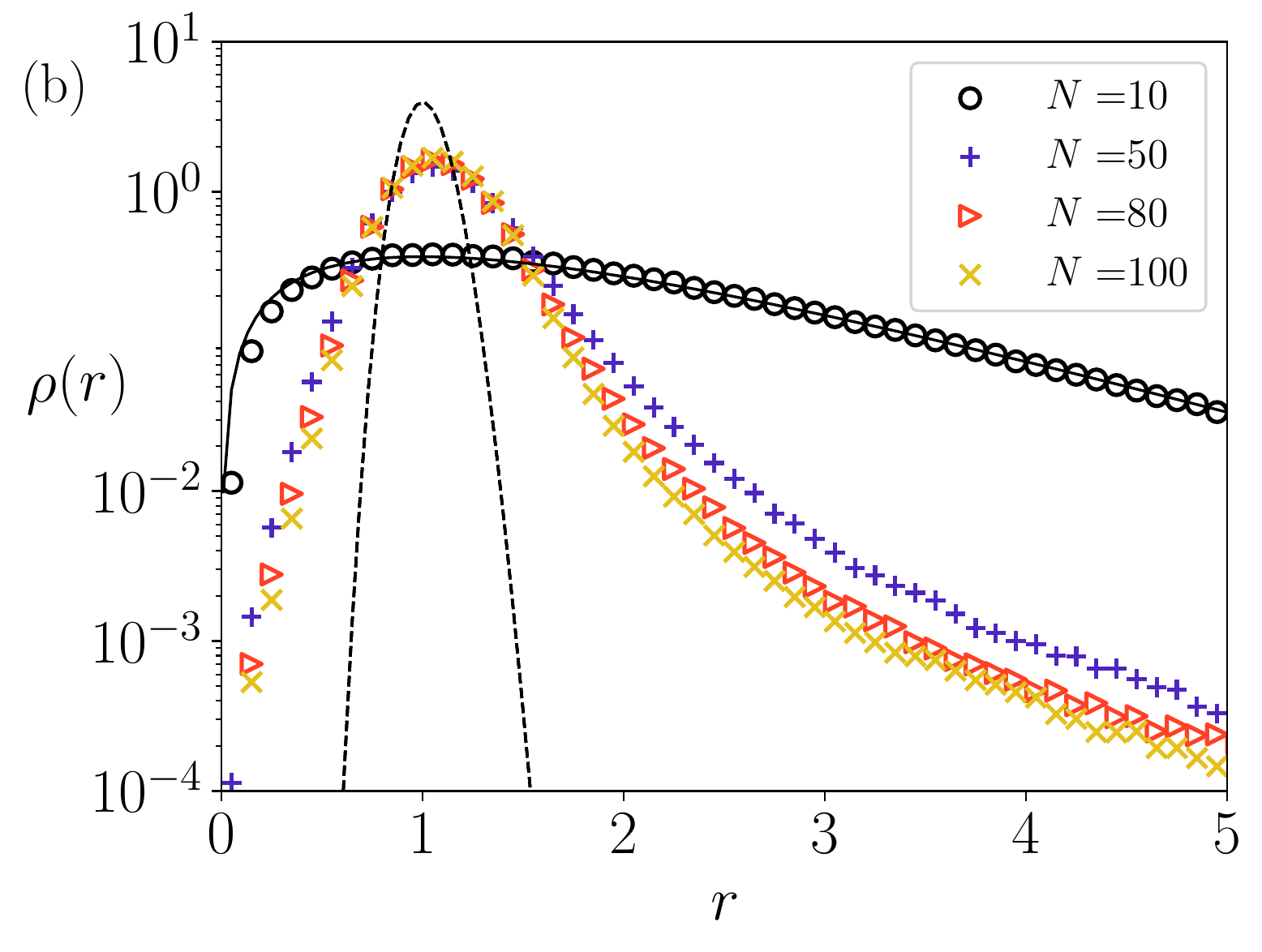}
  \caption{(a): Sample trajectories for a rotating cluster. Total number  of particles $N=50$.
  (b): Spatial densities for different total number $N$ of particles. 
    Symbols according to
    simulations. Dashed line according to Eq.
    \eqref{eq:rho_f}.  Solid line according to Eq. \eqref{eq:rho_under_ss}.
    Parameters: $\mu=1$, $v_0=1$,
    $\kappa=1$, $\sigma^2=0.01$, $r_{\text{sens}}=0.1$.  }
  \label{fig:smallnumbers}
\end{figure}
We present in Fig. \ref{fig:smallnumbers}(a) sample
trajectories for such a rotating cluster. The majority of the
simulated particles $(N=50)$ is part of the cluster, while some
particles follow their own path. The sensing radius is given by
$r_{\text{sens}}=0.1$.  Correspondingly the spatial density does not
follow the undercritical steady state density \eqref{eq:rho_under_ss}
it slowly starts to approach the steady state density given by
\eqref{eq:rho_f}. This can be seen in 
Fig. \ref{fig:smallnumbers}(b). Here the total particle number $N$ is
varied. For $N=10$ the density (black circles) follows the
undercritical or free particles density \eqref{eq:rho_under_ss}
(line), while for $N=50$ the simulation results (blue plus) for the
density already clearly deviates from the undercritical density and
approaches the overcritical density (dashed line). With increasing the
particle numbers the overcritical density is further approached.
Although the particles form a rotating cluster the ensemble average of
several simulations has still rotational symmetry around the home.
\begin{figure}
  \includegraphics[width=0.49\linewidth]{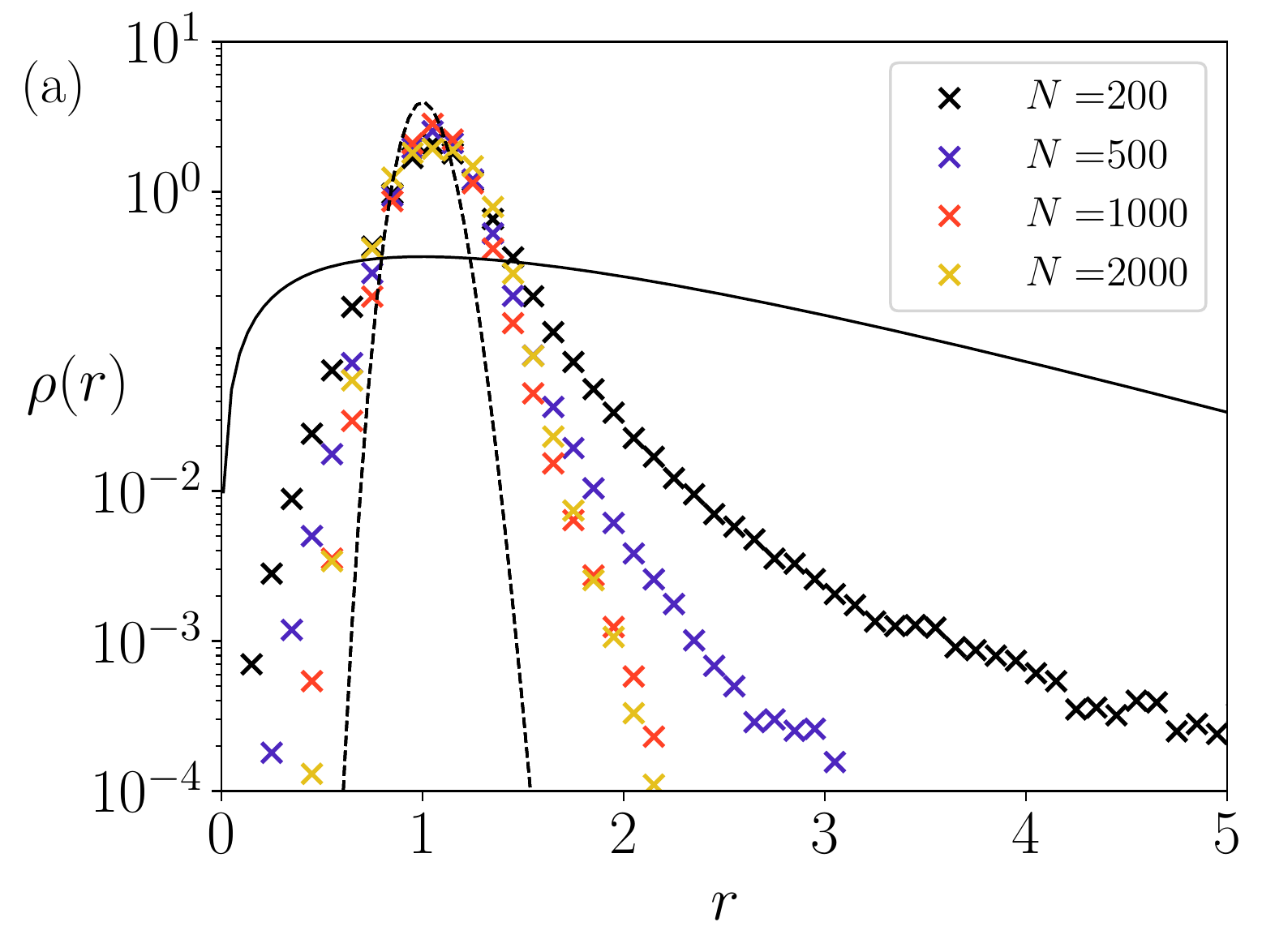}
  \includegraphics[width=0.49\linewidth]{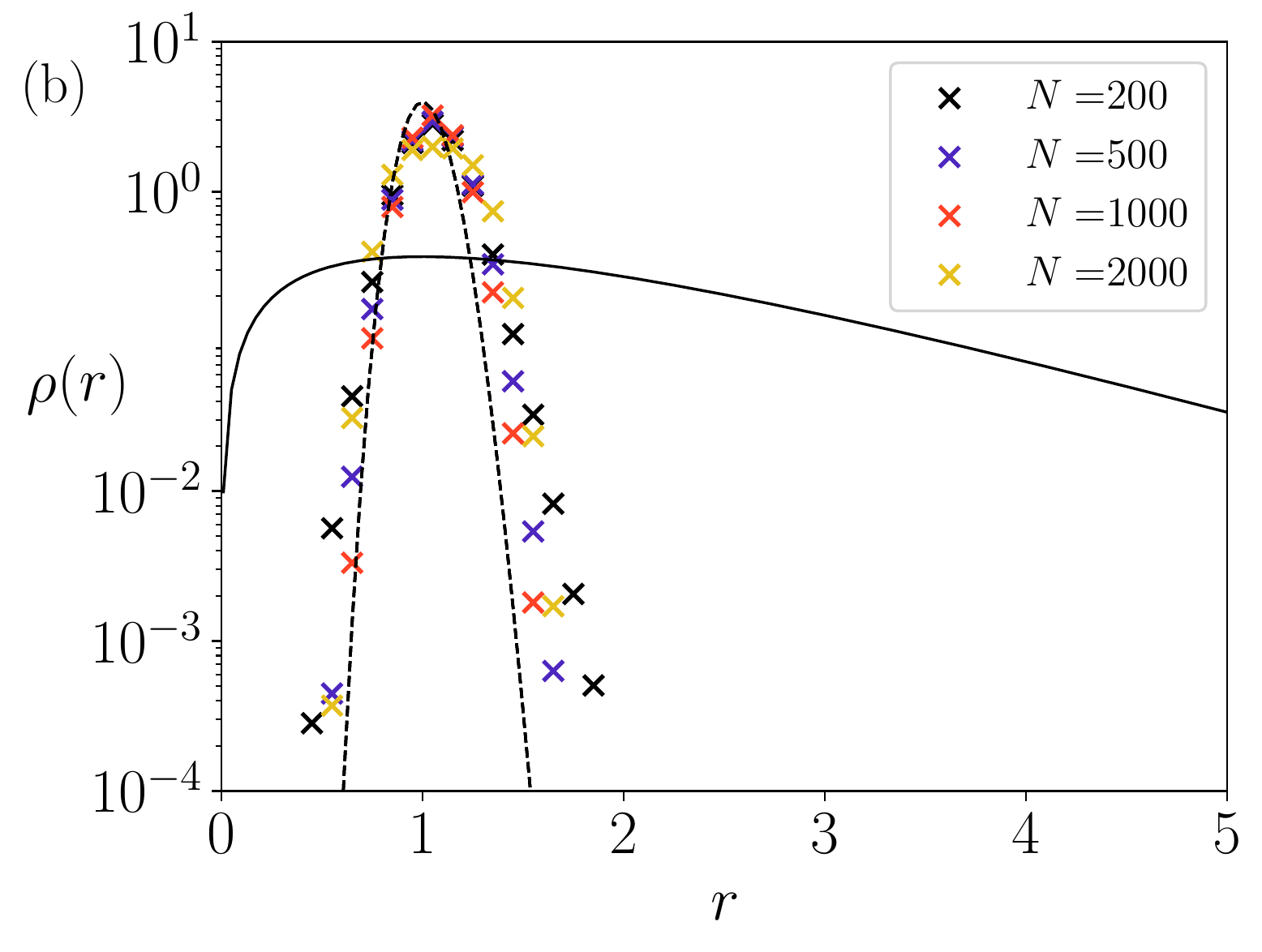}
  \caption{(a): Normalized spatial densities for different total number $N$ of particles. 
  (b): Normalized spatial densities for different total number $N$ of particles
  with only particles with more than 1 neighbors considered. 
    Symbols according to
    simulations. Dashed line according to Eq.
    \eqref{eq:rho_f}.  Solid line according to Eq. \eqref{eq:rho_under_ss}.
    Parameters: $\mu=1$, $v_0=1$,
    $\kappa=1$, $\sigma^2=0.01$, $r_{\text{sens}}=0.05$.  }
  \label{fig:numbers_gt}
\end{figure}
As already mentioned the number of interacting neighbors is important for the validity of the FPE \eqref{eq:fpe_align_2} and its derivations.
In order to clarify this aspect we show in Fig. \ref{fig:numbers_gt} spatial densities for particle numbers $N$ between $200$ and $2000$
but significantly smaller sensing radius than in the previous figures, i.e. $r_{\text{sens}}=0.05$.   
In (a), we show the spatial densities as before. With increasing total particle number the simulation results approach the derived
overcritical density. In (b) however we only consider particles with more than one interacting neighbor and plot their normalized 
density. For these particles the derivations of the FPE \eqref{eq:fpe_align_2}, especially the spatial density \eqref{eq:rho_f} (dashed line)
are valid.

\section{Gaussian decoupling of higher moments}
\label{sec:Gauss}
In this appendix we show results from a Gaussian decoupling of the third order velocity moments. 
Writing  $v_0\cos(z)=u_r  + \delta u_r$ with the first moment is $u_r$, the variance reads as $\sigma_{rr}^2\rho=\mean{\delta u_r^2}$. In an analog way one defines $v_0\sin(z)=u_z+\delta u_z$ and one obtains the variance $\sigma_{zz}^2$ and the covariance. Specifically in the Gaussian approximation,  we will neglect the third central moment of the two velocity components. In particular,  we will put $\mean{\delta u_r^3},\mean{\delta u_z^3} \approx 0$. All other third order moments are expressed by products of the two first moments.

The equations for the mean velocities do not change and are  identical with Eqs. \eqref{eq:trans_r} and \eqref{eq:trans_u}. 
Differences occur for the second moments including the variances. Now the r.h.s. does not vanish as for the von Mises decoupling. We find
\begin{eqnarray}
\frac{\partial}{\partial
  t}  \rho \left(u_r^2+\sigma^2_{rr}\right)&\,\approx\,& -\frac{\partial}{\partial r} \rho u_r \bracket{ u^2_r+3 \sigma^2_{rr}} +\,2\left(
\frac{1}{r}-\frac{1}{r_c}\right) \rho u_r \bracket{u_z^2 +\sigma^2_{zz}} \nonumber \\
&&+ \frac{2\mu}{v_0^2}\rho \bracket{u_r^2\sigma^2_{zz}-u_z^2 \sigma^2_{rr}} +\frac{4\sigma^2}{v_0^2}\left(\frac{\rho v_0^2}{2}-\rho\bracket{u_r^2+\sigma_{rr}}\right).
\label{eq:cov_dyn_r}
\end{eqnarray}
and we neglected the third central  moments $\mean{\delta u_r^3}$.
The orbital energy balance becomes
\begin{eqnarray}
\frac{\partial}{\partial
  t}  \rho \left(u_z^2+\sigma^2_{zz}\right)&\,\approx\,& -\frac{\partial}{\partial r} \rho u_r\bracket{u_z^2+\sigma^2_{zz}}  -\,2\left(
\frac{1}{r}-\frac{1}{r_c}\right)\rho u_r \bracket{u_z^2 +\sigma^2_{zz}}\nonumber \\
&&- \frac{2\mu}{v_0^2}\rho (u_r^2\sigma^2_{zz}-u_z^2 \sigma^2_{rr}) +\frac{4\sigma^2}{v_0^2}\left(\frac{\rho v_0^2}{2}-\rho\bracket{u_z^2+\sigma_{zz}}\right).
\label{eq:cov_dyn_z}
\end{eqnarray}
Eventually we formulate the equation for the mixed second moment
\begin{eqnarray}
\frac{\partial}{\partial
  t}  \rho u_r u_z &\,\approx\,& -\frac{\partial}{\partial r} \rho u_z \bracket{ u^2_r+ \sigma^2_{rr}}  +\,2\left(
\frac{1}{r}-\frac{1}{r_c}\right) \rho u_z \bracket{u_z^2 + 3\sigma^2_{zz} -\bracket{u_r^2+\sigma_{rr}}}\nonumber \\
&&+\frac{2\mu}{v_0^2}\rho u_r u_z \bracket{\sigma^2_{zz}+\sigma^2_{rr}} -\frac{4\sigma^2}{v_0^2}\rho u_r u_z\,.
\label{eq:cov_dyn_rz}
\end{eqnarray}

We now discuss the asymptotic long time behavior of the introduced moments using the Gaussian decoupling. Asymptotically, 
the marginal density becomes stationary $\rho(r,t) \to \rho(r)$. In accordance with the continuity equation and no-flux boundary 
conditions the radial flux $u_r$ has to
vanish $u_r=0$. The equation for the radial energy balance simplifies to
\begin{eqnarray}
\frac{\partial \sigma^2_{rr}}{\partial t}\,=\,\frac{4\sigma^2}{v_0^2}\left(\frac{v_0^2}{2} - \sigma^2_{rr}\left( 1 + \frac{\mu}{2\sigma^2} u_z^2\right)\right)\,.
\label{eq:cov_dyn11}
\end{eqnarray}
Asymptotically, the radial variance becomes also stationary and homogeneous in space. We set the l.h.s. to zero and get the algebraic equation
\begin{eqnarray}
\sigma^2_{rr}\,=\,\frac{v_0^2}{2 \left(1+\frac{\mu}{2 \sigma^2} u_z^2\right)}\,.
\label{eq:as_cov}
\end{eqnarray}

Insertion of this variance yields the stationary  solutions of $u_z$ which obey the equation
\begin{eqnarray}
0= \left(\frac{\mu}{2 \left(1+\frac{\mu}{2 \sigma^2} u_z^2\right)} -\left(\frac{\sigma}{v_0}\right)^2 \right)u_z.
\label{eq:uz_dyn}
\end{eqnarray}
Solutions of this equation exhibit a pitchfork bifurcation. The mean orbital
velocity of the coupled searchers vanishes asymptotically $u_z=0$ which is stable  for a coupling 
strength smaller than the critical value $\mu \leqslant \mu_c$. 
The expression for the critical coupling strength coincides with the value of the von Mises decoupling \eqref{eq:mucrit}. 
Likewise in the former under-critical case, the variances in this state follow from Eqs.\eqref{eq:cov_dyn_z} and \eqref{eq:as_cov} 
\begin{eqnarray}
\sigma^2_{rr}\,=\,\sigma^2_{zz}\,=\,\frac{v_0^2}{2}\,.
\label{eq:as_var}
\end{eqnarray}
The r.h.s of the equation for the mixed second moment \eqref{eq:cov_dyn_rz} vanishes and the equation for the 
radial flux \eqref{eq:trans_r} results in the stationary Smoluchowski-equation \eqref{eq:radial4} for the 
marginal density with the solution \eqref{eq:rho_under_ss}. 

In contrast, the situation differs for coupling strengths $\mu \geqslant \mu_c$ larger than the critical value \eqref{eq:mucrit}. 
The non-vanishing orbital velocity becomes in the Gaussian approximation
\begin{eqnarray}
u_z\,=\,\pm v_0 \sqrt{1-\frac{\mu_c}{\mu}}\,=\,\pm v_0 \sqrt{1-\frac{2\sigma^2}{\mu}}\,.
\label{eq:orb_Gauss}
\end{eqnarray}
In this state the variances coincide again and behave thus differently from the von Mises decoupling and the simulations. Both become
\begin{eqnarray}
\sigma^2_{rr}\,=\,\sigma^2_{zz}\,= \,\frac{v_0^2}{2}\,\frac{\mu_c}{\mu}\,=\,\frac{\sigma^2}{\mu}.
\label{eq:as_var2}
\end{eqnarray}
They vanish for strong coupling where all kinetic energy is contained in the mean orbital motion since $u^2_z=v^2_0$. 

In result, the main difference between the two decoupling scenarios consists in the different partition of kinetic energy at the orbital degree of freedom. 
We point out that the results of the von Mises decoupling fit better with the numerical findings than those from the Gaussian decoupling. 
The radial component of the energy equals in both decoupling scenarios. The orbital variance equals the radial variance in the Gaussian decoupling. 

However, in the von Mises decoupling more energy is spend to the systematic orbital motion $u_z^2$ and less energy to its fluctuation $\sigma_{zz}^2$. It is a consequence of the less broad von Mises distribution compared to its Gaussian counterpart.

In detail, in the Gaussian decoupling we get for the third order moment in the energy balance
\begin{equation}
\frac{\mu}{v_0^2} u_z \mean{v_0^3 \cos^2(z) \sin(z)}\,\approx\, \frac{\mu}{v_0^2} u_z^2 \sigma_{rr}^2\,.
\label{eq:item_gauss}
\end{equation}
The same moment approximated with the von Mises decoupling as 
\begin{equation}
\frac{\mu}{v_0^2} u_z \mean{v_0^3 \cos^2(z) \sin(z)}\,\approx\,\frac{\sigma^2}{v_0^2} \bracket{u_z^2+\sigma_{zz}^2-\sigma_{rr}^2}\,. 
\label{eq:item_mises}
\end{equation}
But, remarkably, the third moment as a solution of the two differently decoupled transport equations possesses the same expression. 
In both cases it becomes asymptotically  
\begin{equation}
\frac{\mu}{v_0^2} u_z \mean{v_0^3 \cos^2(z) \sin(z)}\,\approx\, \frac{\sigma^2}{v_0^2} \bracket{v_0^2-\frac{2\sigma^2}{v_0^2}}\,,
\label{eq:item_mises_gauss}
\end{equation}
But whereas the Gaussian decoupling relies on the asymptotic values of the mean orbital velocity $u_z$ from \eqref{eq:orb_Gauss} and the  variance $\sigma_{rr}$ from \eqref{eq:as_var2}, we made use in the von Mises scheme from $u_z$ given by \eqref{eq:mean_orb_ss} and its variances, $\sigma_{zz}$ from \eqref{eq:orb_var_ss} and $\sigma_{rr}$ from \eqref{eq:rr_over}. 

\begin{figure}[t]
  \includegraphics[width=0.47\linewidth]{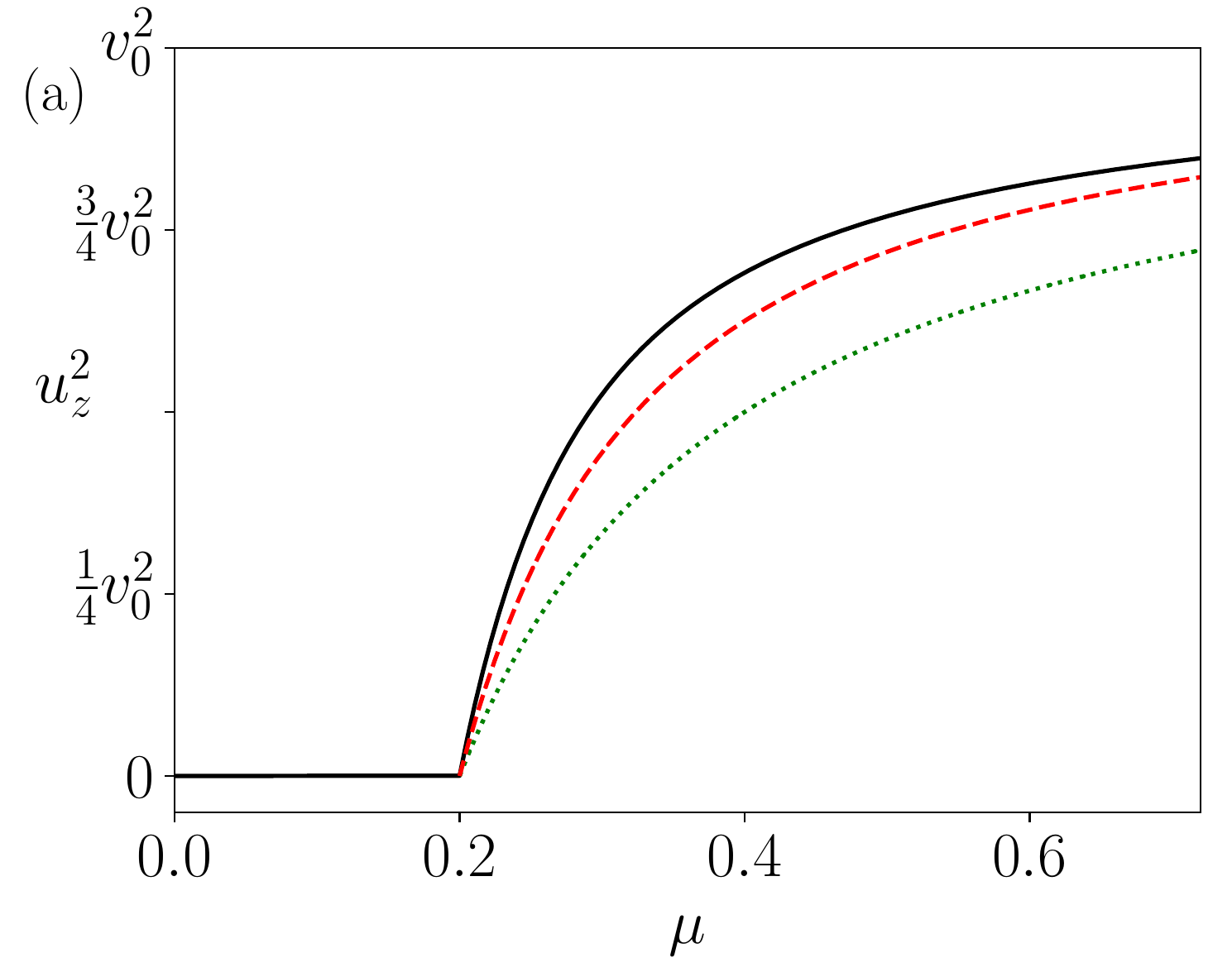}
  \includegraphics[width=0.49\linewidth]{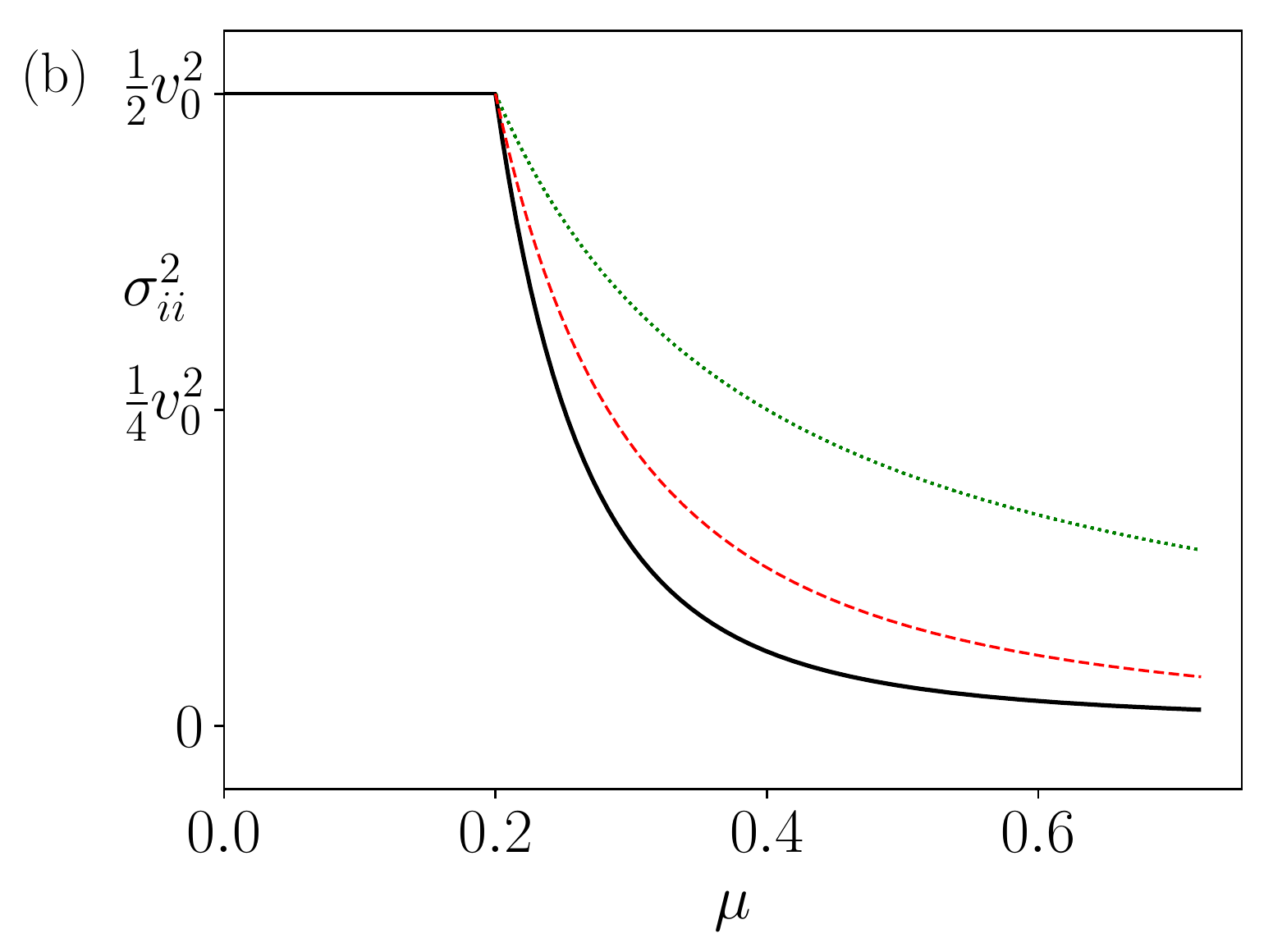}
  \caption{Comparison of the analytical results of the two decoupling schemes. 
  (a): Squared average velocities according to the decoupling using the von Mises distribution Eq.\eqref{eq:consist} (black, line), Eq.\eqref{eq:mean_orb_ss} (red, dashed) 
  and to the Gaussian decoupling \eqref{eq:orb_Gauss} (green, dotted). All nonvanishing mean orbital velocities 
  appear at the same critical value for the bifurcation which is in the particular case $\mu_{\text{crit}}=0.2$.
    (b): Orbital variances obtained from the von Mises density (black, line) and according to 
    Eq.\eqref{eq:orb_var_ss} (red, dashed) and Eq. \eqref{eq:rr_over} (green, dotted). Both Gaussian counterparts coincides with each other, 
    and with variance of the radial velocity fluctuations of the von Mises decoupling scenario. The Gaussian variances 
    does not account that the orbital velocity fluctuations possess a smaller variance. Parameters:
    $N=1000$, $v_0=1$, $\kappa=1$, $\sigma^2=0.1$.  }
  \label{fig:vel_var1}
\end{figure}

With the Gaussian decoupling the stationary marginal radial density $\rho(r)$ obeys the same Smoluchowski equation 
as in the case of the decoupling by help of  the von Mises distribution. Insertion of the steady states for 
$u_z^2, \sigma^2_{rr}$ and $\sigma^2_{zz}$ into Eq.\eqref{eq:trans_r}  results in 
\begin{eqnarray}
  \label{eq:radial_g}
  0  = \frac{\sigma^2}{\mu}\frac{\partial}{\partial r}\left\{\frac{\partial}{\partial r}-\left(\frac{1}{r}-\frac{\kappa}{v_0}\right)\left(\frac{\mu v_0^2}{\sigma^2}-1\right)\right\}\rho(r)\,, 
\end{eqnarray}
which possesses the solution \eqref{eq:rho_f}. 
Thus, the von Mises decoupling improves only the orbital characteristics of the studied problem.

\section{Mechanics of the searcher}
\label{sec:mech}
The search and return mechanism of a single searcher Eqs. \eqref{eq:r_dot} and \eqref{eq:dottheta0} allow a mechanical 
interpretation at the two dimensional plane (see Fig.\ref{fig:schematic}). We formulate the Newton law of mechanics 
\begin{equation}
\frac{{\rm d}}{{\rm d t}}\, \vec{v}\,=\,\gamma\bracket{v_0 \vec{e}_v\,-\,\vec{v}} \,-\,k \vec{e}_r \,.
\label{eq:newton}
\end{equation}
for the two dimensional motion of a particle. The latter is damped by Stokes friction with coefficient $\gamma$ and pushed by an internal motor with strength $\gamma v_0$. The velocity vector is $\vec{v}(t)=v(t) \vec{e}_v$  with speed $v(t)$, heading direction $\phi(t)$ and unit velocity vector $\vec{e}_v(t)=\{\cos(\phi(t)),\sin(\phi(t))\}$. Additionally, a central constant attractive force acts along the unit vector of the position vector $\vec{e}_r(t)=\{\cos(\beta(t),\sin(\beta(t))\}$ 
with orientation given by the angle $\beta(t)$. The force is independent of the distance $r(t)$ from the home and depends only on $\beta(t)$.

Now we allow infinitely large damping $\gamma \to \infty$ as considered in the overdamped limit. Projection on the speed and heading variables gives 
\begin{equation}
\vec{v}(t)\,=\,v_0\,\vec{e}_v(t)\,.
\label{eq:speed1}
\end{equation} 
We assumed that the limit $k/\gamma$ vanishes for large damping. The situation reduces to a microswimmer with constant speed. 

Alternatively, the heading dynamics reads
\begin{equation}
\dot{\phi}\,=\, \frac{k}{v_0}\,\sin(\phi-\beta)\,,
\label{eq:heading}
\end{equation}  
with $\kappa=k/v_0$.  We got Eq. \eqref{eq:dottheta0} and, thus,  the path integration program has been traced back from a mechanical problem given by the Newton law \eqref{eq:newton} and formulating mechanical forces. During the path integration the searcher is always aware of the direction towards the home without having knowledge about the distance.

An alternative interesting model was recently proposed by Waldner and Merkle \cite{Waldner2018}. It considers a central harmonic force and recovers beautiful rosette like trajectories around the home which have been reported several times for dessert ants \cite{Mittelstaedt,Vickerstaff_2005,Ronnacher2008,Collett2013}. Formally, one should 
replace $ k\to k_0*r(t)$ in \eqref{eq:newton} and, subsequently, in\eqref{eq:heading}. 

To take $k$ as a function of the distance might be a fruitful direction to get better agreement with experimentally found trajectories. One could consider general central forces with a distant dependent $k(r)$ in \eqref{eq:newton} and will get the same $k(r)$ in the heading dynamics \eqref{eq:dottheta0}. Some particular cases have been investigated in \cite{Noetel_2018}. Therein, the deterministic case well as the stationary stochastic problem with added angular noise in the $\phi$-dynamics have been analytically solved.

\section*{References}

\begin{thebibliography}{95}%
\makeatletter
\providecommand \@ifxundefined [1]{%
 \@ifx{#1\undefined}
}%
\providecommand \@ifnum [1]{%
 \ifnum #1\expandafter \@firstoftwo
 \else \expandafter \@secondoftwo
 \fi
}%
\providecommand \@ifx [1]{%
 \ifx #1\expandafter \@firstoftwo
 \else \expandafter \@secondoftwo
 \fi
}%
\providecommand \natexlab [1]{#1}%
\providecommand \enquote  [1]{``#1''}%
\providecommand \bibnamefont  [1]{#1}%
\providecommand \bibfnamefont [1]{#1}%
\providecommand \citenamefont [1]{#1}%
\providecommand \href@noop [0]{\@secondoftwo}%
\providecommand \href [0]{\begingroup \@sanitize@url \@href}%
\providecommand \@href[1]{\@@startlink{#1}\@@href}%
\providecommand \@@href[1]{\endgroup#1\@@endlink}%
\providecommand \@sanitize@url [0]{\catcode `\\12\catcode `\$12\catcode
  `\&12\catcode `\#12\catcode `\^12\catcode `\_12\catcode `\%12\relax}%
\providecommand \@@startlink[1]{}%
\providecommand \@@endlink[0]{}%
\providecommand \url  [0]{\begingroup\@sanitize@url \@url }%
\providecommand \@url [1]{\endgroup\@href {#1}{\urlprefix }}%
\providecommand \urlprefix  [0]{URL }%
\providecommand \Eprint [0]{\href }%
\providecommand \doibase [0]{http://dx.doi.org/}%
\providecommand \selectlanguage [0]{\@gobble}%
\providecommand \bibinfo  [0]{\@secondoftwo}%
\providecommand \bibfield  [0]{\@secondoftwo}%
\providecommand \translation [1]{[#1]}%
\providecommand \BibitemOpen [0]{}%
\providecommand \bibitemStop [0]{}%
\providecommand \bibitemNoStop [0]{.\EOS\space}%
\providecommand \EOS [0]{\spacefactor3000\relax}%
\providecommand \BibitemShut  [1]{\csname bibitem#1\endcsname}%
\let\auto@bib@innerbib\@empty
\bibitem [{\citenamefont {Klages}(2017)}]{Klages}%
  \BibitemOpen
  \bibfield  {author} {\bibinfo {author} {\bibfnamefont {R.}~\bibnamefont
  {Klages}},\ }\enquote {\bibinfo {title} {{Search for Food of Birds, Fish and
  Insects}},}\ in\ \href@noop {} {\emph {\bibinfo {booktitle} {{Diffusive
  Spreading in Nature, Technology and Society}}}},\ \bibinfo {editor} {edited
  by\ \bibinfo {editor} {\bibfnamefont {A.}~\bibnamefont {Bunde}}, \bibinfo
  {editor} {\bibfnamefont {J.}~\bibnamefont {Caro}}, \bibinfo {editor}
  {\bibfnamefont {J.}~\bibnamefont {K\"arger}}, \ and\ \bibinfo {editor}
  {\bibfnamefont {G.}~\bibnamefont {Vogl}}}\ (\bibinfo  {publisher} {Springer,
  Cham},\ \bibinfo {year} {2017})\ pp.\ \bibinfo {pages} {49--69}\BibitemShut
  {NoStop}%
\bibitem [{\citenamefont {Mittelstaedt}\ and\ \citenamefont
  {Mittelstaedt}(1980)}]{Mittelstaedt}%
  \BibitemOpen
  \bibfield  {author} {\bibinfo {author} {\bibfnamefont {M.}~\bibnamefont
  {Mittelstaedt}}\ and\ \bibinfo {author} {\bibfnamefont {H.}~\bibnamefont
  {Mittelstaedt}},\ }\href {\doibase 10.1007/BF00450672} {\bibfield  {journal}
  {\bibinfo  {journal} {Naturwissenwschaften}\ }\textbf {\bibinfo {volume}
  {67}},\ \bibinfo {pages} {566} (\bibinfo {year} {1980})}\BibitemShut
  {NoStop}%
\bibitem [{\citenamefont {B\'enichou}\ \emph {et~al.}(2011)\citenamefont
  {B\'enichou}, \citenamefont {Loverdo}, \citenamefont {Moreau},\ and\
  \citenamefont {Voituriez}}]{Bennichou2011}%
  \BibitemOpen
  \bibfield  {author} {\bibinfo {author} {\bibfnamefont {O.}~\bibnamefont
  {B\'enichou}}, \bibinfo {author} {\bibfnamefont {C.}~\bibnamefont {Loverdo}},
  \bibinfo {author} {\bibfnamefont {M.}~\bibnamefont {Moreau}}, \ and\ \bibinfo
  {author} {\bibfnamefont {R.}~\bibnamefont {Voituriez}},\ }\href {\doibase
  10.1103/RevModPhys.83.81} {\bibfield  {journal} {\bibinfo  {journal} {Rev.
  Mod. Phys.}\ }\textbf {\bibinfo {volume} {83}},\ \bibinfo {pages} {81}
  (\bibinfo {year} {2011})}\BibitemShut {NoStop}%
\bibitem [{\citenamefont {Cheng}(1995)}]{Cheng}%
  \BibitemOpen
  \bibfield  {author} {\bibinfo {author} {\bibfnamefont {K.}~\bibnamefont
  {Cheng}},\ }in\ \href {\doibase 10.1016/S0079-7421(08)60370-9} {\emph
  {\bibinfo {booktitle} {{Psychology of Learning and Motivation}}}},\ \bibinfo
  {series} {Psychology of Learning and Motivation}, Vol.~\bibinfo {volume}
  {33}\ (\bibinfo  {publisher} {Academic Press},\ \bibinfo {year} {1995})\ pp.\
  \bibinfo {pages} {1 -- 21}\BibitemShut {NoStop}%
\bibitem [{\citenamefont {Wang}(2003)}]{Wang}%
  \BibitemOpen
  \bibfield  {author} {\bibinfo {author} {\bibfnamefont {R.~F.}\ \bibnamefont
  {Wang}},\ }in\ \href {\doibase 10.1016/S0079-7421(03)01004-1} {\emph
  {\bibinfo {booktitle} {{Cognitive Vision}}}},\ \bibinfo {series} {Psychology
  of Learning and Motivation}, Vol.~\bibinfo {volume} {42}\ (\bibinfo
  {publisher} {Academic Press},\ \bibinfo {year} {2003})\ pp.\ \bibinfo {pages}
  {109 -- 156}\BibitemShut {NoStop}%
\bibitem [{\citenamefont {Seelig}\ and\ \citenamefont
  {Jayaraman}(2015)}]{Seelig}%
  \BibitemOpen
  \bibfield  {author} {\bibinfo {author} {\bibfnamefont {J.~D.}\ \bibnamefont
  {Seelig}}\ and\ \bibinfo {author} {\bibfnamefont {V.}~\bibnamefont
  {Jayaraman}},\ }\href {\doibase 10.1038/nature14446} {\bibfield  {journal}
  {\bibinfo  {journal} {Nature}\ }\textbf {\bibinfo {volume} {521}},\ \bibinfo
  {pages} {186} (\bibinfo {year} {2015})}\BibitemShut {NoStop}%
\bibitem [{\citenamefont {Green}\ \emph {et~al.}(2017)\citenamefont {Green},
  \citenamefont {Adachi}, \citenamefont {Shah}, \citenamefont {Hirokawa},
  \citenamefont {Magani},\ and\ \citenamefont {Maimon}}]{Green}%
  \BibitemOpen
  \bibfield  {author} {\bibinfo {author} {\bibfnamefont {J.}~\bibnamefont
  {Green}}, \bibinfo {author} {\bibfnamefont {A.}~\bibnamefont {Adachi}},
  \bibinfo {author} {\bibfnamefont {K.~K.}\ \bibnamefont {Shah}}, \bibinfo
  {author} {\bibfnamefont {J.~D.}\ \bibnamefont {Hirokawa}}, \bibinfo {author}
  {\bibfnamefont {P.~S.}\ \bibnamefont {Magani}}, \ and\ \bibinfo {author}
  {\bibfnamefont {G.}~\bibnamefont {Maimon}},\ }\href {\doibase
  10.1038/nature22343} {\bibfield  {journal} {\bibinfo  {journal} {Nature}\
  }\textbf {\bibinfo {volume} {546}},\ \bibinfo {pages} {101} (\bibinfo {year}
  {2017})}\BibitemShut {NoStop}%
\bibitem [{\citenamefont {Zeil}(2012)}]{Zeil}%
  \BibitemOpen
  \bibfield  {author} {\bibinfo {author} {\bibfnamefont {J.}~\bibnamefont
  {Zeil}},\ }\href {\doibase 10.1016/j.conb.2011.12.008} {\bibfield  {journal}
  {\bibinfo  {journal} {Cur. Opinion in Neurobiology}\ }\textbf {\bibinfo
  {volume} {22}},\ \bibinfo {pages} {285} (\bibinfo {year} {2012})}\BibitemShut
  {NoStop}%
\bibitem [{\citenamefont {{Wehner}}\ and\ \citenamefont
  {Srinivasan}(1981)}]{Wehner_1981}%
  \BibitemOpen
  \bibfield  {author} {\bibinfo {author} {\bibfnamefont {R.}~\bibnamefont
  {{Wehner}}}\ and\ \bibinfo {author} {\bibfnamefont {M.~V.}\ \bibnamefont
  {Srinivasan}},\ }\href {\doibase 10.1007/BF00605445} {\bibfield  {journal}
  {\bibinfo  {journal} {Journal of Comparative Physiology A}\ }\textbf
  {\bibinfo {volume} {142}},\ \bibinfo {pages} {315} (\bibinfo {year}
  {1981})}\BibitemShut {NoStop}%
\bibitem [{\citenamefont {Ronacher}(2008)}]{Ronnacher2008}%
  \BibitemOpen
  \bibfield  {author} {\bibinfo {author} {\bibfnamefont {B.}~\bibnamefont
  {Ronacher}},\ }\href@noop {} {\bibfield  {journal} {\bibinfo  {journal}
  {Myrmecological News}\ }\textbf {\bibinfo {volume} {11}},\ \bibinfo {pages}
  {53} (\bibinfo {year} {2008})}\BibitemShut {NoStop}%
\bibitem [{\citenamefont {Collett}\ \emph {et~al.}(2013)\citenamefont
  {Collett}, \citenamefont {Chittka},\ and\ \citenamefont
  {Collett}}]{Collett2013}%
  \BibitemOpen
  \bibfield  {author} {\bibinfo {author} {\bibfnamefont {M.}~\bibnamefont
  {Collett}}, \bibinfo {author} {\bibfnamefont {L.}~\bibnamefont {Chittka}}, \
  and\ \bibinfo {author} {\bibfnamefont {T.~S.}\ \bibnamefont {Collett}},\
  }\href@noop {} {\bibfield  {journal} {\bibinfo  {journal} {Current Biology}\
  }\textbf {\bibinfo {volume} {23}},\ \bibinfo {pages} {R789} (\bibinfo {year}
  {2013})}\BibitemShut {NoStop}%
\bibitem [{\citenamefont {{el Jundi}}(2017)}]{ElJundi_2017}%
  \BibitemOpen
  \bibfield  {author} {\bibinfo {author} {\bibfnamefont {B.}~\bibnamefont {{el
  Jundi}}},\ }\href {\doibase 10.1016/j.cub.2017.06.051} {\bibfield  {journal}
  {\bibinfo  {journal} {Current Biology}\ }\textbf {\bibinfo {volume} {27}},\
  \bibinfo {pages} {R748} (\bibinfo {year} {2017})}\BibitemShut {NoStop}%
\bibitem [{\citenamefont {Kim}\ and\ \citenamefont
  {Dickinson}(2017)}]{Kim_Dickinson_2017}%
  \BibitemOpen
  \bibfield  {author} {\bibinfo {author} {\bibfnamefont {I.~S.}\ \bibnamefont
  {Kim}}\ and\ \bibinfo {author} {\bibfnamefont {M.~H.}\ \bibnamefont
  {Dickinson}},\ }\href {\doibase 10.1016/j.cub.2017.06.026} {\bibfield
  {journal} {\bibinfo  {journal} {Current Biology}\ }\textbf {\bibinfo {volume}
  {27}},\ \bibinfo {pages} {2227 } (\bibinfo {year} {2017})}\BibitemShut
  {NoStop}%
\bibitem [{\citenamefont {Wehner}\ \emph {et~al.}(1996)\citenamefont {Wehner},
  \citenamefont {Michel},\ and\ \citenamefont {Antonsen}}]{Wehner_et_al_1996}%
  \BibitemOpen
  \bibfield  {author} {\bibinfo {author} {\bibfnamefont {R.}~\bibnamefont
  {Wehner}}, \bibinfo {author} {\bibfnamefont {B.}~\bibnamefont {Michel}}, \
  and\ \bibinfo {author} {\bibfnamefont {P.}~\bibnamefont {Antonsen}},\
  }\href@noop {} {\bibfield  {journal} {\bibinfo  {journal} {Journal of
  Experimental Biology}\ }\textbf {\bibinfo {volume} {199}},\ \bibinfo {pages}
  {129} (\bibinfo {year} {1996})}\BibitemShut {NoStop}%
\bibitem [{\citenamefont {Vickerstaff}\ and\ \citenamefont
  {Di~Paolo}(2005)}]{Vickerstaff_2005}%
  \BibitemOpen
  \bibfield  {author} {\bibinfo {author} {\bibfnamefont {R.~J.}\ \bibnamefont
  {Vickerstaff}}\ and\ \bibinfo {author} {\bibfnamefont {E.~A.}\ \bibnamefont
  {Di~Paolo}},\ }in\ \href {\doibase 10.1007/11553090_2} {\emph {\bibinfo
  {booktitle} {{Advances in Artificial Life}}}},\ \bibinfo {editor} {edited by\
  \bibinfo {editor} {\bibfnamefont {M.~S.}\ \bibnamefont {Capcarr{\`e}re}},
  \bibinfo {editor} {\bibfnamefont {A.~A.}\ \bibnamefont {Freitas}}, \bibinfo
  {editor} {\bibfnamefont {P.~J.}\ \bibnamefont {Bentley}}, \bibinfo {editor}
  {\bibfnamefont {C.~G.}\ \bibnamefont {Johnson}}, \ and\ \bibinfo {editor}
  {\bibfnamefont {J.}~\bibnamefont {Timmis}}}\ (\bibinfo  {publisher} {Springer
  Berlin Heidelberg},\ \bibinfo {address} {Berlin, Heidelberg},\ \bibinfo
  {year} {2005})\ pp.\ \bibinfo {pages} {221--230}\BibitemShut {NoStop}%
\bibitem [{\citenamefont {Vickerstaff}\ and\ \citenamefont
  {Merkle}(2012)}]{Vickerstaff_2012}%
  \BibitemOpen
  \bibfield  {author} {\bibinfo {author} {\bibfnamefont {R.~J.}\ \bibnamefont
  {Vickerstaff}}\ and\ \bibinfo {author} {\bibfnamefont {T.}~\bibnamefont
  {Merkle}},\ }\href {\doibase 10.1016/j.jtbi.2012.04.034} {\bibfield
  {journal} {\bibinfo  {journal} {Journal of Theoretical Biology}\ }\textbf
  {\bibinfo {volume} {307}},\ \bibinfo {pages} {1 } (\bibinfo {year}
  {2012})}\BibitemShut {NoStop}%
\bibitem [{\citenamefont {Waldner}\ and\ \citenamefont
  {Merkle}(2018)}]{Waldner2018}%
  \BibitemOpen
  \bibfield  {author} {\bibinfo {author} {\bibfnamefont {F.}~\bibnamefont
  {Waldner}}\ and\ \bibinfo {author} {\bibfnamefont {T.}~\bibnamefont
  {Merkle}},\ }\href@noop {} {\bibfield  {journal} {\bibinfo  {journal}
  {Journal of Comparative Physiology A}\ }\textbf {\bibinfo {volume} {204}},\
  \bibinfo {pages} {985} (\bibinfo {year} {2018})}\BibitemShut {NoStop}%
\bibitem [{\citenamefont {Chien}\ and\ \citenamefont
  {Wagstaff}(2017)}]{chien_2017}%
  \BibitemOpen
  \bibfield  {author} {\bibinfo {author} {\bibfnamefont {S.}~\bibnamefont
  {Chien}}\ and\ \bibinfo {author} {\bibfnamefont {K.~L.}\ \bibnamefont
  {Wagstaff}},\ }\href@noop {} {\bibfield  {journal} {\bibinfo  {journal}
  {Science Robotics}\ }\textbf {\bibinfo {volume} {2}} (\bibinfo {year}
  {2017})}\BibitemShut {NoStop}%
\bibitem [{\citenamefont {Hook}\ \emph {et~al.}(2013)\citenamefont {Hook},
  \citenamefont {Tokekar}, \citenamefont {Branson}, \citenamefont {Bajer},
  \citenamefont {Sorensen},\ and\ \citenamefont {Isler}}]{Hook_2013}%
  \BibitemOpen
  \bibfield  {author} {\bibinfo {author} {\bibfnamefont {J.~V.}\ \bibnamefont
  {Hook}}, \bibinfo {author} {\bibfnamefont {P.}~\bibnamefont {Tokekar}},
  \bibinfo {author} {\bibfnamefont {E.}~\bibnamefont {Branson}}, \bibinfo
  {author} {\bibfnamefont {P.~G.}\ \bibnamefont {Bajer}}, \bibinfo {author}
  {\bibfnamefont {P.~W.}\ \bibnamefont {Sorensen}}, \ and\ \bibinfo {author}
  {\bibfnamefont {V.}~\bibnamefont {Isler}},\ }\enquote {\bibinfo {title}
  {{Local-Search Strategy for Active Localization of Multiple Invasive
  Fish}},}\ in\ \href {\doibase 10.1007/978-3-319-00065-7_57} {\emph {\bibinfo
  {booktitle} {{Experimental Robotics: The 13th International Symposium on
  Experimental Robotics}}}},\ \bibinfo {editor} {edited by\ \bibinfo {editor}
  {\bibfnamefont {J.~P.}\ \bibnamefont {Desai}}, \bibinfo {editor}
  {\bibfnamefont {G.}~\bibnamefont {Dudek}}, \bibinfo {editor} {\bibfnamefont
  {O.}~\bibnamefont {Khatib}}, \ and\ \bibinfo {editor} {\bibfnamefont
  {V.}~\bibnamefont {Kumar}}}\ (\bibinfo  {publisher} {Springer International
  Publishing},\ \bibinfo {address} {Heidelberg},\ \bibinfo {year} {2013})\ pp.\
  \bibinfo {pages} {859--873}\BibitemShut {NoStop}%
\bibitem [{\citenamefont {Girdhar}\ \emph {et~al.}(2011)\citenamefont
  {Girdhar}, \citenamefont {Xu}, \citenamefont {Dey}, \citenamefont {Meghjani},
  \citenamefont {Shkurti}, \citenamefont {Rekleitis},\ and\ \citenamefont
  {Dudek}}]{Girdhar_2011}%
  \BibitemOpen
  \bibfield  {author} {\bibinfo {author} {\bibfnamefont {Y.}~\bibnamefont
  {Girdhar}}, \bibinfo {author} {\bibfnamefont {A.}~\bibnamefont {Xu}},
  \bibinfo {author} {\bibfnamefont {B.~B.}\ \bibnamefont {Dey}}, \bibinfo
  {author} {\bibfnamefont {M.}~\bibnamefont {Meghjani}}, \bibinfo {author}
  {\bibfnamefont {F.}~\bibnamefont {Shkurti}}, \bibinfo {author} {\bibfnamefont
  {I.}~\bibnamefont {Rekleitis}}, \ and\ \bibinfo {author} {\bibfnamefont
  {G.}~\bibnamefont {Dudek}},\ }\href {\doibase 10.1109/IROS.2011.6094914}
  {\bibfield  {journal} {\bibinfo  {journal} {IEEE/RSJ}\ ,\ \bibinfo {pages}
  {5048}} (\bibinfo {year} {2011})}\BibitemShut {NoStop}%
\bibitem [{\citenamefont {Leonard}\ \emph {et~al.}(2007)\citenamefont
  {Leonard}, \citenamefont {Paley}, \citenamefont {Lekien}, \citenamefont
  {Sepulchre}, \citenamefont {Fratantoni},\ and\ \citenamefont
  {Davis}}]{Leonard_et_al_2007}%
  \BibitemOpen
  \bibfield  {author} {\bibinfo {author} {\bibfnamefont {N.}~\bibnamefont
  {Leonard}}, \bibinfo {author} {\bibfnamefont {D.}~\bibnamefont {Paley}},
  \bibinfo {author} {\bibfnamefont {F.}~\bibnamefont {Lekien}}, \bibinfo
  {author} {\bibfnamefont {R.}~\bibnamefont {Sepulchre}}, \bibinfo {author}
  {\bibfnamefont {D.}~\bibnamefont {Fratantoni}}, \ and\ \bibinfo {author}
  {\bibfnamefont {R.}~\bibnamefont {Davis}},\ }\href {\doibase
  10.1109/JPROC.2006.887295} {\bibfield  {journal} {\bibinfo  {journal}
  {Proceedings of the IEEE}\ }\textbf {\bibinfo {volume} {95}},\ \bibinfo
  {pages} {48} (\bibinfo {year} {2007})}\BibitemShut {NoStop}%
\bibitem [{\citenamefont {Dubowsky}\ \emph {et~al.}(2005)\citenamefont
  {Dubowsky}, \citenamefont {Iagnemma}, \citenamefont {Liberatore},
  \citenamefont {Lambeth}, \citenamefont {Plante},\ and\ \citenamefont
  {Boston}}]{Dubowsky_2005}%
  \BibitemOpen
  \bibfield  {author} {\bibinfo {author} {\bibfnamefont {S.}~\bibnamefont
  {Dubowsky}}, \bibinfo {author} {\bibfnamefont {K.}~\bibnamefont {Iagnemma}},
  \bibinfo {author} {\bibfnamefont {S.}~\bibnamefont {Liberatore}}, \bibinfo
  {author} {\bibfnamefont {D.}~\bibnamefont {Lambeth}}, \bibinfo {author}
  {\bibfnamefont {J.}~\bibnamefont {Plante}}, \ and\ \bibinfo {author}
  {\bibfnamefont {P.~J.}\ \bibnamefont {Boston}},\ }\href {\doibase
  doi.org/10.1063/1.1867276} {\bibfield  {journal} {\bibinfo  {journal} {Space
  Technology International Forum}\ ,\ \bibinfo {pages} {1449}} (\bibinfo {year}
  {2005})}\BibitemShut {NoStop}%
\bibitem [{\citenamefont {Duarte}\ \emph {et~al.}(2016)\citenamefont {Duarte},
  \citenamefont {Costa}, \citenamefont {Gomes}, \citenamefont {Rodrigues},
  \citenamefont {Silva}, \citenamefont {Oliveira},\ and\ \citenamefont
  {Christensen}}]{Duarte_et_al_2016}%
  \BibitemOpen
  \bibfield  {author} {\bibinfo {author} {\bibfnamefont {M.}~\bibnamefont
  {Duarte}}, \bibinfo {author} {\bibfnamefont {V.}~\bibnamefont {Costa}},
  \bibinfo {author} {\bibfnamefont {J.}~\bibnamefont {Gomes}}, \bibinfo
  {author} {\bibfnamefont {T.}~\bibnamefont {Rodrigues}}, \bibinfo {author}
  {\bibfnamefont {F.}~\bibnamefont {Silva}}, \bibinfo {author} {\bibfnamefont
  {S.~M.}\ \bibnamefont {Oliveira}}, \ and\ \bibinfo {author} {\bibfnamefont
  {A.~L.}\ \bibnamefont {Christensen}},\ }\href {\doibase
  10.1371/journal.pone.0151834} {\bibfield  {journal} {\bibinfo  {journal}
  {PLOS ONE}\ }\textbf {\bibinfo {volume} {11}},\ \bibinfo {pages} {1}
  (\bibinfo {year} {2016})}\BibitemShut {NoStop}%
\bibitem [{\citenamefont {Nirmal}\ and\ \citenamefont {Lyons}(2016)}]{Nirmal}%
  \BibitemOpen
  \bibfield  {author} {\bibinfo {author} {\bibfnamefont {P.}~\bibnamefont
  {Nirmal}}\ and\ \bibinfo {author} {\bibfnamefont {D.}~\bibnamefont {Lyons}},\
  }\href {\doibase doi:10.1017/S026357471500034X} {\bibfield  {journal}
  {\bibinfo  {journal} {Robotica}\ }\textbf {\bibinfo {volume} {34}},\ \bibinfo
  {pages} {2741} (\bibinfo {year} {2016})}\BibitemShut {NoStop}%
\bibitem [{\citenamefont {M\"oller}\ \emph {et~al.}(2001)\citenamefont
  {M\"oller}, \citenamefont {Lambrinos}, \citenamefont {Roggendorf},\ and\
  \citenamefont {Wehner}}]{Moeller}%
  \BibitemOpen
  \bibfield  {author} {\bibinfo {author} {\bibfnamefont {R.}~\bibnamefont
  {M\"oller}}, \bibinfo {author} {\bibfnamefont {D.}~\bibnamefont {Lambrinos}},
  \bibinfo {author} {\bibfnamefont {T.}~\bibnamefont {Roggendorf}}, \ and\
  \bibinfo {author} {\bibfnamefont {R.~P.~R.}\ \bibnamefont {Wehner}},\
  }\enquote {\bibinfo {title} {{Insect Strategies of Visual Homing in Mobile
  Robots}},}\ in\ \href@noop {} {\emph {\bibinfo {booktitle} {{Biorobotics.
  Methods and Applications}}}},\ \bibinfo {editor} {edited by\ \bibinfo
  {editor} {\bibfnamefont {B.}~\bibnamefont {Webb}}\ and\ \bibinfo {editor}
  {\bibfnamefont {T.~R.}\ \bibnamefont {Consi}}}\ (\bibinfo  {publisher} {AAAI
  Press / MIT Press},\ \bibinfo {year} {2001})\ pp.\ \bibinfo {pages}
  {37--66}\BibitemShut {NoStop}%
\bibitem [{\citenamefont {Noetel}\ \emph
  {et~al.}(2018{\natexlab{a}})\citenamefont {Noetel}, \citenamefont {Freitas},
  \citenamefont {Macau},\ and\ \citenamefont {Schimansky-Geier}}]{Noetel_2018}%
  \BibitemOpen
  \bibfield  {author} {\bibinfo {author} {\bibfnamefont {J.}~\bibnamefont
  {Noetel}}, \bibinfo {author} {\bibfnamefont {V.~L.~S.}\ \bibnamefont
  {Freitas}}, \bibinfo {author} {\bibfnamefont {E.~E.~N.}\ \bibnamefont
  {Macau}}, \ and\ \bibinfo {author} {\bibfnamefont {L.}~\bibnamefont
  {Schimansky-Geier}},\ }\href {\doibase 10.1103/PhysRevE.98.022128} {\bibfield
   {journal} {\bibinfo  {journal} {Phys. Rev. E}\ }\textbf {\bibinfo {volume}
  {98}},\ \bibinfo {pages} {022128} (\bibinfo {year}
  {2018}{\natexlab{a}})}\BibitemShut {NoStop}%
\bibitem [{\citenamefont {Noetel}\ \emph
  {et~al.}(2018{\natexlab{b}})\citenamefont {Noetel}, \citenamefont {Freitas},
  \citenamefont {Macau},\ and\ \citenamefont
  {Schimansky-Geier}}]{Noetel_2018c}%
  \BibitemOpen
  \bibfield  {author} {\bibinfo {author} {\bibfnamefont {J.}~\bibnamefont
  {Noetel}}, \bibinfo {author} {\bibfnamefont {V.~L.~S.}\ \bibnamefont
  {Freitas}}, \bibinfo {author} {\bibfnamefont {E.~E.~N.}\ \bibnamefont
  {Macau}}, \ and\ \bibinfo {author} {\bibfnamefont {L.}~\bibnamefont
  {Schimansky-Geier}},\ }\href@noop {} {\bibfield  {journal} {\bibinfo
  {journal} {Chaos}\ }\textbf {\bibinfo {volume} {28}},\ \bibinfo {pages}
  {106302} (\bibinfo {year} {2018}{\natexlab{b}})}\BibitemShut {NoStop}%
\bibitem [{\citenamefont {Okubo}(1986)}]{Okubo1986}%
  \BibitemOpen
  \bibfield  {author} {\bibinfo {author} {\bibfnamefont {A.}~\bibnamefont
  {Okubo}},\ }\href@noop {} {\bibfield  {journal} {\bibinfo  {journal}
  {Advances in Biophysics}\ }\textbf {\bibinfo {volume} {22}},\ \bibinfo
  {pages} {1} (\bibinfo {year} {1986})}\BibitemShut {NoStop}%
\bibitem [{\citenamefont {Gorbonos}\ \emph {et~al.}(2016)\citenamefont
  {Gorbonos}, \citenamefont {Ianconescu}, \citenamefont {Puckett},
  \citenamefont {Ni}, \citenamefont {Ouellette},\ and\ \citenamefont
  {Gov}}]{Gorbonos2016}%
  \BibitemOpen
  \bibfield  {author} {\bibinfo {author} {\bibfnamefont {D.}~\bibnamefont
  {Gorbonos}}, \bibinfo {author} {\bibfnamefont {R.}~\bibnamefont
  {Ianconescu}}, \bibinfo {author} {\bibfnamefont {J.~G.}\ \bibnamefont
  {Puckett}}, \bibinfo {author} {\bibfnamefont {R.}~\bibnamefont {Ni}},
  \bibinfo {author} {\bibfnamefont {N.~T.}\ \bibnamefont {Ouellette}}, \ and\
  \bibinfo {author} {\bibfnamefont {N.~S.}\ \bibnamefont {Gov}},\ }\href
  {\doibase 10.1088/1367-2630/18/7/073042} {\bibfield  {journal} {\bibinfo
  {journal} {New Journal of Physics}\ }\textbf {\bibinfo {volume} {18}},\
  \bibinfo {pages} {073042} (\bibinfo {year} {2016})}\BibitemShut {NoStop}%
\bibitem [{\citenamefont {Reynolds}\ \emph {et~al.}(2017)\citenamefont
  {Reynolds}, \citenamefont {Sinhuber},\ and\ \citenamefont
  {Ouellette}}]{Reynolds2017}%
  \BibitemOpen
  \bibfield  {author} {\bibinfo {author} {\bibfnamefont {A.~M.}\ \bibnamefont
  {Reynolds}}, \bibinfo {author} {\bibfnamefont {M.}~\bibnamefont {Sinhuber}},
  \ and\ \bibinfo {author} {\bibfnamefont {N.~T.}\ \bibnamefont {Ouellette}},\
  }\href {\doibase 10.1140/epje/i2017-11531-7} {\bibfield  {journal} {\bibinfo
  {journal} {The European Physical Journal E}\ }\textbf {\bibinfo {volume}
  {40}},\ \bibinfo {pages} {46} (\bibinfo {year} {2017})}\BibitemShut {NoStop}%
\bibitem [{\citenamefont {Reynolds}(2018)}]{Reynolds2018}%
  \BibitemOpen
  \bibfield  {author} {\bibinfo {author} {\bibfnamefont {A.~M.}\ \bibnamefont
  {Reynolds}},\ }\href {\doibase 10.1098/rsif.2017.0806} {\bibfield  {journal}
  {\bibinfo  {journal} {Journal of The Royal Society Interface}\ }\textbf
  {\bibinfo {volume} {15}} (\bibinfo {year} {2018}),\
  10.1098/rsif.2017.0806}\BibitemShut {NoStop}%
\bibitem [{\citenamefont {Schweitzer}\ and\ \citenamefont
  {Schimansky-Geier}(1994)}]{Schweitzer1994}%
  \BibitemOpen
  \bibfield  {author} {\bibinfo {author} {\bibfnamefont {F.}~\bibnamefont
  {Schweitzer}}\ and\ \bibinfo {author} {\bibfnamefont {L.}~\bibnamefont
  {Schimansky-Geier}},\ }\href {\doibase
  https://doi.org/10.1016/0378-4371(94)90312-3} {\bibfield  {journal} {\bibinfo
   {journal} {Physica A}\ }\textbf {\bibinfo {volume} {206}},\ \bibinfo {pages}
  {359 } (\bibinfo {year} {1994})}\BibitemShut {NoStop}%
\bibitem [{\citenamefont {Chavanis}(2014)}]{Chavanis2014}%
  \BibitemOpen
  \bibfield  {author} {\bibinfo {author} {\bibfnamefont {P.-H.}\ \bibnamefont
  {Chavanis}},\ }\href {\doibase 10.1140/epjb/e2014-40586-6} {\bibfield
  {journal} {\bibinfo  {journal} {The European Physical Journal B}\ }\textbf
  {\bibinfo {volume} {87}},\ \bibinfo {pages} {120} (\bibinfo {year}
  {2014})}\BibitemShut {NoStop}%
\bibitem [{\citenamefont {Delcourt}\ \emph {et~al.}(2016)\citenamefont
  {Delcourt}, \citenamefont {Bode},\ and\ \citenamefont
  {Deno{\'el}}}]{Delcourt}%
  \BibitemOpen
  \bibfield  {author} {\bibinfo {author} {\bibfnamefont {J.}~\bibnamefont
  {Delcourt}}, \bibinfo {author} {\bibfnamefont {N.}~\bibnamefont {Bode}}, \
  and\ \bibinfo {author} {\bibfnamefont {M.}~\bibnamefont {Deno{\'el}}},\
  }\href@noop {} {\bibfield  {journal} {\bibinfo  {journal} {The Quarterly
  Review of Biology}\ }\textbf {\bibinfo {volume} {91}},\ \bibinfo {pages} {1}
  (\bibinfo {year} {2016})}\BibitemShut {NoStop}%
\bibitem [{\citenamefont {Ordemann}(2002)}]{Ordemann2002}%
  \BibitemOpen
  \bibfield  {author} {\bibinfo {author} {\bibfnamefont {A.}~\bibnamefont
  {Ordemann}},\ }\href@noop {} {\bibfield  {journal} {\bibinfo  {journal}
  {Biol. Physicist}\ }\textbf {\bibinfo {volume} {2}},\ \bibinfo {pages} {5}
  (\bibinfo {year} {2002})}\BibitemShut {NoStop}%
\bibitem [{\citenamefont {Ordemann}\ \emph
  {et~al.}(2003{\natexlab{a}})\citenamefont {Ordemann}, \citenamefont
  {Balazsi},\ and\ \citenamefont {Moss}}]{Ordemann}%
  \BibitemOpen
  \bibfield  {author} {\bibinfo {author} {\bibfnamefont {A.}~\bibnamefont
  {Ordemann}}, \bibinfo {author} {\bibfnamefont {G.}~\bibnamefont {Balazsi}}, \
  and\ \bibinfo {author} {\bibfnamefont {F.}~\bibnamefont {Moss}},\ }\href
  {\doibase 10.1016/S0378-4371(03)00204-8} {\bibfield  {journal} {\bibinfo
  {journal} {Physica A: Statistical Mechanics and its Applications}\ }\textbf
  {\bibinfo {volume} {325}},\ \bibinfo {pages} {260 } (\bibinfo {year}
  {2003}{\natexlab{a}})}\BibitemShut {NoStop}%
\bibitem [{\citenamefont {Ordemann}\ \emph
  {et~al.}(2003{\natexlab{b}})\citenamefont {Ordemann}, \citenamefont
  {Balaszi},\ and\ \citenamefont {Moss}}]{Ordemann2003Nova}%
  \BibitemOpen
  \bibfield  {author} {\bibinfo {author} {\bibfnamefont {A.}~\bibnamefont
  {Ordemann}}, \bibinfo {author} {\bibfnamefont {G.}~\bibnamefont {Balaszi}}, \
  and\ \bibinfo {author} {\bibfnamefont {F.}~\bibnamefont {Moss}},\ }\href@noop
  {} {\bibfield  {journal} {\bibinfo  {journal} {Nova Acta Leopoldina}\
  }\textbf {\bibinfo {volume} {88}},\ \bibinfo {pages} {87} (\bibinfo {year}
  {2003}{\natexlab{b}})}\BibitemShut {NoStop}%
\bibitem [{\citenamefont {Erdmann}\ \emph {et~al.}(2004)\citenamefont
  {Erdmann}, \citenamefont {Ebeling}, \citenamefont {Schimansky-Geier},
  \citenamefont {Ordemann},\ and\ \citenamefont {Moss}}]{ErdmannDaphnia}%
  \BibitemOpen
  \bibfield  {author} {\bibinfo {author} {\bibfnamefont {U.}~\bibnamefont
  {Erdmann}}, \bibinfo {author} {\bibfnamefont {W.}~\bibnamefont {Ebeling}},
  \bibinfo {author} {\bibfnamefont {L.}~\bibnamefont {Schimansky-Geier}},
  \bibinfo {author} {\bibfnamefont {A.}~\bibnamefont {Ordemann}}, \ and\
  \bibinfo {author} {\bibfnamefont {F.}~\bibnamefont {Moss}},\ }\href@noop {}
  {\bibfield  {journal} {\bibinfo  {journal} {arXiv:q-bio}\ }\textbf {\bibinfo
  {volume} {325}},\ \bibinfo {pages} {0404018} (\bibinfo {year}
  {2004})}\BibitemShut {NoStop}%
\bibitem [{\citenamefont {Garcia}\ \emph {et~al.}(2007)\citenamefont {Garcia},
  \citenamefont {Moss}, \citenamefont {Nihongi}, \citenamefont {Strickler},
  \citenamefont {G\"oller}, \citenamefont {Erdmann}, \citenamefont
  {Schimansky-Geier},\ and\ \citenamefont {Sokolov}}]{Garcia_daphnia}%
  \BibitemOpen
  \bibfield  {author} {\bibinfo {author} {\bibfnamefont {R.}~\bibnamefont
  {Garcia}}, \bibinfo {author} {\bibfnamefont {F.}~\bibnamefont {Moss}},
  \bibinfo {author} {\bibfnamefont {A.}~\bibnamefont {Nihongi}}, \bibinfo
  {author} {\bibfnamefont {J.}~\bibnamefont {Strickler}}, \bibinfo {author}
  {\bibfnamefont {S.}~\bibnamefont {G\"oller}}, \bibinfo {author}
  {\bibfnamefont {U.}~\bibnamefont {Erdmann}}, \bibinfo {author} {\bibfnamefont
  {L.}~\bibnamefont {Schimansky-Geier}}, \ and\ \bibinfo {author}
  {\bibfnamefont {I.}~\bibnamefont {Sokolov}},\ }\href {\doibase
  10.1016/j.mbs.2006.11.014} {\bibfield  {journal} {\bibinfo  {journal}
  {Mathematical Biosciences}\ }\textbf {\bibinfo {volume} {207}},\ \bibinfo
  {pages} {165 } (\bibinfo {year} {2007})}\BibitemShut {NoStop}%
\bibitem [{\citenamefont {Dees}\ \emph {et~al.}(2008)\citenamefont {Dees},
  \citenamefont {Bahar},\ and\ \citenamefont {Moss}}]{Dees2008}%
  \BibitemOpen
  \bibfield  {author} {\bibinfo {author} {\bibfnamefont {N.}~\bibnamefont
  {Dees}}, \bibinfo {author} {\bibfnamefont {S.}~\bibnamefont {Bahar}}, \ and\
  \bibinfo {author} {\bibfnamefont {F.}~\bibnamefont {Moss}},\ }\href@noop {}
  {\bibfield  {journal} {\bibinfo  {journal} {Physical Biology}\ }\textbf
  {\bibinfo {volume} {5}},\ \bibinfo {pages} {044001} (\bibinfo {year}
  {2008})}\BibitemShut {NoStop}%
\bibitem [{\citenamefont {Erdmann}\ and\ \citenamefont
  {Ebeling}(2009)}]{Erdmann2003}%
  \BibitemOpen
  \bibfield  {author} {\bibinfo {author} {\bibfnamefont {U.}~\bibnamefont
  {Erdmann}}\ and\ \bibinfo {author} {\bibfnamefont {W.}~\bibnamefont
  {Ebeling}},\ }\href@noop {} {\bibfield  {journal} {\bibinfo  {journal}
  {Fluctuation and Noise Letters}\ }\textbf {\bibinfo {volume} {3}},\ \bibinfo
  {pages} {L145} (\bibinfo {year} {2009})}\BibitemShut {NoStop}%
\bibitem [{\citenamefont {Vollmer}\ \emph {et~al.}(2006)\citenamefont
  {Vollmer}, \citenamefont {Vegh}, \citenamefont {Lange},\ and\ \citenamefont
  {Eckhardt}}]{Vollmer}%
  \BibitemOpen
  \bibfield  {author} {\bibinfo {author} {\bibfnamefont {J.}~\bibnamefont
  {Vollmer}}, \bibinfo {author} {\bibfnamefont {A.}~\bibnamefont {Vegh}},
  \bibinfo {author} {\bibfnamefont {C.}~\bibnamefont {Lange}}, \ and\ \bibinfo
  {author} {\bibfnamefont {B.}~\bibnamefont {Eckhardt}},\ }\href@noop {}
  {\bibfield  {journal} {\bibinfo  {journal} {Physical Review E}\ }\textbf
  {\bibinfo {volume} {73}},\ \bibinfo {pages} {061924} (\bibinfo {year}
  {2006})}\BibitemShut {NoStop}%
\bibitem [{\citenamefont {Mach}\ and\ \citenamefont
  {Schweitzer}(2007)}]{Mach_Schweitzer2007}%
  \BibitemOpen
  \bibfield  {author} {\bibinfo {author} {\bibfnamefont {R.}~\bibnamefont
  {Mach}}\ and\ \bibinfo {author} {\bibfnamefont {F.}~\bibnamefont
  {Schweitzer}},\ }\href@noop {} {\bibfield  {journal} {\bibinfo  {journal}
  {Bulletin of Mathematical Biology}\ }\textbf {\bibinfo {volume} {69}},\
  \bibinfo {pages} {539} (\bibinfo {year} {2007})}\BibitemShut {NoStop}%
\bibitem [{\citenamefont {Levine}\ \emph {et~al.}(2001)\citenamefont {Levine},
  \citenamefont {Rappel},\ and\ \citenamefont {Cohen}}]{Levine2001}%
  \BibitemOpen
  \bibfield  {author} {\bibinfo {author} {\bibfnamefont {H.}~\bibnamefont
  {Levine}}, \bibinfo {author} {\bibfnamefont {W.-J.}\ \bibnamefont {Rappel}},
  \ and\ \bibinfo {author} {\bibfnamefont {I.}~\bibnamefont {Cohen}},\
  }\href@noop {} {\bibfield  {journal} {\bibinfo  {journal} {Physical Review
  E}\ }\textbf {\bibinfo {volume} {63}},\ \bibinfo {pages} {017101} (\bibinfo
  {year} {2001})}\BibitemShut {NoStop}%
\bibitem [{\citenamefont {Strefler}\ \emph {et~al.}(2008)\citenamefont
  {Strefler}, \citenamefont {Erdmann},\ and\ \citenamefont
  {Schimansky-Geier}}]{Strefler2008}%
  \BibitemOpen
  \bibfield  {author} {\bibinfo {author} {\bibfnamefont {J.}~\bibnamefont
  {Strefler}}, \bibinfo {author} {\bibfnamefont {U.}~\bibnamefont {Erdmann}}, \
  and\ \bibinfo {author} {\bibfnamefont {L.}~\bibnamefont {Schimansky-Geier}},\
  }\href@noop {} {\bibfield  {journal} {\bibinfo  {journal} {Physical Review
  E}\ }\textbf {\bibinfo {volume} {78}},\ \bibinfo {pages} {031927} (\bibinfo
  {year} {2008})}\BibitemShut {NoStop}%
\bibitem [{\citenamefont {Thouma}\ \emph {et~al.}(2010)\citenamefont {Thouma},
  \citenamefont {Shreim},\ and\ \citenamefont {Klushin}}]{Thouma}%
  \BibitemOpen
  \bibfield  {author} {\bibinfo {author} {\bibfnamefont {J.}~\bibnamefont
  {Thouma}}, \bibinfo {author} {\bibfnamefont {A.}~\bibnamefont {Shreim}}, \
  and\ \bibinfo {author} {\bibfnamefont {L.}~\bibnamefont {Klushin}},\
  }\href@noop {} {\bibfield  {journal} {\bibinfo  {journal} {Physical Review
  E}\ }\textbf {\bibinfo {volume} {81}},\ \bibinfo {pages} {066106} (\bibinfo
  {year} {2010})}\BibitemShut {NoStop}%
\bibitem [{\citenamefont {Okubo}\ and\ \citenamefont
  {Levin}(2002)}]{Okubo2002}%
  \BibitemOpen
  \bibfield  {author} {\bibinfo {author} {\bibfnamefont {A.}~\bibnamefont
  {Okubo}}\ and\ \bibinfo {author} {\bibfnamefont {S.}~\bibnamefont {Levin}},\
  }\href@noop {} {\emph {\bibinfo {title} {{Diffusion and Ecological Problems:
  Modern Perspectives}}}}\ (\bibinfo  {publisher} {Springer},\ \bibinfo
  {address} {Berlin, 2nd ed.},\ \bibinfo {year} {2002})\ \bibinfo {note}
  {interdisciplinary Applied Mathematics Vol.14}\BibitemShut {NoStop}%
\bibitem [{\citenamefont {Vicsek}\ \emph {et~al.}(1995)\citenamefont {Vicsek},
  \citenamefont {Czir\'ok}, \citenamefont {Ben-Jacob}, \citenamefont {Cohen},\
  and\ \citenamefont {Shochet}}]{Vicsek1995}%
  \BibitemOpen
  \bibfield  {author} {\bibinfo {author} {\bibfnamefont {T.}~\bibnamefont
  {Vicsek}}, \bibinfo {author} {\bibfnamefont {A.}~\bibnamefont {Czir\'ok}},
  \bibinfo {author} {\bibfnamefont {E.}~\bibnamefont {Ben-Jacob}}, \bibinfo
  {author} {\bibfnamefont {I.}~\bibnamefont {Cohen}}, \ and\ \bibinfo {author}
  {\bibfnamefont {O.}~\bibnamefont {Shochet}},\ }\href {\doibase
  10.1103/PhysRevLett.75.1226} {\bibfield  {journal} {\bibinfo  {journal}
  {Phys. Rev. Lett.}\ }\textbf {\bibinfo {volume} {75}},\ \bibinfo {pages}
  {1226} (\bibinfo {year} {1995})}\BibitemShut {NoStop}%
\bibitem [{\citenamefont {Chat{\'e }}\ \emph
  {et~al.}(2008{\natexlab{a}})\citenamefont {Chat{\'e }}, \citenamefont
  {Ginelli}, \citenamefont {Gr{\'e}goire},\ and\ \citenamefont
  {Raynaud}}]{Chate2008a}%
  \BibitemOpen
  \bibfield  {author} {\bibinfo {author} {\bibfnamefont {H.}~\bibnamefont
  {Chat{\'e }}}, \bibinfo {author} {\bibfnamefont {F.}~\bibnamefont {Ginelli}},
  \bibinfo {author} {\bibfnamefont {G.}~\bibnamefont {Gr{\'e}goire}}, \ and\
  \bibinfo {author} {\bibfnamefont {F.}~\bibnamefont {Raynaud}},\ }\href@noop
  {} {\bibfield  {journal} {\bibinfo  {journal} {Physical Review E}\ }\textbf
  {\bibinfo {volume} {77}},\ \bibinfo {pages} {046113} (\bibinfo {year}
  {2008}{\natexlab{a}})}\BibitemShut {NoStop}%
\bibitem [{\citenamefont {Chat{\'e }}\ \emph
  {et~al.}(2008{\natexlab{b}})\citenamefont {Chat{\'e }}, \citenamefont
  {Ginelli}, \citenamefont {Gr{\'e}goire}, \citenamefont {Peruani},\ and\
  \citenamefont {Raynaud}}]{Chate2008b}%
  \BibitemOpen
  \bibfield  {author} {\bibinfo {author} {\bibfnamefont {H.}~\bibnamefont
  {Chat{\'e }}}, \bibinfo {author} {\bibfnamefont {F.}~\bibnamefont {Ginelli}},
  \bibinfo {author} {\bibfnamefont {G.}~\bibnamefont {Gr{\'e}goire}}, \bibinfo
  {author} {\bibfnamefont {F.}~\bibnamefont {Peruani}}, \ and\ \bibinfo
  {author} {\bibfnamefont {F.}~\bibnamefont {Raynaud}},\ }\href@noop {}
  {\bibfield  {journal} {\bibinfo  {journal} {European Physical Journal B}\
  }\textbf {\bibinfo {volume} {64}},\ \bibinfo {pages} {451} (\bibinfo {year}
  {2008}{\natexlab{b}})}\BibitemShut {NoStop}%
\bibitem [{\citenamefont {Peruani}\ \emph {et~al.}(2008)\citenamefont
  {Peruani}, \citenamefont {Deutsch},\ and\ \citenamefont
  {B\"ar}}]{Peruani2008}%
  \BibitemOpen
  \bibfield  {author} {\bibinfo {author} {\bibfnamefont {F.}~\bibnamefont
  {Peruani}}, \bibinfo {author} {\bibfnamefont {A.}~\bibnamefont {Deutsch}}, \
  and\ \bibinfo {author} {\bibfnamefont {M.}~\bibnamefont {B\"ar}},\
  }\href@noop {} {\bibfield  {journal} {\bibinfo  {journal} {European Physical
  Journal Special Topics}\ }\textbf {\bibinfo {volume} {157}},\ \bibinfo
  {pages} {111} (\bibinfo {year} {2008})}\BibitemShut {NoStop}%
\bibitem [{\citenamefont {Romanczuk}\ \emph {et~al.}(2012)\citenamefont
  {Romanczuk}, \citenamefont {B{\"a}r}, \citenamefont {Ebeling}, \citenamefont
  {Lindner},\ and\ \citenamefont {Schimansky-Geier}}]{Romanczuk}%
  \BibitemOpen
  \bibfield  {author} {\bibinfo {author} {\bibfnamefont {P.}~\bibnamefont
  {Romanczuk}}, \bibinfo {author} {\bibfnamefont {M.}~\bibnamefont {B{\"a}r}},
  \bibinfo {author} {\bibfnamefont {W.}~\bibnamefont {Ebeling}}, \bibinfo
  {author} {\bibfnamefont {B.}~\bibnamefont {Lindner}}, \ and\ \bibinfo
  {author} {\bibfnamefont {L.}~\bibnamefont {Schimansky-Geier}},\ }\href@noop
  {} {\bibfield  {journal} {\bibinfo  {journal} {European Physical Journal
  Special Topics}\ }\textbf {\bibinfo {volume} {202}},\ \bibinfo {pages} {1}
  (\bibinfo {year} {2012})}\BibitemShut {NoStop}%
\bibitem [{\citenamefont {Vicsek}\ and\ \citenamefont
  {Zafeiris}(2012)}]{Vicsek_2012}%
  \BibitemOpen
  \bibfield  {author} {\bibinfo {author} {\bibfnamefont {T.}~\bibnamefont
  {Vicsek}}\ and\ \bibinfo {author} {\bibfnamefont {A.}~\bibnamefont
  {Zafeiris}},\ }\href {\doibase
  http://dx.doi.org/10.1016/j.physrep.2012.03.004} {\bibfield  {journal}
  {\bibinfo  {journal} {Physics Reports}\ }\textbf {\bibinfo {volume} {517}},\
  \bibinfo {pages} {71 } (\bibinfo {year} {2012})}\BibitemShut {NoStop}%
\bibitem [{\citenamefont {Gro\ss{}mann}\ \emph {et~al.}(2014)\citenamefont
  {Gro\ss{}mann}, \citenamefont {Romanczuk}, \citenamefont {B\"ar},\ and\
  \citenamefont {Schimansky-Geier}}]{vortex}%
  \BibitemOpen
  \bibfield  {author} {\bibinfo {author} {\bibfnamefont {R.}~\bibnamefont
  {Gro\ss{}mann}}, \bibinfo {author} {\bibfnamefont {P.}~\bibnamefont
  {Romanczuk}}, \bibinfo {author} {\bibfnamefont {M.}~\bibnamefont {B\"ar}}, \
  and\ \bibinfo {author} {\bibfnamefont {L.}~\bibnamefont {Schimansky-Geier}},\
  }\href {\doibase 10.1103/PhysRevLett.113.258104} {\bibfield  {journal}
  {\bibinfo  {journal} {Phys. Rev. Lett.}\ }\textbf {\bibinfo {volume} {113}},\
  \bibinfo {pages} {258104} (\bibinfo {year} {2014})}\BibitemShut {NoStop}%
\bibitem [{\citenamefont {Kuramoto}(1984)}]{Kuramoto_1984}%
  \BibitemOpen
  \bibfield  {author} {\bibinfo {author} {\bibfnamefont {Y.}~\bibnamefont
  {Kuramoto}},\ }\href@noop {} {\emph {\bibinfo {title} {{Chemical
  Oscillations, Waves, and Turbulence}}}}\ (\bibinfo  {publisher}
  {Springer-Verlag},\ \bibinfo {address} {New York},\ \bibinfo {year}
  {1984})\BibitemShut {NoStop}%
\bibitem [{\citenamefont {Pikovsky}\ \emph {et~al.}(2001)\citenamefont
  {Pikovsky}, \citenamefont {Rosenblum},\ and\ \citenamefont
  {Kurths}}]{Pikovsky_et_al_2001}%
  \BibitemOpen
  \bibfield  {author} {\bibinfo {author} {\bibfnamefont {A.}~\bibnamefont
  {Pikovsky}}, \bibinfo {author} {\bibfnamefont {M.}~\bibnamefont {Rosenblum}},
  \ and\ \bibinfo {author} {\bibfnamefont {J.}~\bibnamefont {Kurths}},\
  }\href@noop {} {\emph {\bibinfo {title} {{Synchronization: A universal
  concept in nonlinear sciences}}}}\ (\bibinfo  {publisher} {Pres Syndicate of
  the University of Cambridge},\ \bibinfo {address} {Cambridge},\ \bibinfo
  {year} {2001})\BibitemShut {NoStop}%
\bibitem [{\citenamefont {Lauga}\ and\ \citenamefont
  {Powers}(2009)}]{Lauga2009}%
  \BibitemOpen
  \bibfield  {author} {\bibinfo {author} {\bibfnamefont {E.}~\bibnamefont
  {Lauga}}\ and\ \bibinfo {author} {\bibfnamefont {T.}~\bibnamefont {Powers}},\
  }\href@noop {} {\bibfield  {journal} {\bibinfo  {journal} {Reports on
  Progress in Physics}\ }\textbf {\bibinfo {volume} {72}},\ \bibinfo {pages}
  {096601} (\bibinfo {year} {2009})}\BibitemShut {NoStop}%
\bibitem [{\citenamefont {Downton}\ and\ \citenamefont {Stark}(2009)}]{Stark}%
  \BibitemOpen
  \bibfield  {author} {\bibinfo {author} {\bibfnamefont {M.~T.}\ \bibnamefont
  {Downton}}\ and\ \bibinfo {author} {\bibfnamefont {H.}~\bibnamefont
  {Stark}},\ }\href {http://stacks.iop.org/0953-8984/21/i=20/a=204101}
  {\bibfield  {journal} {\bibinfo  {journal} {Journal of Physics: Condensed
  Matter}\ }\textbf {\bibinfo {volume} {21}},\ \bibinfo {pages} {204101}
  (\bibinfo {year} {2009})}\BibitemShut {NoStop}%
\bibitem [{\citenamefont {Schwarzendahl}\ and\ \citenamefont
  {Mazza}(2018)}]{Mazza2018}%
  \BibitemOpen
  \bibfield  {author} {\bibinfo {author} {\bibfnamefont {F.}~\bibnamefont
  {Schwarzendahl}}\ and\ \bibinfo {author} {\bibfnamefont {M.}~\bibnamefont
  {Mazza}},\ }\href@noop {} {\bibfield  {journal} {\bibinfo  {journal} {Soft
  Matter}\ }\textbf {\bibinfo {volume} {14}},\ \bibinfo {pages} {4666}
  (\bibinfo {year} {2018})}\BibitemShut {NoStop}%
\bibitem [{\citenamefont {Schwarzendahl}\ and\ \citenamefont
  {Mazza}(2019)}]{Mazza2019}%
  \BibitemOpen
  \bibfield  {author} {\bibinfo {author} {\bibfnamefont {F.}~\bibnamefont
  {Schwarzendahl}}\ and\ \bibinfo {author} {\bibfnamefont {M.}~\bibnamefont
  {Mazza}},\ }\href@noop {} {\bibfield  {journal} {\bibinfo  {journal} {The
  Journal of Chemical Physics}\ }\textbf {\bibinfo {volume} {150}},\ \bibinfo
  {pages} {184902} (\bibinfo {year} {2019})}\BibitemShut {NoStop}%
\bibitem [{\citenamefont {Russel}\ and\ \citenamefont
  {Norvig}(2010)}]{Gallistel1990}%
  \BibitemOpen
  \bibfield  {author} {\bibinfo {author} {\bibfnamefont {S.~J.}\ \bibnamefont
  {Russel}}\ and\ \bibinfo {author} {\bibfnamefont {P.}~\bibnamefont
  {Norvig}},\ }\href@noop {} {\emph {\bibinfo {title} {{The Organization of
  Learning}}}}\ (\bibinfo  {publisher} {The MIT Press, Cambridge MA, US},\
  \bibinfo {year} {2010})\BibitemShut {NoStop}%
\bibitem [{\citenamefont {Mittelstaedt}(1962)}]{Mittelstaedt_bico}%
  \BibitemOpen
  \bibfield  {author} {\bibinfo {author} {\bibfnamefont {H.}~\bibnamefont
  {Mittelstaedt}},\ }\href {\doibase 10.1146/annurev.en.07.010162.001141}
  {\bibfield  {journal} {\bibinfo  {journal} {Annual Review of Entomology}\
  }\textbf {\bibinfo {volume} {7}},\ \bibinfo {pages} {177} (\bibinfo {year}
  {1962})}\BibitemShut {NoStop}%
\bibitem [{\citenamefont {Forucassie}\ and\ \citenamefont
  {Traniello}(1994)}]{Fourcassie1994}%
  \BibitemOpen
  \bibfield  {author} {\bibinfo {author} {\bibfnamefont {V.}~\bibnamefont
  {Forucassie}}\ and\ \bibinfo {author} {\bibfnamefont {J.}~\bibnamefont
  {Traniello}},\ }\href@noop {} {\bibfield  {journal} {\bibinfo  {journal}
  {Animal Behavior}\ }\textbf {\bibinfo {volume} {48}},\ \bibinfo {pages} {69}
  (\bibinfo {year} {1994})}\BibitemShut {NoStop}%
\bibitem [{\citenamefont {Freska}\ and\ \citenamefont
  {Mark}(1999)}]{Freska1999}%
  \BibitemOpen
  \bibfield  {author} {\bibinfo {author} {\bibfnamefont {C.}~\bibnamefont
  {Freska}}\ and\ \bibinfo {author} {\bibfnamefont {D.}~\bibnamefont {Mark}},\
  }\href@noop {} {\emph {\bibinfo {title} {{Spatial Information Theory}}}},\
  Vol.\ \bibinfo {volume} {1661}\ (\bibinfo  {publisher} {Springer, Berlin},\
  \bibinfo {year} {1999})\BibitemShut {NoStop}%
\bibitem [{\citenamefont {Vickerstaff}\ and\ \citenamefont
  {Cheung}(2010)}]{Vickerstaff2010}%
  \BibitemOpen
  \bibfield  {author} {\bibinfo {author} {\bibfnamefont {R.}~\bibnamefont
  {Vickerstaff}}\ and\ \bibinfo {author} {\bibfnamefont {A.}~\bibnamefont
  {Cheung}},\ }\href@noop {} {\bibfield  {journal} {\bibinfo  {journal}
  {Journal of Theoretical Biology}\ }\textbf {\bibinfo {volume} {263}},\
  \bibinfo {pages} {242} (\bibinfo {year} {2010})}\BibitemShut {NoStop}%
\bibitem [{\citenamefont {Hoffmann}(1983)}]{Hoffmann1983}%
  \BibitemOpen
  \bibfield  {author} {\bibinfo {author} {\bibfnamefont {G.}~\bibnamefont
  {Hoffmann}},\ }\href@noop {} {\bibfield  {journal} {\bibinfo  {journal}
  {Behavioral Ecology and Sociobiology}\ }\textbf {\bibinfo {volume} {13}},\
  \bibinfo {pages} {81} (\bibinfo {year} {1983})}\BibitemShut {NoStop}%
\bibitem [{\citenamefont {Wehner}\ \emph {et~al.}(2002)\citenamefont {Wehner},
  \citenamefont {Gallizzi}, \citenamefont {Frei},\ and\ \citenamefont
  {Vesely}}]{Wehner2002Cal}%
  \BibitemOpen
  \bibfield  {author} {\bibinfo {author} {\bibfnamefont {R.}~\bibnamefont
  {Wehner}}, \bibinfo {author} {\bibfnamefont {K.}~\bibnamefont {Gallizzi}},
  \bibinfo {author} {\bibfnamefont {C.}~\bibnamefont {Frei}}, \ and\ \bibinfo
  {author} {\bibfnamefont {M.}~\bibnamefont {Vesely}},\ }\href@noop {}
  {\bibfield  {journal} {\bibinfo  {journal} {J. Comp. Physiol. A}\ }\textbf
  {\bibinfo {volume} {188}},\ \bibinfo {pages} {683} (\bibinfo {year}
  {2002})}\BibitemShut {NoStop}%
\bibitem [{\citenamefont {Collett}(2010)}]{Collett2010}%
  \BibitemOpen
  \bibfield  {author} {\bibinfo {author} {\bibfnamefont {M.}~\bibnamefont
  {Collett}},\ }\href@noop {} {\bibfield  {journal} {\bibinfo  {journal}
  {Proceedings of the National Academy of Science USA}\ }\textbf {\bibinfo
  {volume} {107}},\ \bibinfo {pages} {11638} (\bibinfo {year}
  {2010})}\BibitemShut {NoStop}%
\bibitem [{\citenamefont {Capaldi}\ \emph {et~al.}(2000)\citenamefont
  {Capaldi}, \citenamefont {Smith}, \citenamefont {Fahrbach}, \citenamefont
  {Farris}, \citenamefont {Reynolds}, \citenamefont {Edwards}, \citenamefont
  {Martin}, \citenamefont {Robinson}, \citenamefont {Poppy},\ and\
  \citenamefont {Riley}}]{Capaldi2000}%
  \BibitemOpen
  \bibfield  {author} {\bibinfo {author} {\bibfnamefont {E.}~\bibnamefont
  {Capaldi}}, \bibinfo {author} {\bibfnamefont {J.}~\bibnamefont {Smith},
  \bibfnamefont {A.D. ad~Osborne}}, \bibinfo {author} {\bibfnamefont
  {S.}~\bibnamefont {Fahrbach}}, \bibinfo {author} {\bibfnamefont
  {S.}~\bibnamefont {Farris}}, \bibinfo {author} {\bibfnamefont
  {D.}~\bibnamefont {Reynolds}}, \bibinfo {author} {\bibfnamefont
  {A.}~\bibnamefont {Edwards}}, \bibinfo {author} {\bibfnamefont
  {A.}~\bibnamefont {Martin}}, \bibinfo {author} {\bibfnamefont
  {G.}~\bibnamefont {Robinson}}, \bibinfo {author} {\bibfnamefont
  {G.}~\bibnamefont {Poppy}}, \ and\ \bibinfo {author} {\bibfnamefont
  {J.}~\bibnamefont {Riley}},\ }\href@noop {} {\bibfield  {journal} {\bibinfo
  {journal} {Nature}\ }\textbf {\bibinfo {volume} {403}},\ \bibinfo {pages}
  {537} (\bibinfo {year} {2000})}\BibitemShut {NoStop}%
\bibitem [{\citenamefont {Osborne}\ \emph {et~al.}(2013)\citenamefont
  {Osborne}, \citenamefont {Smith}, \citenamefont {Clark}, \citenamefont
  {Reynolds}, \citenamefont {Barron}, \citenamefont {Lim},\ and\ \citenamefont
  {Reynolds}}]{Osborne2013}%
  \BibitemOpen
  \bibfield  {author} {\bibinfo {author} {\bibfnamefont {J.}~\bibnamefont
  {Osborne}}, \bibinfo {author} {\bibfnamefont {A.}~\bibnamefont {Smith}},
  \bibinfo {author} {\bibfnamefont {S.}~\bibnamefont {Clark}}, \bibinfo
  {author} {\bibfnamefont {D.}~\bibnamefont {Reynolds}}, \bibinfo {author}
  {\bibfnamefont {M.}~\bibnamefont {Barron}}, \bibinfo {author} {\bibfnamefont
  {K.}~\bibnamefont {Lim}}, \ and\ \bibinfo {author} {\bibfnamefont
  {A.}~\bibnamefont {Reynolds}},\ }\href@noop {} {\bibfield  {journal}
  {\bibinfo  {journal} {PLOS One}\ }\textbf {\bibinfo {volume} {8}},\ \bibinfo
  {pages} {e78681} (\bibinfo {year} {2013})}\BibitemShut {NoStop}%
\bibitem [{\citenamefont {Reynolds}\ \emph
  {et~al.}(2007{\natexlab{a}})\citenamefont {Reynolds}, \citenamefont {Smith},
  \citenamefont {Greggers}, \citenamefont {Reynolds},\ and\ \citenamefont
  {Riley}}]{Reynolds2007b}%
  \BibitemOpen
  \bibfield  {author} {\bibinfo {author} {\bibfnamefont {A.}~\bibnamefont
  {Reynolds}}, \bibinfo {author} {\bibfnamefont {A.}~\bibnamefont {Smith}},
  \bibinfo {author} {\bibfnamefont {U.}~\bibnamefont {Greggers}}, \bibinfo
  {author} {\bibfnamefont {D.}~\bibnamefont {Reynolds}}, \ and\ \bibinfo
  {author} {\bibfnamefont {J.}~\bibnamefont {Riley}},\ }\href@noop {}
  {\bibfield  {journal} {\bibinfo  {journal} {Ecology}\ }\textbf {\bibinfo
  {volume} {88}},\ \bibinfo {pages} {1955} (\bibinfo {year}
  {2007}{\natexlab{a}})}\BibitemShut {NoStop}%
\bibitem [{\citenamefont {Reynolds}\ \emph
  {et~al.}(2007{\natexlab{b}})\citenamefont {Reynolds}, \citenamefont {Smith},
  \citenamefont {Reynolds}, \citenamefont {Carreck},\ and\ \citenamefont
  {Osborne}}]{Reynolds2007}%
  \BibitemOpen
  \bibfield  {author} {\bibinfo {author} {\bibfnamefont {A.}~\bibnamefont
  {Reynolds}}, \bibinfo {author} {\bibfnamefont {A.}~\bibnamefont {Smith}},
  \bibinfo {author} {\bibfnamefont {D.}~\bibnamefont {Reynolds}}, \bibinfo
  {author} {\bibfnamefont {N.}~\bibnamefont {Carreck}}, \ and\ \bibinfo
  {author} {\bibfnamefont {J.}~\bibnamefont {Osborne}},\ }\href@noop {}
  {\bibfield  {journal} {\bibinfo  {journal} {Journal of Experimental Biology}\
  }\textbf {\bibinfo {volume} {210}},\ \bibinfo {pages} {3763} (\bibinfo {year}
  {2007}{\natexlab{b}})}\BibitemShut {NoStop}%
\bibitem [{\citenamefont {Lenz}\ \emph {et~al.}(2013)\citenamefont {Lenz},
  \citenamefont {Chechkin},\ and\ \citenamefont {Klages}}]{Klages2013}%
  \BibitemOpen
  \bibfield  {author} {\bibinfo {author} {\bibfnamefont {F.}~\bibnamefont
  {Lenz}}, \bibinfo {author} {\bibfnamefont {A.}~\bibnamefont {Chechkin}}, \
  and\ \bibinfo {author} {\bibfnamefont {R.}~\bibnamefont {Klages}},\
  }\href@noop {} {\bibfield  {journal} {\bibinfo  {journal} {PLOS One}\
  }\textbf {\bibinfo {volume} {8}},\ \bibinfo {pages} {e59036} (\bibinfo {year}
  {2013})}\BibitemShut {NoStop}%
\bibitem [{\citenamefont {Collett}(2000)}]{Collett2000}%
  \BibitemOpen
  \bibfield  {author} {\bibinfo {author} {\bibfnamefont {T.}~\bibnamefont
  {Collett}},\ }\href@noop {} {\bibfield  {journal} {\bibinfo  {journal}
  {Nature}\ }\textbf {\bibinfo {volume} {403}},\ \bibinfo {pages} {488}
  (\bibinfo {year} {2000})}\BibitemShut {NoStop}%
\bibitem [{\citenamefont {Makinson}\ \emph {et~al.}(2019)\citenamefont
  {Makinson}, \citenamefont {Woodgate}, \citenamefont {Reynolds}, \citenamefont
  {Capaldi}, \citenamefont {Clint},\ and\ \citenamefont
  {Chittka}}]{Makinson2018}%
  \BibitemOpen
  \bibfield  {author} {\bibinfo {author} {\bibfnamefont {J.}~\bibnamefont
  {Makinson}}, \bibinfo {author} {\bibfnamefont {J.}~\bibnamefont {Woodgate}},
  \bibinfo {author} {\bibfnamefont {A.}~\bibnamefont {Reynolds}}, \bibinfo
  {author} {\bibfnamefont {E.}~\bibnamefont {Capaldi}}, \bibinfo {author}
  {\bibfnamefont {J.}~\bibnamefont {Clint}}, \ and\ \bibinfo {author}
  {\bibfnamefont {L.}~\bibnamefont {Chittka}},\ }\href@noop {} {\bibfield
  {journal} {\bibinfo  {journal} {Scientific Reports}\ }\textbf {\bibinfo
  {volume} {9}},\ \bibinfo {pages} {4651} (\bibinfo {year} {2019})}\BibitemShut
  {NoStop}%
\bibitem [{\citenamefont {Mikhailov}\ and\ \citenamefont
  {Meink{\"o}hn}(1997)}]{mikhailov}%
  \BibitemOpen
  \bibfield  {author} {\bibinfo {author} {\bibfnamefont {A.}~\bibnamefont
  {Mikhailov}}\ and\ \bibinfo {author} {\bibfnamefont {D.}~\bibnamefont
  {Meink{\"o}hn}},\ }in\ \href@noop {} {\emph {\bibinfo {booktitle}
  {{Stochastic Dynamics}}}},\ \bibinfo {editor} {edited by\ \bibinfo {editor}
  {\bibfnamefont {L.}~\bibnamefont {Schimansky-Geier}}\ and\ \bibinfo {editor}
  {\bibfnamefont {T.}~\bibnamefont {P{\"o}schel}}}\ (\bibinfo  {publisher}
  {Springer Berlin Heidelberg},\ \bibinfo {address} {Berlin, Heidelberg},\
  \bibinfo {year} {1997})\ pp.\ \bibinfo {pages} {334--345}\BibitemShut
  {NoStop}%
\bibitem [{\citenamefont {N\"otel}\ \emph {et~al.}(2017)\citenamefont
  {N\"otel}, \citenamefont {Sokolov},\ and\ \citenamefont
  {Schimansky-Geier}}]{Noetel:2017}%
  \BibitemOpen
  \bibfield  {author} {\bibinfo {author} {\bibfnamefont {J.}~\bibnamefont
  {N\"otel}}, \bibinfo {author} {\bibfnamefont {I.~M.}\ \bibnamefont
  {Sokolov}}, \ and\ \bibinfo {author} {\bibfnamefont {L.}~\bibnamefont
  {Schimansky-Geier}},\ }\href
  {http://stacks.iop.org/1751-8121/50/i=3/a=034003} {\bibfield  {journal}
  {\bibinfo  {journal} {Journal of Physics A: Mathematical and Theoretical}\
  }\textbf {\bibinfo {volume} {50}},\ \bibinfo {pages} {034003} (\bibinfo
  {year} {2017})}\BibitemShut {NoStop}%
\bibitem [{\citenamefont {Stratonovich}(1967)}]{stratonovich}%
  \BibitemOpen
  \bibfield  {author} {\bibinfo {author} {\bibfnamefont {R.}~\bibnamefont
  {Stratonovich}},\ }\href@noop {} {\emph {\bibinfo {title} {{Topics in the
  Theory of Random Noise}}}},\ Vol.~\bibinfo {volume} {II}\ (\bibinfo
  {publisher} {Gordon and Breach, Science publisher},\ \bibinfo {address} {New
  York},\ \bibinfo {year} {1967})\ p.\ \bibinfo {pages} {234ff}\BibitemShut
  {NoStop}%
\bibitem [{\citenamefont {K{\"u}rsten}\ and\ \citenamefont
  {Ihle}(2017)}]{ihle}%
  \BibitemOpen
  \bibfield  {author} {\bibinfo {author} {\bibfnamefont {R.}~\bibnamefont
  {K{\"u}rsten}}\ and\ \bibinfo {author} {\bibfnamefont {T.}~\bibnamefont
  {Ihle}},\ }\href@noop {} {\bibfield  {journal} {\bibinfo  {journal} {Journal
  of Statistical Mechanics: Theory and Experiment}\ }\textbf {\bibinfo {volume}
  {2017/3}},\ \bibinfo {pages} {033202} (\bibinfo {year} {2017})}\BibitemShut
  {NoStop}%
\bibitem [{\citenamefont {Slater}(1959)}]{Slater1959}%
  \BibitemOpen
  \bibfield  {author} {\bibinfo {author} {\bibfnamefont {J.}~\bibnamefont
  {Slater}},\ }\href@noop {} {\bibfield  {journal} {\bibinfo  {journal}
  {Physical Review}\ }\textbf {\bibinfo {volume} {81}},\ \bibinfo {pages} {385}
  (\bibinfo {year} {1959})}\BibitemShut {NoStop}%
\bibitem [{\citenamefont {Balescu}(1960)}]{Balescu1960}%
  \BibitemOpen
  \bibfield  {author} {\bibinfo {author} {\bibfnamefont {R.}~\bibnamefont
  {Balescu}},\ }\href@noop {} {\bibfield  {journal} {\bibinfo  {journal} {The
  Physics of Fluids}\ }\textbf {\bibinfo {volume} {3}},\ \bibinfo {pages} {52}
  (\bibinfo {year} {1960})}\BibitemShut {NoStop}%
\bibitem [{\citenamefont {Mukamel}\ \emph {et~al.}(1978)\citenamefont
  {Mukamel}, \citenamefont {Proccacia},\ and\ \citenamefont
  {Ross}}]{Mukamel1978}%
  \BibitemOpen
  \bibfield  {author} {\bibinfo {author} {\bibfnamefont {S.}~\bibnamefont
  {Mukamel}}, \bibinfo {author} {\bibfnamefont {I.}~\bibnamefont {Proccacia}},
  \ and\ \bibinfo {author} {\bibfnamefont {J.}~\bibnamefont {Ross}},\
  }\href@noop {} {\bibfield  {journal} {\bibinfo  {journal} {The Journal of
  Chemical Physics}\ }\textbf {\bibinfo {volume} {68}},\ \bibinfo {pages}
  {1205} (\bibinfo {year} {1978})}\BibitemShut {NoStop}%
\bibitem [{\citenamefont {Stanley}(1971)}]{Stanley1971}%
  \BibitemOpen
  \bibfield  {author} {\bibinfo {author} {\bibfnamefont {H.}~\bibnamefont
  {Stanley}},\ }\href@noop {} {\emph {\bibinfo {title} {{Mean Field Theory of
  Magnetic Phase Transitions}}}}\ (\bibinfo  {publisher} {Phenomena. Oxford
  University Press},\ \bibinfo {year} {1971})\BibitemShut {NoStop}%
\bibitem [{\citenamefont {Kadanoff}(2009)}]{Kadanoff2009}%
  \BibitemOpen
  \bibfield  {author} {\bibinfo {author} {\bibfnamefont {L.~P.}\ \bibnamefont
  {Kadanoff}},\ }\href@noop {} {\bibfield  {journal} {\bibinfo  {journal}
  {Jounral of Statistical Physics}\ ,\ \bibinfo {pages} {777}} (\bibinfo {year}
  {2009})}\BibitemShut {NoStop}%
\bibitem [{\citenamefont {den Broeck}\ \emph {et~al.}(1997)\citenamefont {den
  Broeck}, \citenamefont {Parrondo}, \citenamefont {Toral},\ and\ \citenamefont
  {Kawai}}]{Vandenbroeck1997}%
  \BibitemOpen
  \bibfield  {author} {\bibinfo {author} {\bibfnamefont {C.~V.}\ \bibnamefont
  {den Broeck}}, \bibinfo {author} {\bibfnamefont {J.}~\bibnamefont
  {Parrondo}}, \bibinfo {author} {\bibfnamefont {R.}~\bibnamefont {Toral}}, \
  and\ \bibinfo {author} {\bibfnamefont {R.}~\bibnamefont {Kawai}},\
  }\href@noop {} {\bibfield  {journal} {\bibinfo  {journal} {Physical Review
  E}\ }\textbf {\bibinfo {volume} {55}},\ \bibinfo {pages} {4084} (\bibinfo
  {year} {1997})}\BibitemShut {NoStop}%
\bibitem [{\citenamefont {Sagu{\'e}s}\ \emph {et~al.}(2007)\citenamefont
  {Sagu{\'e}s}, \citenamefont {Sancho},\ and\ \citenamefont
  {García-Ojalvo}}]{Sagues2007}%
  \BibitemOpen
  \bibfield  {author} {\bibinfo {author} {\bibfnamefont {F.}~\bibnamefont
  {Sagu{\'e}s}}, \bibinfo {author} {\bibfnamefont {J.}~\bibnamefont {Sancho}},
  \ and\ \bibinfo {author} {\bibfnamefont {J.}~\bibnamefont {García-Ojalvo}},\
  }\href@noop {} {\bibfield  {journal} {\bibinfo  {journal} {Review of Modern
  Physics}\ }\textbf {\bibinfo {volume} {79}},\ \bibinfo {pages} {829}
  (\bibinfo {year} {2007})}\BibitemShut {NoStop}%
\bibitem [{\citenamefont {Strogatz}(2000)}]{Strogatz_2000}%
  \BibitemOpen
  \bibfield  {author} {\bibinfo {author} {\bibfnamefont {S.~H.}\ \bibnamefont
  {Strogatz}},\ }\href {\doibase 10.1016/S0167-2789(00)00094-4} {\bibfield
  {journal} {\bibinfo  {journal} {Physica D}\ }\textbf {\bibinfo {volume}
  {143}},\ \bibinfo {pages} {1} (\bibinfo {year} {2000})}\BibitemShut {NoStop}%
\bibitem [{\citenamefont {Bonilla}\ and\ \citenamefont
  {Trenado}(2018)}]{Bonilla2018}%
  \BibitemOpen
  \bibfield  {author} {\bibinfo {author} {\bibfnamefont {L.}~\bibnamefont
  {Bonilla}}\ and\ \bibinfo {author} {\bibfnamefont {C.}~\bibnamefont
  {Trenado}},\ }\href@noop {} {\bibfield  {journal} {\bibinfo  {journal}
  {Physical Review E}\ }\textbf {\bibinfo {volume} {98}},\ \bibinfo {pages}
  {062603} (\bibinfo {year} {2018})}\BibitemShut {NoStop}%
\bibitem [{\citenamefont {Erdmann}\ \emph {et~al.}(2005)\citenamefont
  {Erdmann}, \citenamefont {Ebeling},\ and\ \citenamefont
  {Mikhailov}}]{Erdmann2005}%
  \BibitemOpen
  \bibfield  {author} {\bibinfo {author} {\bibfnamefont {U.}~\bibnamefont
  {Erdmann}}, \bibinfo {author} {\bibfnamefont {W.}~\bibnamefont {Ebeling}}, \
  and\ \bibinfo {author} {\bibfnamefont {A.}~\bibnamefont {Mikhailov}},\
  }\href@noop {} {\bibfield  {journal} {\bibinfo  {journal} {Physical Review
  E}\ }\textbf {\bibinfo {volume} {71}},\ \bibinfo {pages} {051904} (\bibinfo
  {year} {2005})}\BibitemShut {NoStop}%
\bibitem [{\citenamefont {Enculescu}\ and\ \citenamefont
  {Stark}(2011)}]{Enculescu}%
  \BibitemOpen
  \bibfield  {author} {\bibinfo {author} {\bibfnamefont {M.}~\bibnamefont
  {Enculescu}}\ and\ \bibinfo {author} {\bibfnamefont {H.}~\bibnamefont
  {Stark}},\ }\href {\doibase 10.1103/PhysRevLett.82.209} {\bibfield  {journal}
  {\bibinfo  {journal} {Physical Review Letters}\ }\textbf {\bibinfo {volume}
  {107}},\ \bibinfo {pages} {058301} (\bibinfo {year} {2011})}\BibitemShut
  {NoStop}%
\bibitem [{\citenamefont {Attanasi}\ \emph
  {et~al.}(2014{\natexlab{a}})\citenamefont {Attanasi}, \citenamefont
  {Cavagna}, \citenamefont {Del~Castello}, \citenamefont {Giardina},
  \citenamefont {Melillo}, \citenamefont {Parisi}, \citenamefont {Pohl},
  \citenamefont {Rossaro}, \citenamefont {Shen}, \citenamefont {Silvestri},\
  and\ \citenamefont {Viale}}]{Attanasi2014}%
  \BibitemOpen
  \bibfield  {author} {\bibinfo {author} {\bibfnamefont {A.}~\bibnamefont
  {Attanasi}}, \bibinfo {author} {\bibfnamefont {A.}~\bibnamefont {Cavagna}},
  \bibinfo {author} {\bibfnamefont {L.}~\bibnamefont {Del~Castello}}, \bibinfo
  {author} {\bibfnamefont {S.}~\bibnamefont {Giardina}}, \bibinfo {author}
  {\bibfnamefont {S.}~\bibnamefont {Melillo}}, \bibinfo {author} {\bibfnamefont
  {L.}~\bibnamefont {Parisi}}, \bibinfo {author} {\bibfnamefont
  {O.}~\bibnamefont {Pohl}}, \bibinfo {author} {\bibfnamefont {B.}~\bibnamefont
  {Rossaro}}, \bibinfo {author} {\bibfnamefont {E.}~\bibnamefont {Shen}},
  \bibinfo {author} {\bibfnamefont {E.}~\bibnamefont {Silvestri}}, \ and\
  \bibinfo {author} {\bibfnamefont {M.}~\bibnamefont {Viale}},\ }\href@noop {}
  {\bibfield  {journal} {\bibinfo  {journal} {Physical Review Letters}\
  }\textbf {\bibinfo {volume} {113}},\ \bibinfo {pages} {238102} (\bibinfo
  {year} {2014}{\natexlab{a}})}\BibitemShut {NoStop}%
\bibitem [{\citenamefont {Attanasi}\ \emph
  {et~al.}(2014{\natexlab{b}})\citenamefont {Attanasi}, \citenamefont
  {Cavagna}, \citenamefont {Del~Castello}, \citenamefont {Giardina},
  \citenamefont {Melillo}, \citenamefont {Parisi}, \citenamefont {Pohl},
  \citenamefont {Rossaro}, \citenamefont {Shen}, \citenamefont {Silvestri},\
  and\ \citenamefont {Viale}}]{Attanasi2014b}%
  \BibitemOpen
  \bibfield  {author} {\bibinfo {author} {\bibfnamefont {A.}~\bibnamefont
  {Attanasi}}, \bibinfo {author} {\bibfnamefont {A.}~\bibnamefont {Cavagna}},
  \bibinfo {author} {\bibfnamefont {L.}~\bibnamefont {Del~Castello}}, \bibinfo
  {author} {\bibfnamefont {S.}~\bibnamefont {Giardina}}, \bibinfo {author}
  {\bibfnamefont {S.}~\bibnamefont {Melillo}}, \bibinfo {author} {\bibfnamefont
  {L.}~\bibnamefont {Parisi}}, \bibinfo {author} {\bibfnamefont
  {O.}~\bibnamefont {Pohl}}, \bibinfo {author} {\bibfnamefont {B.}~\bibnamefont
  {Rossaro}}, \bibinfo {author} {\bibfnamefont {E.}~\bibnamefont {Shen}},
  \bibinfo {author} {\bibfnamefont {E.}~\bibnamefont {Silvestri}}, \ and\
  \bibinfo {author} {\bibfnamefont {M.}~\bibnamefont {Viale}},\ }\href@noop {}
  {\bibfield  {journal} {\bibinfo  {journal} {PLOS Computational Biology}\
  }\textbf {\bibinfo {volume} {10}},\ \bibinfo {pages} {e1003697} (\bibinfo
  {year} {2014}{\natexlab{b}})}\BibitemShut {NoStop}%
\bibitem [{\citenamefont {Kramers}(1940)}]{Kramers}%
  \BibitemOpen
  \bibfield  {author} {\bibinfo {author} {\bibfnamefont {H.}~\bibnamefont
  {Kramers}},\ }\href {\doibase 10.1016/S0031-8914(40)90098-2} {\bibfield
  {journal} {\bibinfo  {journal} {Physica}\ }\textbf {\bibinfo {volume} {7}},\
  \bibinfo {pages} {284 } (\bibinfo {year} {1940})}\BibitemShut {NoStop}%
\bibitem [{\citenamefont {H'walisz}\ \emph {et~al.}(1989)\citenamefont
  {H'walisz}, \citenamefont {Jung}, \citenamefont {H\"anggi}, \citenamefont
  {Talkner},\ and\ \citenamefont {Schimansky-Geier}}]{Hwalisz}%
  \BibitemOpen
  \bibfield  {author} {\bibinfo {author} {\bibfnamefont {L.}~\bibnamefont
  {H'walisz}}, \bibinfo {author} {\bibfnamefont {P.}~\bibnamefont {Jung}},
  \bibinfo {author} {\bibfnamefont {P.}~\bibnamefont {H\"anggi}}, \bibinfo
  {author} {\bibfnamefont {P.}~\bibnamefont {Talkner}}, \ and\ \bibinfo
  {author} {\bibfnamefont {L.}~\bibnamefont {Schimansky-Geier}},\ }\href
  {\doibase 10.1007/BF01453798} {\bibfield  {journal} {\bibinfo  {journal}
  {Zeitschrift f\"ur Physik B Condensed Matter}\ }\textbf {\bibinfo {volume}
  {77}},\ \bibinfo {pages} {471} (\bibinfo {year} {1989})}\BibitemShut
  {NoStop}%
\bibitem [{\citenamefont {Bonilla}\ and\ \citenamefont
  {Trenado}(2019)}]{Bonilla2019}%
  \BibitemOpen
  \bibfield  {author} {\bibinfo {author} {\bibfnamefont {L.}~\bibnamefont
  {Bonilla}}\ and\ \bibinfo {author} {\bibfnamefont {C.}~\bibnamefont
  {Trenado}},\ }\href@noop {} {\bibfield  {journal} {\bibinfo  {journal}
  {Physical Review E}\ }\textbf {\bibinfo {volume} {99}},\ \bibinfo {pages}
  {012612} (\bibinfo {year} {2019})}\BibitemShut {NoStop}%
\end{thebibliography}
%

\end{document}